\documentclass[a4paper,11pt]{article}
\pdfoutput=1 

\usepackage{jheppub} 

\usepackage[T1]{fontenc} 

\usepackage{amsmath,amssymb,epsfig,ulem,color,slashed,multirow,wasysym}

\allowdisplaybreaks  

\def \beq{\begin{equation}}
\def \eeq{\end{equation}}
\def \bea{\begin{eqnarray}}
\def \eea{\end{eqnarray}}

\title{\boldmath  Simplified dark matter top-quark \\ interactions at the LHC}

\author[a,b]{Ulrich Haisch}

\author[a]{and Emanuele Re}

\affiliation[a]{Rudolf Peierls Centre for Theoretical Physics,
    University of Oxford, \\ OX1 3NP Oxford, United Kingdom}
    
\affiliation[b]{CERN, Theory Division, \\ CH-1211 Geneva 23, Switzerland}

\emailAdd{u.haisch1@physics.ox.ac.uk}
\emailAdd{Emanuele.Re@physics.ox.ac.uk}

\abstract{Stringent limits on the interactions between dark matter~(DM) and the standard model~(SM) can be set by studying how initial-state or final-state particles  recoil against missing transverse energy~$(\slashed{E}_T)$. In this work, we improve, extend and correct  LHC constraints on the interactions between DM and top quarks that are mediated by the exchange of spin-$0$ $s$-channel resonances. A comparison of the LHC run-1 sensitivity of the two main search channels is presented,  which shows that mono-jet searches are typically more restrictive than the $\slashed{E}_T + \bar t t$ searches. We furthermore explore the reach of the~$14 \, {\rm TeV}$~LHC.  The collider constraints are compared to the restrictions arising from  direct and indirect detection  as well as the DM relic abundance, and we also reflect on effective field theory~(EFT) interpretations of the LHC exclusions.}

\preprint{OUTP-15-02P \\ \hspace*{\fill} CERN-PH-TH-2015-201}

\begin{document} 

\maketitle

\flushbottom

\section{Introduction}
\label{sec:introduction}

Searches for $\slashed{E}_T$ signatures represent one of the main focus of the ATLAS and CMS collaborations in their hunt for physics beyond the SM, because of their possible connection to DM. These searches can be classified based on the type of SM particles which recoils against the DM pair. In LHC run-1, ATLAS and CMS have examined a variety  of $\slashed{E}_T$ signatures involving  jets of hadrons, gauge bosons, top and bottom quarks as well as the Higgs boson in the final state (for a recent review of the experimental status see for instance~\cite{Askew:2014kqa}). 

The existing LHC studies are in most cases performed in the context of an  EFT which describes the physics of heavy particles mediating the interactions between DM and SM fields, assuming that the mediators are heavy enough so that they can be removed as active degrees of freedom.  It has however been realised early on \cite{Bai:2010hh,Fox:2011pm}, that an EFT description of~$\slashed{E}_T$ signatures is deemed to fail if the masses of the mediators are within kinematic reach, which can  cause the sensitivity of the LHC searches to change significantly. In order to correctly account for both off-shell and on-shell effects in DM pair production different   simplified models have been put forward, in which the contact interactions present in the EFT are resolved into single-particle $s$-channel or $t$-channel exchanges. By specifying the spin and gauge quantum numbers of DM and the mediators and requiring the interactions to be minimal flavour violating~(MFV)~\cite{D'Ambrosio:2002ex}, the parameter spaces remain low dimensional,  which in turn allows for a simple translation of bounds between experiments and theories (cf.~\cite{Abdallah:2014hon} for a  comprehensive overview on simplified DM models).

In the present work we focus on the DM pair production from quark or gluon initial states, where the production proceeds via the exchange of spin-$0$ $s$-channel mediators. Under the assumption of  MFV, the most relevant DM-SM couplings in this class of simplified models are those that involve  top quarks.  Two main strategies have been  exploited at the LHC to search for scalar and pseudo-scalar interactions of this type. The first possibility consists in looking for a $pp \to {\slashed E}_T + j$ signal \cite{Haisch:2012kf}, where the mediators that pair produce DM  are radiated from top-quark loops, while the second possibility relies on detecting the top-quark decay products that arise from the tree-level  reaction $pp \to {\slashed E}_T + \bar t t$ \cite{Cheung:2010zf, Lin:2013sca}. Further progress in characterising the LHC signatures associated to DM top-quark interactions has been made in \cite{Haisch:2013fla, Fox:2012ru, Buckley:2014fba, Harris:2014hgav1} for the mono-jet signal and in the articles \cite{Artoni:2013zba, CMS:2014mxa, CMS:2014pvf, Aad:2014vea, Buckley:2014fba} for final states involving top-quark pairs.  

Our  goal is  to  refine, to extend  and to correct existing LHC constraints. As a first benchmark, we examine the constraints on the EFT that stem from LHC run-1 mono-jet searches as well as the different $\slashed{E}_T +\bar t t$ channels with  di-leptonic~\cite{CMS:2014mxa}, single-leptonic~\cite{CMS:2014pvf, Aad:2014vea} and fully hadronic~\cite{Aad:2014vea} top-quark decays. This exercise serves not only  as an independent cross-check of the $\slashed{E}_T +\bar t t$  analyses performed by ATLAS and CMS, but also as a validation of our Monte Carlo (MC) chain. We find that at present the strongest constraints that the LHC can place on effective DM top-quark interactions arise from mono-jet searches, and that this strategy is expected to remain  the most powerful one also at future LHC runs. Our EFT results at both $8  \, {\rm TeV}$ and $14 \, {\rm TeV}$ are compared to the exact exclusions limits in the simplified DM models. This allows us to determine under which circumstances an EFT interpretation of the collider bounds is justified. By scanning the parameter space of the simplified models, we furthermore show that in both the scalar and the pseudo-scalar case the ATLAS and CMS searches cannot presently exclude parameters arising from purely weakly-coupled theories. As far as a comparison is possible, this finding agrees qualitatively with the conclusions drawn in~\cite{Buckley:2014fba,Harris:2014hgav1}. Our analysis  reveals in addition that the $\slashed{E}_T + j$ searches generically exclude more parameter space than the $\slashed{E}_T +\bar t t$ searches. We finally discuss the interplay of the various DM searches, including direct and indirect detection as well as the constraints  from the observed relic abundance. 

The outline of this paper is as follows: in Section~\ref{sec:preliminaries} we discuss the structure of the simplified DM models and  the corresponding EFTs. This section  contains in addition the formulas needed to calculate the DM-nucleon scattering cross sections as well as the DM relic density. The results of our phenomenological analyses of the DM top-quark interactions are presented in Section~\ref{sec:phenomenology}. In this section we discuss the present and possible future bounds that result from the different LHC search strategies, comparing the obtained limits to those arising from direct and indirect detection as well as the requirement not to overclose the Universe. In this context, special attention is payed to the differences in the results and the conclusions drawn when the calculations are performed in the simplified model framework or the EFT.  We conclude and provide an outlook in Section~\ref{sec:conclusions}. Additional material that might be of particular interest for the practitioner is relegated to Appendix~\ref{app:effectiveMC}. 

\section{Preliminaries}
\label{sec:preliminaries}

In the following we introduce the simplified models for the DM-SM interactions that results from the exchange of a colourless scalar or pseudo-scalar mediator (Section~\ref{sec:interactions}) and discuss the relevant operators in the corresponding EFT~(Section~\ref{sec:eft}). To make our article self-contained we furthermore collect the formulas necessary to calculate the DM-nucleon scattering cross sections  and the DM relic abundance~(Section~\ref{sec:directdetection}). 

\subsection{\boldmath Simplified models}
\label{sec:interactions}

The relevant interactions between DM and quarks involving the exchange of a colourless scalar~($S$) or pseudo-scalar~($P$) mediator are parameterised as follows 
\begin{equation} \label{eq:LSP}
\mathcal{L} \, \supset \, g_{\rm DM}^S \left(  \bar \chi   \chi  \right ) S +  g^S_{\rm SM} \sum_q  \, \frac{m_q}{v} \left(  \bar q q  \right ) S  +  i g_{\rm DM}^P \left(  \bar \chi  \gamma_5 \chi  \right ) P + i g^P_{\rm SM} \sum_q  \, \frac{m_q}{v}  \left(  \bar q   \gamma_5 q \right )P \,,
\end{equation}
where the sum is over all quarks and $v \simeq 246 \, {\rm GeV}$ denotes the Higgs vacuum expectation value. In writing (\ref{eq:LSP}) we have assumed  that the couplings of the mediators to quarks are proportional to the associated SM Yukawa couplings. This is motivated by the hypothesis of MFV, which curbs the size of dangerous flavour-changing neutral current processes \cite{D'Ambrosio:2002ex} and  automatically leads to a stable DM candidate~\cite{Batell:2011tc}. Notice that the MFV hypothesis  allows the mediator quark couplings to be scaled by separate factors  $g^{S,P}_{d}$  and $g^{S,P}_{u}$ for down-type quarks and up-type quarks, respectively. For simplicity, we have ignored this possibility when writing ${\cal L}$ and choose the same scaling factors $g^{S,P}_{d}= g^{S,P}_{u} = g^{S,P}_{\rm SM}$ for all quarks. While the DM particle $\chi$ in (\ref{eq:LSP}) is understood to be a Dirac fermion, extending our discussion to Majorana DM or the case of a complex/real scalar is straightforward~\cite{Haisch:2012kf}. In order to avoid the severe experimental bounds from the electric dipole moment of the neutron~(cf.~\cite{UH}), we take the spin-0 mediators $S, P$ to be CP eigenstates and in addition assume that the couplings $g_{\rm DM}^{S,P}$ and $g^{S,P}_{\rm SM}$ are all real. 

Further constraints on our simplified DM models can in principle arise  from existing and future LHC resonance searches in $\bar t t$ final states. Including the one-loop process $gg \to S,P \to \bar t t$, one finds \cite{Haisch:2013fla} that for weakly-coupled models the total $\bar t t$ cross section is changed by only~${\cal O} (1\%)$.  Such small effects are likely to remain unnoticed given that the theoretical uncertainty on the total $\bar t t$ cross section is at the level of 5\% at the LHC~\cite{Czakon:2013goa}. A di-jet signal arises in the simplified models (\ref{eq:LSP}) first at the two-loop level via $gg \to S,P \to gg$. The strong loop suppression renders the contributions of $S,P$ exchange to di-jet production  unobservable at the LHC~\cite{Haisch:2013fla}.  Since the SM portion of the Lagrangian~(\ref{eq:LSP}) is not a electroweak singlet additional restrictions also stem from the fact that  the mediators~$S,P$  necessarily have  portal couplings involving the Higgs field. The resulting modifications in Higgs phenomenology are, however, model dependent and we do not study them in what follows. If the mediators have weak-scale masses, the couplings in (\ref{eq:LSP}) to light quarks are, to the best of our knowledge, unconstrained by direct and indirect collider searches that do not involve large amounts of $\slashed{E}_T$. For $M_{S,P} \lesssim 10 \, {\rm GeV}$ important constraints can however arise from quark flavour physics \cite{Dolan:2014ska}.

The signal strength in DM pair production does not only depend on the couplings $g_{\rm DM}^{S,P}$ and $g^{S,P}_{\rm SM}$ and masses $m_\chi$ and $M_{S,P}$, but also on the total decay widths $\Gamma_{S,P}$ of the mediators~$S,P$. In the case of the scalar mediator, one finds the following results for the partial decay widths (see~e.g.~\cite{Haisch:2013fla})
\begin{equation} \label{eq:partial}
\begin{split}
\Gamma \hspace{0.25mm} (S \to \bar \chi \chi) & = \left ( g_{\rm DM}^S \right )^2 \frac{M_S}{8 \pi }   \, \bigg ( 1 - \frac{4 m_\chi^2}{M_S^2} \bigg )^{3/2} \, \theta \hspace{0.25mm} (M_S - 2 m_\chi ) \,, \\[2mm]
\Gamma \hspace{0.25mm} (S \to \bar q q) & = \left ( g_{\rm SM}^S \right)^2 \, \frac{3 m_q^2 \hspace{0.25mm} M_S}{8 \pi v^2 }  \, \bigg ( 1 - \frac{4 m_q^2}{M_S^2} \bigg )^{3/2} \, \theta  \hspace{0.25mm} (M_S - 2 m_q )\,, \\[1mm]
\Gamma \hspace{0.25mm} (S \to gg) & = \left ( g_{\rm SM}^S \right)^2 \frac{\alpha_s^2}{2 \pi^3 v^2 M_S} \, \left |  \sum_q m_q^2 \, F_S \left ( \frac{4 m_q^2}{M_S^2} \right ) \right |^2 \,,
\end{split}
\end{equation}
where $\theta  \hspace{0.25mm} (x)$ denotes the Heaviside step function defined by $\theta  \hspace{0.25mm} (x) = 0$ for $x < 0$ and $\theta  \hspace{0.25mm} (x) = 1$ for $x \geq 0$, while 
\begin{equation} \label{eq:FS}
F_S (x) = 1 + (1 - x) \arctan^2 \left ( \frac{1}{\sqrt{x-1}} \right )\,.
\end{equation}
The analogue expressions for the pseudo-scalar mediator are obtained from (\ref{eq:partial}) by the replacements $S \to P$ and  $3/2 \to 1/2$ in the exponents, and the relevant form factor reads  
\begin{equation} \label{eq:FP}
F_P (x) =  \arctan^2 \left ( \frac{1}{\sqrt{x-1}} \right ) \,.
\end{equation}
At the loop level the mediators can decay not only to gluons but also to pairs of photons and other final states if these are kinematical accessible. The decay rates  $\Gamma  \hspace{0.25mm} (S \to gg)$ and $\Gamma  \hspace{0.25mm} (P \to gg)$ are however always larger than the other loop-induced partial widths, and in consequence the total decay widths $\Gamma_{S}$ and $\Gamma_{P}$ are well approximated by the corresponding sum of the individual partial decay widths involving DM, quark or gluon pairs. Notice finally that if $M_{S,P} > 2 m_t$ and $g_{\rm SM}^{S,P} \gtrsim g_{\rm DM}^{S,P}$, the total width of $S,P$ is dominated by the partial widths to top quarks due their large mass or Yukawa coupling.

\subsection{EFT description}
\label{sec:eft}

\begin{figure}[!t]
\begin{center}
\includegraphics[width=0.65\textwidth]{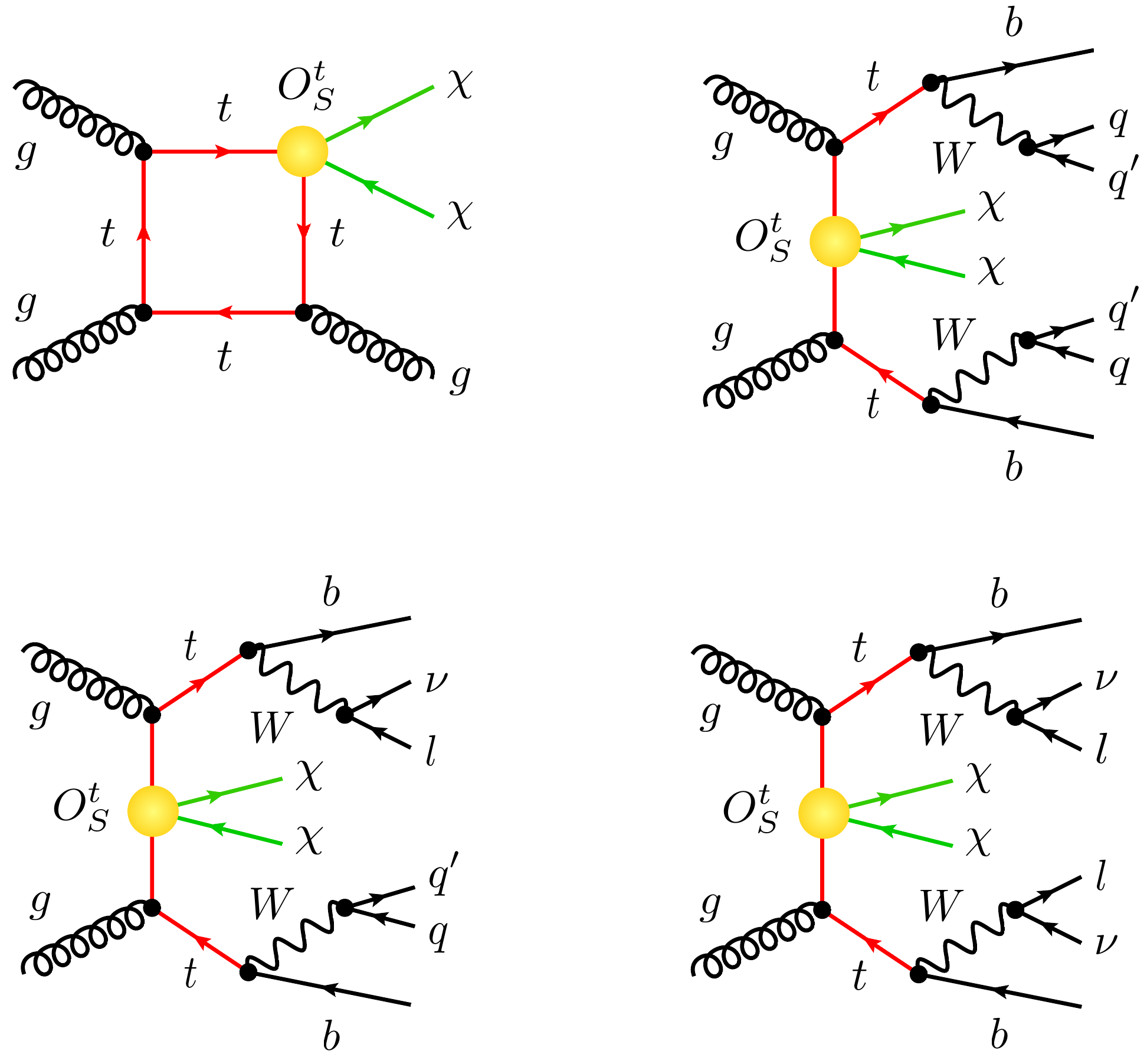} 
\vspace{2mm}
\caption{\label{fig:diagrams}  Examples of Feynman diagrams with an insertion of  $O_S^t$ that contribute to a $\slashed{E}_T + j$~(top left), $\slashed{E}_T +\bar t t \, (\to  2j \bar b b)$~(top right), $\slashed{E}_T +\bar t t \, (\to  jbl \nu)$~(bottom left)  or a $\slashed{E}_T +\bar t t \, (\to \bar bl^-\bar \nu b l^+ \nu)$~(bottom right) signal. The operator insertions are indicated by  yellow blobs, while the regular SM vertices are represented by black dots. }
\end{center}
\end{figure}

If the mediator masses $M_{S,P}$ are large compared to the other scales involved in a given process, one can describe the underlying partonic reaction by means of an EFT. Integrating out the scalar and the pseudo-scalar mediator then gives rise to 
\begin{equation} \label{eq:OSP}
O_S^q = \frac{m_q}{\Lambda^3_S}\, \bar{\chi} \chi  \, \bar{q} q  \, , \qquad 
O_P^q = \frac{m_q}{\Lambda^3_P} \, \bar{\chi} \gamma_ 5 \chi \, \bar{q} \gamma_5 q  \, ,
\end{equation}
at tree level as well as contact terms consisting of four DM or quark fields. In the case of the scalar operators $O_S^q$ the suppression scale $\Lambda_S$ is related to the mediator mass~$M_S$ and the fundamental couplings $g_{\rm DM}^S$ and $g_{\rm SM}^S$ by
\begin{equation} \label{eq:Lambda}
\Lambda_S = \left ( \frac{v M_S^2}{g_{\rm SM}^S \hspace{0.25mm} g_{\rm DM}^S} \right )^{1/3} \,,
\end{equation}
and an analogous expression with $S \to P$ holds for the pseudo-scalar operators $O_P^q$. 

Integrating out the top quark generates an effective interaction between DM and gluons. At the one-loop level, one obtains 
\begin{equation} \label{eq:OGG}
O_G = \frac{\alpha_s}{12 \pi \hspace{0.25mm} \Lambda_S^3}  \, \bar \chi \chi \, G_{\mu\nu}^a G^{a, \mu \nu} \,, \qquad 
O_{\widetilde G}  = \frac{\alpha_s}{8 \pi \hspace{0.25mm} \Lambda_P^3} \, \bar \chi \gamma_5 \chi \, G_{\mu\nu}^a \widetilde G^{a, \mu \nu}  \,, 
\end{equation}
by employing the Shifman-Vainshtein-Zakharov relations \cite{Shifman:1978zn}. Here $G_{\mu \nu}^a$ denotes the gluon field strength tensor and $\widetilde G^{a, \mu \nu} = 1/2 \hspace{0.5mm} \epsilon^{\mu\nu\lambda\rho} \hspace{0.25mm} G_{\lambda\rho}^a$ its dual. At the bottom-quark threshold and the charm-quark threshold one has to integrate out the corresponding heavy quark by again applying (\ref{eq:OGG}). Note that this matching procedure is crucial to obtain the correct DM-nucleon scattering cross section associated with effective scalar DM-quark interactions. 

Constraints from mono-jet searches on the scalar and pseudo-scalar DM-quark interactions $O_{S,P}^q$ involving the light quark flavours  as well as the gluonic operators $O_{G}$ have been discussed in detail in \cite{Fox:2012ru, Haisch:2013ata}, while analyses of the bounds on $O_{S,P}^b$ arising from $\slashed{E}_T + \bar b b$ final states have been carried out in \cite{Cheung:2010zf, Lin:2013sca, Artoni:2013zba, Aad:2014vea}. We  instead focus on the effective interactions~$O_{S,P}^t$ containing top quarks. In Figure~\ref{fig:diagrams} we show representative graphs with an insertion of $O_S^t$ corresponding to the different dedicated  search strategies that have so far been exploited to constrain DM top-quark interactions at the LHC. From top left to bottom right these are mono-jet searches \cite{Haisch:2012kf,  Fox:2012ru, Buckley:2014fba,  Harris:2014hgav1},  $\slashed{E}_T +\bar t t$ with fully hadronic top-quark decays~\cite{Aad:2014vea}, $\slashed{E}_T +\bar t t$ where one top quark decays hadronically and the other one semi-leptonically~\cite{Lin:2013sca, CMS:2014pvf, Aad:2014vea, Buckley:2014fba} and  $\slashed{E}_T +\bar t t$ with di-lepton final states \cite{CMS:2014mxa, Buckley:2014fba}. Future prospects and opportunities of the searches for DM heavy-quark interactions have been discussed in~\cite{Haisch:2013fla, Artoni:2013zba}. 

\subsection{DM-nucleon scattering and relic density}
\label{sec:directdetection}

In the case of the scalar operators $O_S^q$ the cross section for elastic Dirac scattering on a nucleon is spin-independent (SI) and given by 
\beq \label{eq:sigmaSIN}
\left ( \sigma_{\rm SI}^N \right )_S \simeq \frac{m_{\rm red}^2 \hspace{0.25mm} m_N^2 \hspace{0.25mm} f_N^2}{\pi \Lambda_S^6}  \,,
\eeq
where $m_{\rm red} = m_\chi m_N/(m_\chi + m_N)$ denotes the reduced mass of the DM-nucleon system, $m_N \simeq 0.939 \, {\rm GeV}$ is the average nucleon mass and  $f_N \simeq 0.30$~(see \cite{Crivellin:2013ipa} for a recent evaluation) is the effective DM-nucleon coupling. For the pseudo-scalar operators  $O_P^q$, on the other hand, the DM direct detection cross section is spin-dependent  and momentum-suppressed by $q^4/m_N^4$.  Existing direct detection experiments are hence not sensitive to effective pseudo-scalar DM-quark interactions.  

In the presence of the effective interactions (\ref{eq:OSP}) two different annihilation channels contribute to the total annihilation cross section. Tree-level annihilation into quarks will be dominant for $m_\chi > m_t$, while annihilation into gluons via heavy-quark loops can give a relevant contribution for lower DM masses.   Performing an expansion in the DM velocity~$v_\chi$, the total annihilation cross section for $O_S^q$ and $O_P^q$ take the form $\left ( \sigma \hspace{0.25mm} v_\chi \right)_S = b_S \hspace{0.25mm} v_\chi^2 + {\cal O} (v_\chi^4)$ and $\left ( \sigma \hspace{0.25mm} v_\chi \right)_P = a_P + {\cal O} (v_\chi^2)$ where (see for instance \cite{Haisch:2012kf})
\beq \label{eq:annihilation}
\begin{split}
b_{S} & = \sum_q 
\frac{3  m_q^2  \hspace{0.25mm} m_\chi^2}{8 \pi \Lambda^6_S}   \left( 1 - \frac{m_q^2}{m_\chi^2} \right)^{3/2} \, \theta (m_\chi - m_q) + \frac{\alpha_s^2}{8 \pi^3 \Lambda^6_S} \, \left | \sum_q m_q^2 \, F_S \left ( \frac{m_q^2}{m_\chi^2} \right ) \right |^2  \, , \\[2mm]
a_{P} & = \sum_q 
\frac{3  m_q^2  \hspace{0.25mm} m_\chi^2}{2 \pi \Lambda^6_P}   \left( 1 - \frac{m_q^2}{m_\chi^2} \right)^{1/2} \, \theta (m_\chi - m_q) + \frac{\alpha_s^2}{2 \pi^3 \Lambda^6_P} \, \left | \sum_q m_q^2 \, F_P \left ( \frac{m_q^2}{m_\chi^2} \right ) \right |^2  \, ,
\end{split}
\eeq
and the analytic expressions for the form factors $F_S(x)$ and $F_P(x)$ have already been given in~(\ref{eq:FS}) and~(\ref{eq:FP}), respectively. Note that  the total annihilation cross section associated to~$O_S^q$ is $p$-wave suppressed, while DM annihilation proceeds via $s$-wave for $O_P^q$. 

Since $\gamma$-rays are an unavoidable product of hadronisation,  it follows from (\ref{eq:annihilation}) that for pseudo-scalar interactions  indirect detection experiments can provide relevant constraints on the parameter space.   The corresponding velocity-averaged DM annihilation cross section is given by $\left \langle \sigma \hspace{0.25mm} v_\chi \right \rangle_P \simeq a_P/2$. Another indirect constraint results from the requirement not to overclose the Universe. In terms of the coefficients (\ref{eq:annihilation}) and the observed DM abundance $\Omega_{\chi} h^2 \simeq 0.11$~\cite{Hinshaw:2012aka}, the predicted DM relic density can then be expressed as~\cite{Kolb:1990vq}
\begin{equation} \label{eq:Omegah2}
\left (\Omega_{\chi} h^2 \right )_{S} \simeq \frac{3.2 \cdot 10^{-8} \, \mathrm{GeV}^{-2}}{b_{S}} \; \Omega_{\chi} h^2 \,, \qquad 
\left (\Omega_{\chi} h^2 \right )_{P} \simeq  \frac{3.8 \cdot 10^{-9} \, \mathrm{GeV}^{-2}}{a_{P}} \; \Omega_{\chi} h^2 \,.
\end{equation}

\section{Phenomenology}
\label{sec:phenomenology}

In this section we study in detail the phenomenology of the  DM top-quarks interactions induced by spin-$0$ $s$-channel exchange. After describing the main features of our MC simulations of the individual $\slashed{E}_T $ signals (Section~\ref{sec:MC}), we turn to the EFT and discuss the present (Section~\ref{sec:EFTpresent}) and the possible future bounds (Section~\ref{sec:EFTfuture}) on the suppression scales $\Lambda_{S,P}$ that result from the different search channels. We then investigate the  accuracy of the heavy top-quark approximation for the case of the mono-jet cross sections~(Section~\ref{sec:EFTwithouttop}).  The current  (Section~\ref{sec:fullpresent}) and projected (Section~\ref{sec:fullfuture}) limits on the parameters space of the simplified models are examined subsequently. In this context,  we discuss the complementarity and the interplay of the various DM search strategies, including the constraints from direct detection, DM-induced $\gamma$-ray emission from dwarf spheroidal satellite galaxies of the Milky Way and the relic abundance, and assess the quality of  the EFT interpretations of  $\slashed{E}_T$ searches at the LHC. 

\subsection{MC simulations}
\label{sec:MC}

Our predictions for the mono-jet cross section are obtained using the  {\tt POWHEG~BOX} \cite{Alioli:2010xd} and include leading order (LO) fixed-order contributions,  parton-shower  effects and hadronisation corrections~(LOPS). The needed partonic one-loop amplitudes are taken from the~{\tt MCFM}~\cite{MCFM} implementation of the process $pp \to H/A + j \to \tau^+ \tau^- + j$, which is based on the analytical results of \cite{Ellis:1987xu} for the scalar Higgs case~($H)$ and \cite{Spira:1995rr} for the pseudo-scalar Higgs case~$(A)$. Our new MC implementation has been validated by calculating the partonic mono-jet cross sections both in the context of  simplified models  and the EFT, finding perfect agreement with existing numerical results~\cite{Haisch:2012kf, Fox:2012ru}. To determine the cross sections for the different $\slashed{E}_T  + \bar t t$ signals, we have implemented the Lagrangian densities~(\ref{eq:LSP})   in~{\tt FeynRules~2}~\cite{Alloul:2013bka}, generating a~{\tt UFO} output \cite{Degrande:2011ua}. The actual event generation is performed at LO with~{\tt MadGraph~5}~\cite{Alwall:2011uj}. Our MC chain has again been successfully validated against the results of previous studies~\cite{Lin:2013sca, Artoni:2013zba, Aad:2014vea, CMS:2014pvf}.  Parton-shower (PS) effects and hadronisation corrections have in all cases been included by means of {\tt PYTHIA~6}~\cite{Sjostrand:2006za} and jets reconstructed using the anti-$k_t$ cluster algorithm~\cite{Cacciari:2008gp} implemented in~{\tt FastJet~3}~\cite{Cacciari:2011ma}. A detector simulation has not been performed, since even without it we  are able to reproduce the  relevant $\slashed{E}_T $ search results of  ATLAS and CMS within errors.

In contrast to the recent theoretical analysis \cite{Buckley:2014fba} we do not multiply our LOPS predictions by a $K$ factor  to mimic the impact of next-to-leading order (NLO) corrections. In the case of the mono-jet signal this is motivated by the observation that the infinite top-quark mass limit is a bad approximation if the $p_T$ cut on the jet is large and/or DM is heavy~\cite{Haisch:2012kf}. However, only in the case of~$m_t \to \infty$ are the NLO corrections to $pp \to H/A + j$ production known (see~e.g.~\cite{Ravindran:2002dc,Field:2002pb}), while an exact ${\cal O} (\alpha_s^4)$ calculation with  resolved top-quark loops is at the moment unavailable. Until such a computation is at hand, we believe it is more conservative not to take $K \simeq 1.6$  from  $pp \to H/A + j$ production and to apply it in  the mono-jet cross section computation. In the case of $pp \to H/A + \bar t t$ production, on the other hand, the exact top-quark mass dependence  is known at ${\cal O} (\alpha_s^3)$ already for some time~(cf.~\cite{Beenakker:2002nc, Dawson:2002tg, Frederix:2011zi}), and NLO effects turn out to be small, leading to $K \simeq 1.2$. Given that the NLO effects are not flat over the entire phase space and that the experimental cuts imposed in $pp \to H/A + \bar t t$  and $pp \to \slashed{E}_T + \bar t t$ are not identical, we again  prefer to be safe and not to include a $K$ factor  in our results for the $\slashed{E}_T  + \bar t t$ cross sections. 

The predictions for all $\slashed{E}_T $ signals are obtained using {\tt MSTW2008LO}  parton distribution functions~\cite{Martin:2009iq} and the corresponding reference value for the strong coupling constant. In the case of mono-jet production, we define $\mu =  \mu_R = \mu_F = \xi H_T/2$ and evaluate this scale on an event-by-event basis. Here $\mu_R$ and $\mu_F$ denotes the renormalisation scale and factorisation scale, respectively, and $H_T = \sqrt{m_{\bar \chi \chi}^2 + p_{T,j_1}^2} +  p_{T,j_1}$. The invariant mass of the DM pair is denoted by  $m_{\bar \chi \chi}$  and $p_{T,j_1}$ corresponds to the transverse momentum of the hardest jet. In the case of the $\slashed{E}_T + \bar t t$ processes, we have instead employed the dynamical scale $\mu =  \mu_R = \mu_F = \xi \left ( m_t + m_{\bar \chi \chi}/2 \right)$.  In order to assess the theoretical uncertainties that plague the calculated cross sections, we study the scale ambiguities by varying the parameter~$\xi$ in the standard range $[1/2, 2]$.  Numerically, we find that the predictions for the mono-jet  cross sections  calculated in this way vary in the ballpark of~$\pm 40\%$, while in the case of the $\slashed{E}_T + \bar t t$ processes, slightly smaller variations of around~$\pm 35\%$ are obtained. 

\subsection{Status of EFT limits}
\label{sec:EFTpresent}

In the following we list the various cuts and the values of the fiducial cross section ($\sigma_{\rm fid}$) of each individual~$\slashed{E}_T$ channel. This information will then be used  to set limits on the suppression scales $\Lambda_{S,P}$ that appear in the effective interactions (\ref{eq:OSP}) involving top quarks. 

\subsubsection*{\boldmath Mono-jet channel}

In order to derive the most stringent constraints from existing $\slashed{E}_T + j$ searches, we employ the latest CMS results \cite{Khachatryan:2014rra}, which make use of $19.7 \, {\rm  fb}^{-1}$ of~$8 \, {\rm TeV}$ data. The relevant selection cuts are 
\begin{equation} \label{eq:monojetpresent}
p_{T,j_1}  > 110 \, {\rm GeV} \,,  \quad |\eta_{j_1}| < 2.4 \,, \quad 
p_{T,j_2}  > 30 \, {\rm GeV} \,, \quad \; \, |\eta_{j_2}| < 4.5  \,,  \quad 
 \Delta \phi _{j_1j_2} < 2.5 \,,
\end{equation}
where $ \Delta \phi_{j_1j_2}$ is the azimuthal separation of the two leading jets, which are reconstructed using a radius parameter of~$R = 0.5$.  Another important selection criterion is the imposed jet veto \cite{Haisch:2013ata}, which rejects  events  if they contain a tertiary  jet with $p_{T,j_3} > 30 \, {\rm GeV}$ and $|\eta_{j_3}| < 4.5$. The~CMS measurement is performed in seven distinct~$\slashed{E}_T $ regions, and we find  that in the case of the operators $O_{S,P}^t$ the highest sensitivity is obtained for~$\slashed{E}_T  > 450 \, {\rm GeV}$. The corresponding~95\%~confidence level (CL) limit on the fiducial cross section reads
\begin{equation}  \label{eq:monofidpresent}
\sigma_{\rm fid} (pp \to \slashed{E}_T +  j ) < 7.8 \, {\rm fb} \,.
\end{equation}

\subsubsection*{\boldmath Fully hadronic $\slashed{E}_T + \bar t t$ channel}

The recent ATLAS search \cite{Aad:2014vea} looks for a $\slashed{E}_T + \bar t t  \, (\to  2j \bar b b)$ signal.  In this analysis based on $20.3 \, {\rm fb}^{-1}$ of $8 \, {\rm TeV}$ data, jets are clustered with $R = 0.4$.  To pass the trigger either five jets with $p_{T,j} > 55 \, {\rm GeV}$ or four jets with $p_{T,j} > 45 \, {\rm GeV}$ one of which is identified as a bottom-quark ($b$) jet are required. Events are only selected if they have at least five reconstructed jets, out  of which  two or more are $b$-tagged, and they  fulfil 
\beq \label{eq:hadroniccuts}
\slashed{E}_T > 200 \, {\rm GeV} \,, \qquad  |\eta_{j}| < 2.5 \,, \qquad \Delta \phi_{b_1 \slashed{E}_T} > 1.6 \,. 
\eeq
Here $b_1$ denotes the $b$-jet with the highest transverse momentum. The $b$-jet tagging efficiency is taken to be 70\% here and in what follows. Based on this selection requirements, the ATLAS collaboration is able to set the following 95\% CL limit 
\begin{equation}  \label{eq:hadronicpresent}
\sigma_{\rm fid} (pp \to \slashed{E}_T +  2 j \bar b b) <  \; 2.0 \, {\rm fb} \,.
\end{equation}

\subsubsection*{\boldmath Single-lepton $\slashed{E}_T +\bar t t$ channel}

For what concerns the $\slashed{E}_T +\bar t t \, (\to  jbl \nu)$ mode, we  again rely on the ATLAS results~\cite{Aad:2014vea} (see also \cite{Aad:2014kra}), since  this search turns out to be slightly more constraining than the dedicated CMS analysis~\cite{CMS:2014pvf}. As we have explicitly verified, comparable  limits on the suppression scales~$\Lambda_{S,P}$ can also be obtained by recasting the searches~\cite{ATLAS:2012maq,Chatrchyan:2013xna} for top-squark pair production in the single-lepton final state.  The trigger employed in \cite{Aad:2014vea} requires exactly one  lepton~($l = e,\mu$) with $p_{T,l} > 25 \, {\rm GeV}$ and $|\eta_l|<2.5$ as well as  four or more jets, where one jet is $b$-tagged and all satisfy~$|\eta_j|<2.5$. Events are selected, if they pass the cuts 
\beq \label{eq:singleleptoncutspresent}
\begin{split}
 \slashed{E}_T > 270 \,{\rm GeV} \,, \qquad p_{T,j_1} > 80 \, {\rm GeV} \,, \qquad p_{T,j_2} > 70 \, {\rm GeV} \,, \qquad  p_{T,j_3} > 50 \, {\rm GeV} \,, \\[2mm]  p_{T,j_4} > 25 \, {\rm GeV} \,, \qquad p_{T,b_1} > 60 \, {\rm GeV} \,, \qquad m_{jjj} < 360 \, {\rm GeV} \,, \qquad \Delta \phi_{f\slashed{E}_T} > 0.6 \,. \hspace{0.5mm}
\end{split}
\eeq
Here $m_{jjj}$ is the three-jet invariant mass \cite{Aad:2014kra} and $f=l,j_1,j_2$. Furthermore, the angular  separation between the lepton and the leading jet ($b$-jet) has to satisfy $\Delta R_{lj_1} < 2.75$ ($\Delta R_{l b_1} < 3.0$), the transverse mass $m_T$ formed by $p_{T,l}$ and $\slashed{E}_T$ has to exceed $130 \,{\rm GeV}$ and ${\slashed E_T}/\sqrt{\sum_{n=1}^4 p_{T,j_n}} > 9 \, \sqrt{\rm GeV}$  is required. The kinematic invariant $a m_{T2}$ \cite{Barr:2009jv, Konar:2009qr, Bai:2012gs} has to fulfil $a m_{T2} > 190 \,{\rm GeV}$. These requirements lead to the following 95\% CL limit on the fiducial cross section 
\begin{equation}  \label{eq:singleleptonpresent}
\sigma_{\rm fid} (pp \to \slashed{E}_T +  jbl ) < 0.5 \, {\rm fb} \,.
\end{equation}

\subsubsection*{\boldmath Di-lepton $\slashed{E}_T +\bar t t$ channel}

We finally consider the results of the CMS search for a $\slashed{E}_T +\bar t t \, (\to \bar bl^- \bar \nu b l^+ \nu) $ signal \cite{CMS:2014mxa},  performed on a $8 \, {\rm TeV}$ data sample that corresponds to an integrated luminosity of~$19.7 \, {\rm fb}^{-1}$. The basic selection requirements are  $p_{T,j} >30 \, {\rm GeV}$, $|\eta_j|<5$, $p_{T,l} > 20 \, {\rm GeV}$,  $|\eta_l|<2.4$, $m_{ll} > 20 \, {\rm GeV}$, $|m_{ll} - 91 \, {\rm GeV}| > 15 \, {\rm GeV}$  and the jet radius is $R=0.5$. In addition, the following four  cuts 
\beq \label{eq:dileptoncuts}
\slashed{E}_T > 320 \, {\rm GeV} \,, \quad  \sum_{n=1}^2  p_{T,j_n}  < 400 \, {\rm GeV}  \,, \quad \sum_{n=1}^2 p_{T,l_n}   > 120 \, {\rm GeV} \,, \quad \Delta \phi_{l_1 l_2} < 2\,, 
\eeq
are applied to separate  signal from background. The relevant 95\% CL limit on the fiducial cross section is 
\begin{equation}  \label{eq:dileptonfidpresent}
\sigma_{\rm fid} ( pp \to \slashed{E}_T + \bar bl^- b l^+) <  0.15 \, {\rm fb} \,.
\end{equation}

\subsubsection*{Comparison of current constraints}

\begin{figure}[!t]
\begin{center}
\includegraphics[width=0.45\textwidth]{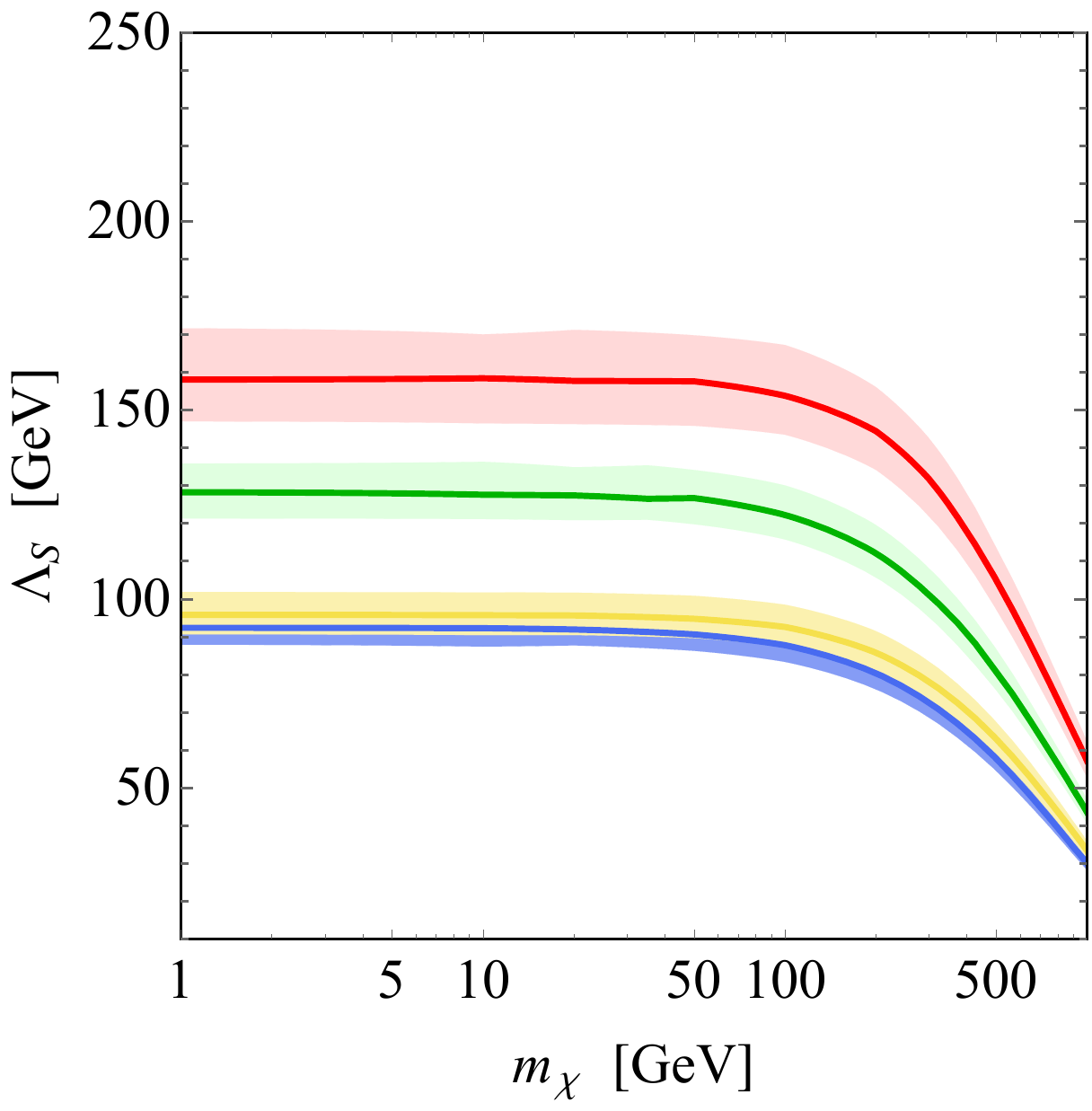} 
\qquad 
\includegraphics[width=0.45\textwidth]{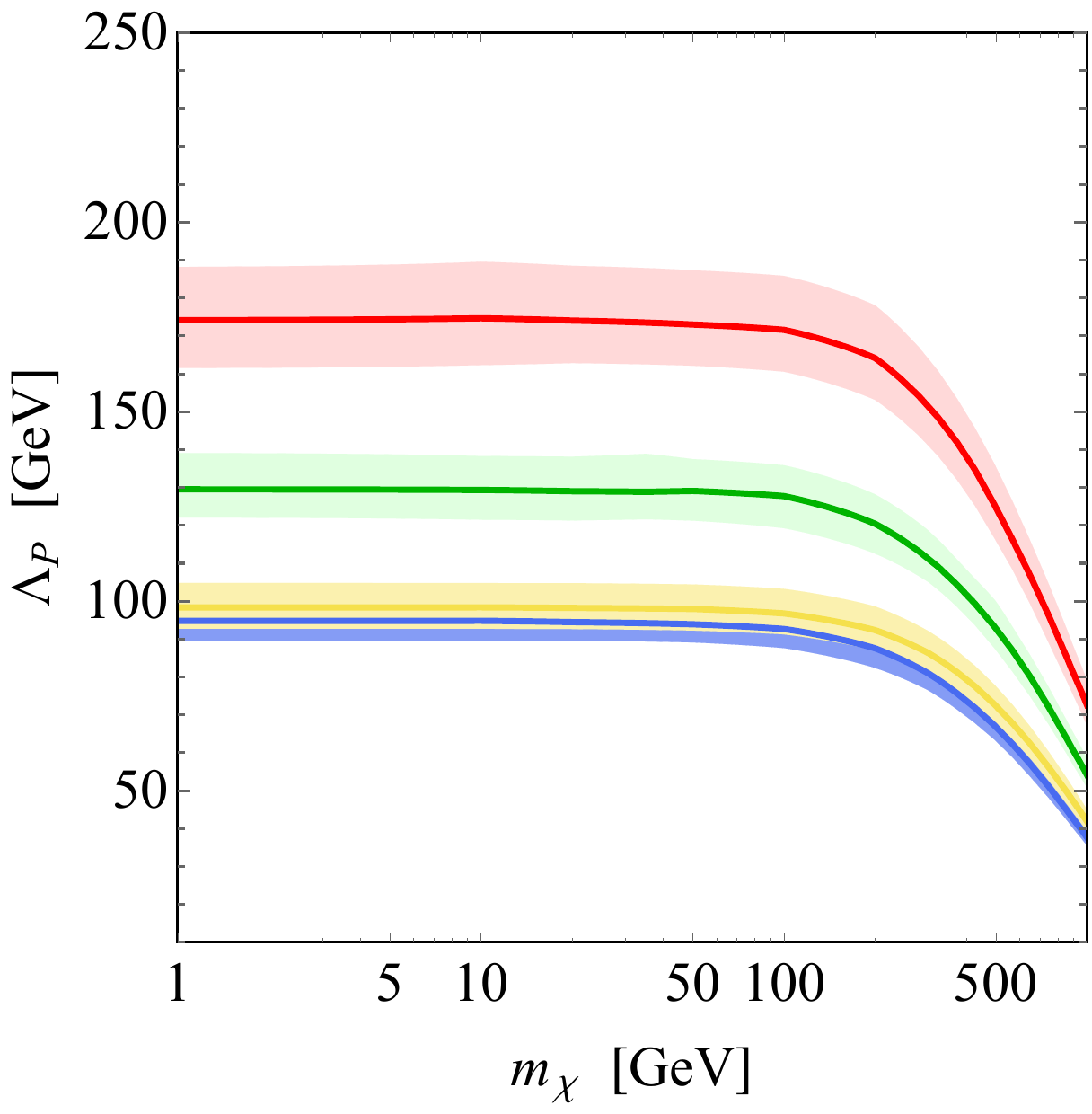} 
\vspace{2mm}
\caption{\label{fig:EFTpresent} Lower 95\% CL limits on the suppression scales $\Lambda_S$ (left) and $\Lambda_P$ (right) that derive from the mono-jet (red), $\slashed{E}_T +\bar t t \, (\to  2j \bar b b)$~(blue), $\slashed{E}_T +\bar t t \, (\to  jbl \nu)$~(green)  and $\slashed{E}_T +\bar t t \, (\to \bar bl^-\bar \nu b l^+ \nu)$~(yellow) searches after  LHC run-1. The widths of the bands reflect scale  uncertainties.}
\end{center}
\end{figure}

The two panels in Figure~\ref{fig:EFTpresent} show the 95\% CL bounds on the suppression scales $\Lambda_S$~(left) and~$\Lambda_P$~(right)   that derive from the individual search strategies discussed before. The widths of the coloured bands illustrate the impact of scale variations. The corresponding relative uncertainties amount to around~$\pm 8\%$ and $\pm6\%$ in the case of the mono-jet and the $\slashed{E}_T + \bar t t$ signals, respectively. We see that for both the scalar and the pseudo-scalar operator $O_S^t$ and $O_P^t$, the bound (\ref{eq:monofidpresent}) on the mono-jet cross section provides to the best constraints at the moment. Numerically, we obtain $\Lambda_S \gtrsim 145 \, {\rm GeV}$ ($\Lambda_P\gtrsim 160 \, {\rm GeV}$) for $m_\chi \lesssim 100 \, {\rm GeV}$, if theoretical uncertainties are included. We add that the  LOPS mono-jet cross sections, on which our limits are based,  are  by roughly $45\%$ smaller than  the corresponding LO fixed-order results. A effect of  similar size has been observed in the case of the operators~$O_{G}$~\cite{Haisch:2013ata}, which implies that the impact  of the jet veto can be  well modelled by working in the EFT~(\ref{eq:OGG}), where both the mediators $S,P$ and the top quark have been integrated out. 

Turning to the bounds arising from the $\slashed{E}_T + \bar t t$ channels, we first observe that our limits are in full agreement with the ones reported in \cite{CMS:2014mxa,Aad:2014vea}.   This shows indirectly that the reconstruction efficiencies are close to 100\%  in these analyses. It furthermore suggests that the signal reconstruction is independent of $m_\chi$ and does not depend on whether one considers the insertion of $O_S^t$ or $O_P^t$. Second one sees that the strongest constraints stem from the single-lepton limit (\ref{eq:singleleptonpresent}). This search allows to set a lower bound of $\Lambda_{S,P} \gtrsim 120 \,{\rm GeV}$ if DM  is lighter than about $100 \, {\rm GeV}$. The corresponding limit for  both the di-lepton and the fully hadronic $\slashed{E}_T + \bar t t$  channel amounts to $\Lambda_{S,P} \gtrsim 90 \,{\rm GeV}$. Notice that in the case of the $\slashed{E}_T + \bar t t$  modes the constraints on the suppression scale entering the operators $O_S^t$ and $O_P^t$ are very similar for light DM, while in the  mono-jet case the limits on $\Lambda_P$ are by roughly $10\%$ stronger than those on $\Lambda_S$. This feature can be understood by observing that the ratio of the multiplicative factors appearing in the operators $O_{\widetilde G}$ and~$O_{G}$ reads~$3/2 \, \Lambda_P^3/\Lambda_S^3$. From~(\ref{eq:OGG}) one would hence expect that for a given mono-jet cross section the restrictions on $\Lambda_P$ are by a factor of~$(3/2)^{1/3} \simeq 1.14$ better than the limits on $\Lambda_S$, and this is to very good approximation what one finds.  Finally, realise that for $m_\chi \gtrsim 100 \, {\rm GeV}$  the limits on $\Lambda_S$ all fall off faster than the bounds on  $\Lambda_P$. This property is related to the fact that the~$S \to \bar \chi \chi$ squared amplitude is proportional to $m_{\bar \chi \chi}^2 - 4 m_\chi^2$, while for $P \to \bar \chi \chi$ one  instead has $m_{\bar \chi \chi}^2$. 

\subsection{Prospects of EFT limits}
\label{sec:EFTfuture}

\begin{figure}[!t]
\begin{center}
\includegraphics[width=0.45\textwidth]{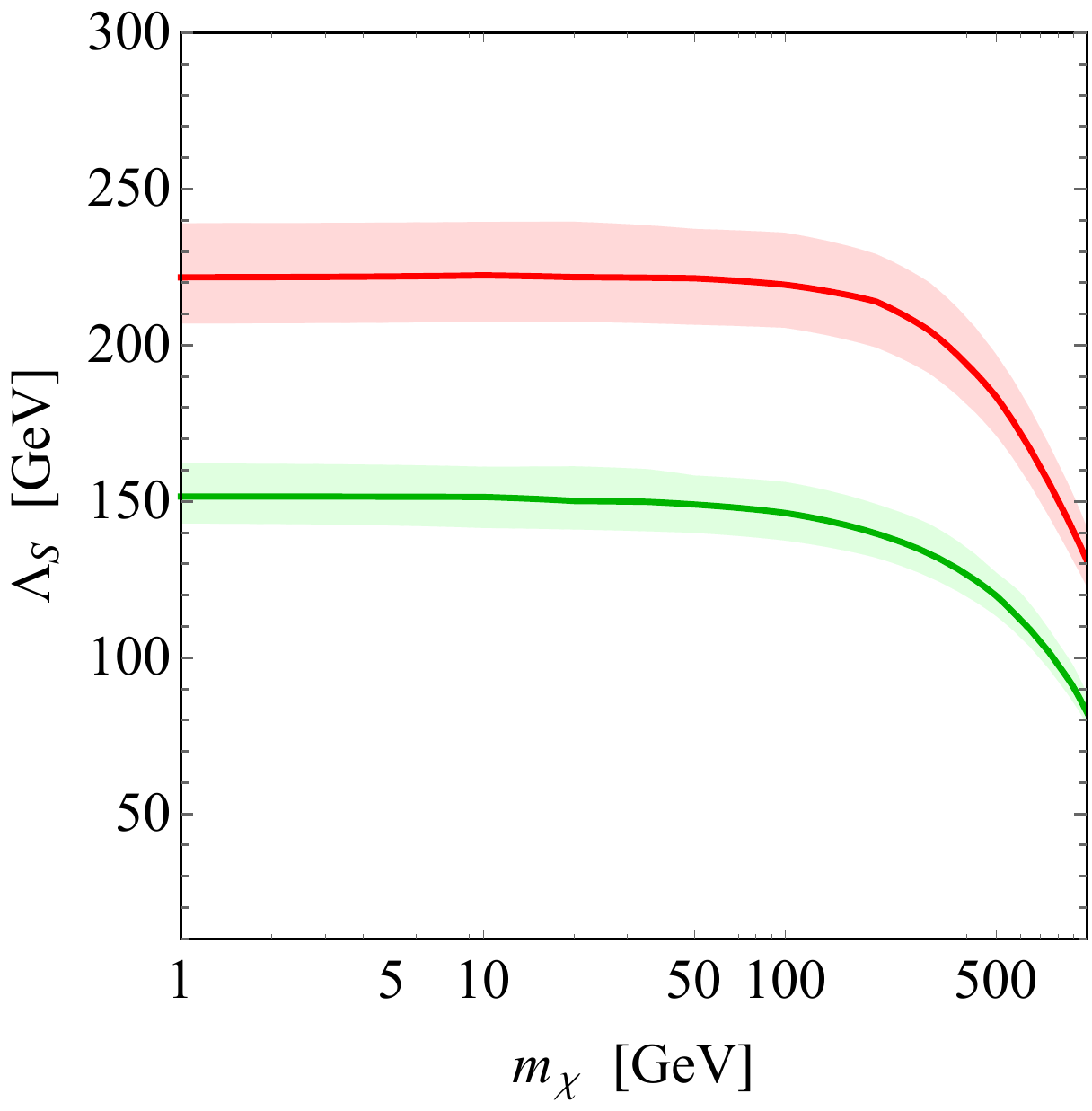} 
\qquad 
\includegraphics[width=0.45\textwidth]{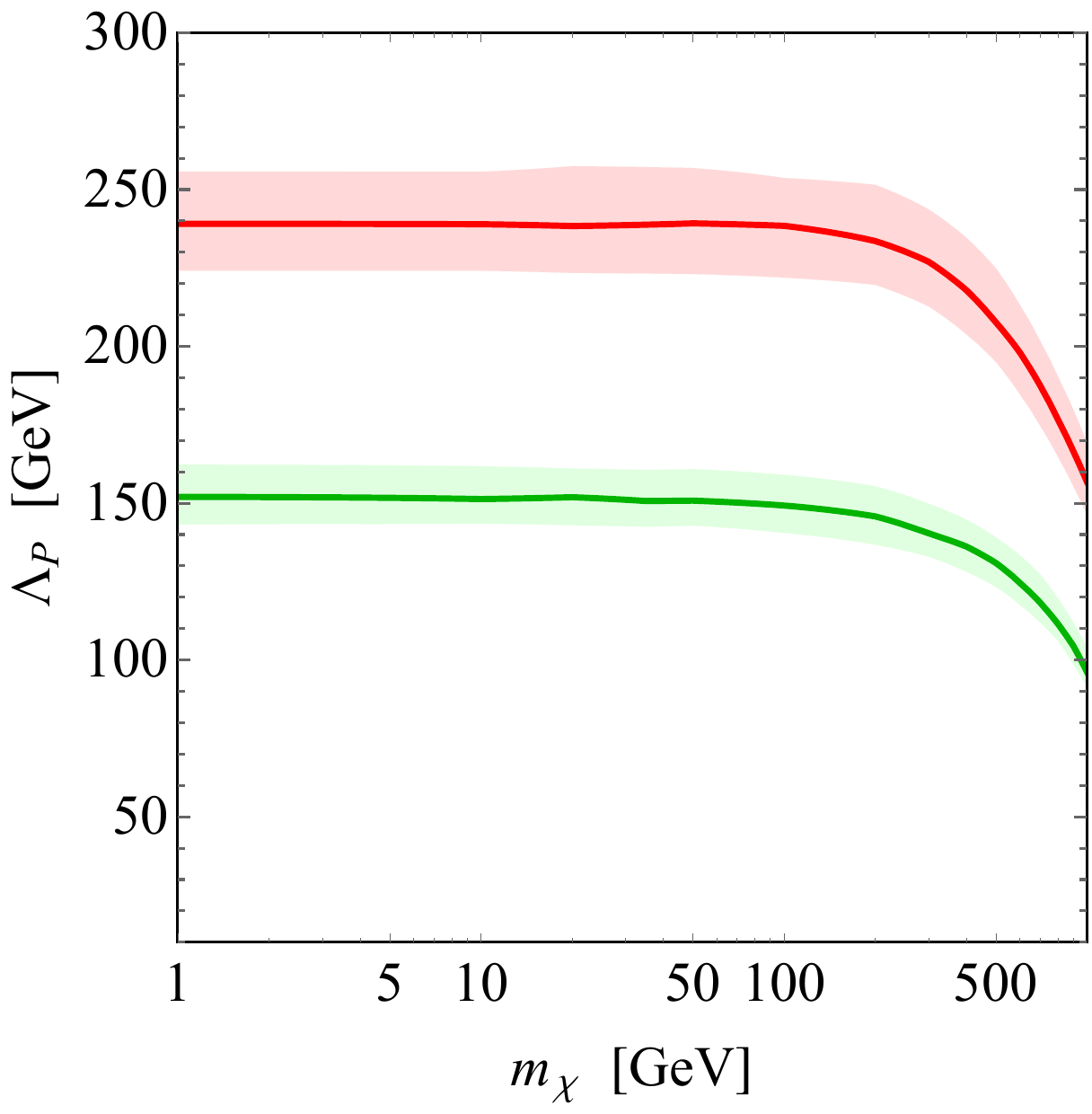} 
\vspace{2mm}
\caption{\label{fig:EFTfuture} Projected lower 95\% CL limits on  $\Lambda_S$ (left) and $\Lambda_P$ (right) from mono-jet~(red) and  $\slashed{E}_T +\bar t t \, (\to  jbl \nu)$~(green)  searches at the $14 \, {\rm TeV}$ LHC. The shown predictions assume an integrated luminosity of $25 \, {\rm fb}^{-1}$. Scale  uncertainties are indicated by the widths of the coloured bands.}
\end{center}
\end{figure}

It is also worthwhile to investigate how the reach on the suppression scales $\Lambda_{S,P}$ might improve at $14 \, {\rm TeV}$. As we have seen the mono-jet channel and the $\slashed{E}_T + \bar t t$ single-lepton mode provide at  present the two most stringent constraints, and this situation is unlikely to change at future LHC runs. We therefore focus on these two search strategies, discussing the relevant experimental cuts and the associated SM backgrounds for each signal in turn. 

\subsubsection*{\boldmath Mono-jet channel}

In the case of the mono-jet signal, we  apply the event selection criteria that have been used in the sensitivity study by ATLAS~\cite{ATL-COM-PHYS-2014-549}. They are given by 
\begin{equation} \label{eq:monojetfuture}
p_{T,j_1}  > 300 \, {\rm GeV} \,,  \quad |\eta_{j_1}| < 2.0 \,, \quad 
p_{T,j_2}  > 50 \, {\rm GeV} \,, \quad \; \, |\eta_{j_2}| < 3.6  \,, \quad 
\Delta \phi _{j \slashed{E}_T } > 0.5 \,,
\end{equation}
and jets are defined with $R = 0.4$. Events with a third jet of $p_{T,j_3}  > 50 \, {\rm GeV}$ and  $ |\eta_{j_3}| < 3.6$ are vetoed and $\slashed{E}_T > 800 \, {\rm GeV}$ is employed, since out of the three $\slashed{E}_T$ thresholds discussed in the ATLAS study this cut provides the strongest restrictions. Notice that compared to~(\ref{eq:monojetpresent}) the $p_{T, j_1}$, $p_{T, j_2}$ and $\slashed{E}_T $ requirements are increased to avoid pile-up and to enhance the signal-over-background ratio. In order to determine the limits on~$\Lambda_{S,P}$, we follow~\cite{ATL-COM-PHYS-2014-549} and take 
\beq \label{eq:fidZnunu}
\sigma_{\rm fid}^{\rm SM} \big (pp \to \slashed{E}_T+ j  \big )= 5.5 \, {\rm fb} \,, 
\eeq
assuming a total systematic uncertainty on the SM background of 5\%. 

\subsubsection*{\boldmath Single-lepton $\slashed{E}_T +\bar t t$ channel}

Our forecast for  the $\slashed{E}_T +\bar t t \, (\to  jbl \nu)$ channel   is based on the cuts introduced in the ATLAS benchmark study of top-squark pair production  \cite{ATL-PHYS-PUB-2013-011}. Specifically, we  require a single  lepton with $p_{T,l} > 25 \, {\rm GeV}$, $|\eta_l|<2.5$ and in addition four or more jets with one $b$-tag all satisfying~$|\eta_j|<2.5$. Jets are found using $R = 0.4$ and we impose 
\beq \label{eq:singleleptoncutsfuture}
\begin{split}
 \slashed{E}_T > 550 \,{\rm GeV} \,, \qquad p_{T,j_1} > 80 \, {\rm GeV} \,, \qquad p_{T,j_2} > 60 \, {\rm GeV} \,, \qquad  p_{T,j_3} > 40 \, {\rm GeV} \,, \\[2mm]  p_{T,j_4} > 25 \, {\rm GeV} \,, \qquad 130 \, {\rm GeV}  < m_{jjj} < 205 \, {\rm GeV} \,, \qquad \Delta \phi_{f\slashed{E}_T} > 0.8 \,, \hspace{0.8cm} 
\end{split}
\eeq
where $f=j_1,j_2$.  The requirement on the transverse mass calculated from $p_{T,l}$ and $\slashed{E}_T$ is $m_T > 350 \, {\rm GeV}$ and we ask for ${\slashed E_T}/\sqrt{\sum_{n=1}^4 p_{T,j_n}} > 15 \, \sqrt{\rm GeV}$. Compared to (\ref{eq:singleleptoncutspresent}) the $\slashed{E}_T$ and $m_T$ selections are significantly stronger in (\ref{eq:singleleptoncutsfuture}), which allows to better disentangle signal from background. For the above cuts, the total SM background amounts to \cite{ATL-PHYS-PUB-2013-011}
\begin{equation}  \label{eq:singleleptonfuturebackground}
\sigma_{\rm fid}^{\rm SM} (pp \to \slashed{E}_T +  jbl ) = \, 0.13 \, {\rm fb} \,,
\end{equation}
and has a total uncertainty of 7\%. Notice that the cuts in (\ref{eq:singleleptoncutsfuture}) are not fully optimised for our purposes. The future constraints that we derive using them should therefore be conservative. 

\subsubsection*{Comparison of future constraints}

In Figure~\ref{fig:EFTfuture} we present our projection of the 95\% CL limits on the suppression scales $\Lambda_S$~(left panel) and $\Lambda_P$~(right panel). In the mono-jet case, we observe that with $25 \, {\rm fb^{-1}}$ of data, corresponding to the first year of running after the LHC upgrade to $14 \, {\rm TeV}$, one may be able  to set a  bound of   $\Lambda_S \gtrsim 205 \, {\rm GeV}$ ($\Lambda_P \gtrsim 220  \, {\rm GeV}$) for $m_\chi \lesssim 100 \, {\rm GeV}$. Compared to the present mono-jet limits, this corresponds to improvements by a  factor of~1.4.  With $300 \, {\rm fb}^{-1}$ of accumulated data, we arrive  instead at $\Lambda_{S} \gtrsim  230 \, {\rm GeV}$ ($\Lambda_{P} \gtrsim 250 \, {\rm GeV}$). Collecting~$3000  \, {\rm fb}^{-1}$ will not allow to notably improve these limits, which shows that at $14 \, {\rm TeV}$ the reach of the $\slashed{E}_T  + j$ channel is not statistically limited, but  limited by the systematic uncertainties associated to the imperfect understanding of irreducible SM backgrounds. Finding ways to overcome these limitations will be crucial to exploit the full physics potential of mono-jet searches to be carried out at later stages of the LHC. In the case of the $\slashed{E}_T +\bar t t$ search in the single-lepton final state, one observes  that in the first year of data taking at $14 \, {\rm TeV}$ the present bounds can be improved by a factor of $1.2$ only.  Such an improvement will allow to exclude scales $\Lambda_{S,P} \gtrsim 140 \, {\rm GeV}$ for $m_\chi \lesssim 100 \, {\rm GeV}$. Since the fiducial cross section (\ref{eq:singleleptonfuturebackground}) is compared to (\ref{eq:fidZnunu}) very small, the single-lepton $\slashed{E}_T +\bar t t$ channel will show his true potential only after the LHC has collected   enough statistics. We find that with $300 \, {\rm fb}^{-1}$ ($3000  \, {\rm fb}^{-1}$) of integrated luminosity, $\slashed{E}_T +\bar t t$ searches  should be able to exclude scales $\Lambda_{S,P} \gtrsim 190 \, {\rm GeV}$ ($\Lambda_{S,P} \gtrsim 210 \, {\rm GeV}$) if  DM is light. Notice that the ATLAS sensitivity study \cite{ATL-PHYS-PUB-2013-011} observed a quite similar luminosity dependence of mass limits in the case of top-squark searches. 

\subsection{Infinite top-quark mass limit}
\label{sec:EFTwithouttop}

\begin{figure}[!t]
\begin{center}
\includegraphics[width=0.45\textwidth]{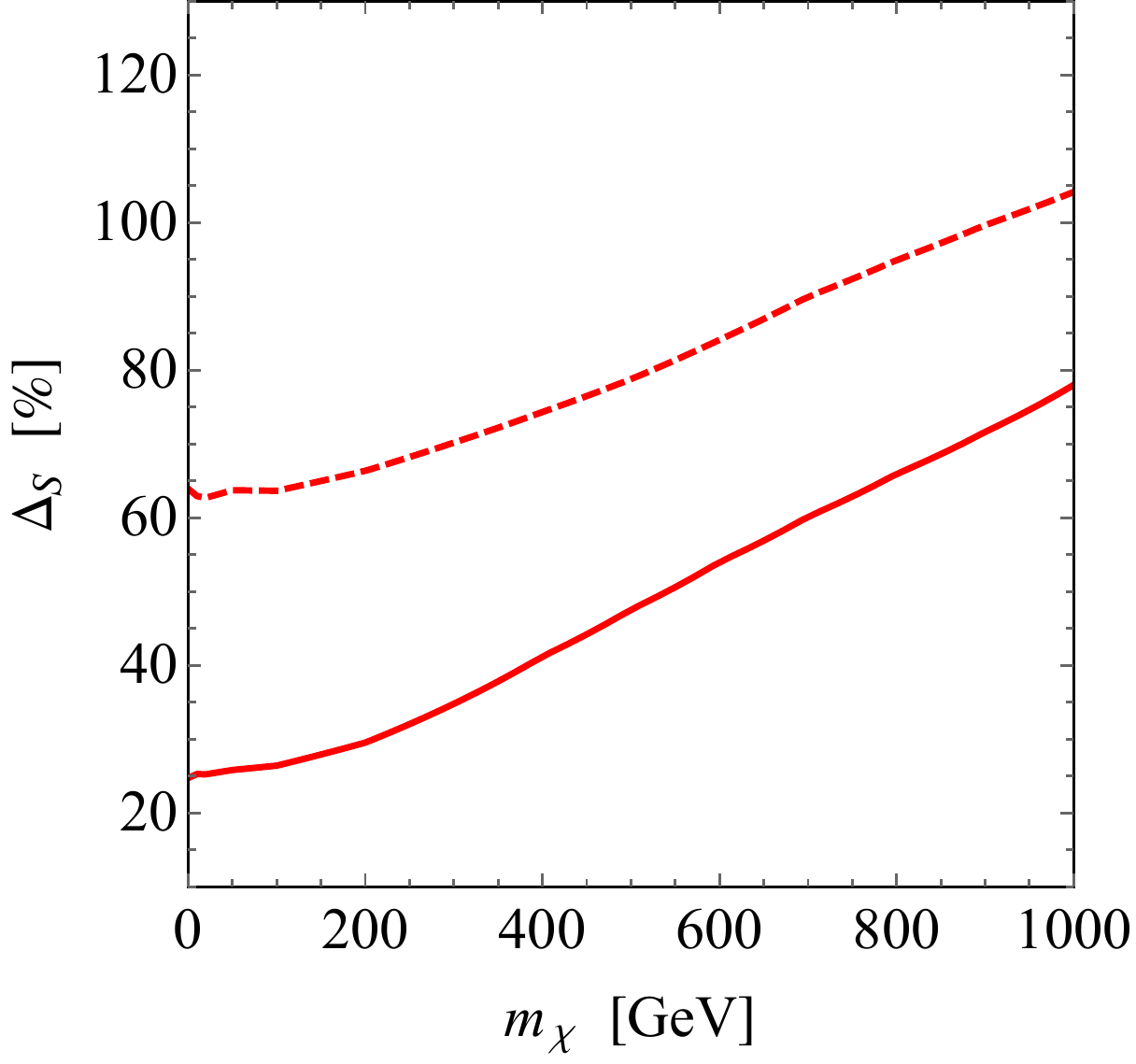} 
\qquad 
\includegraphics[width=0.45\textwidth]{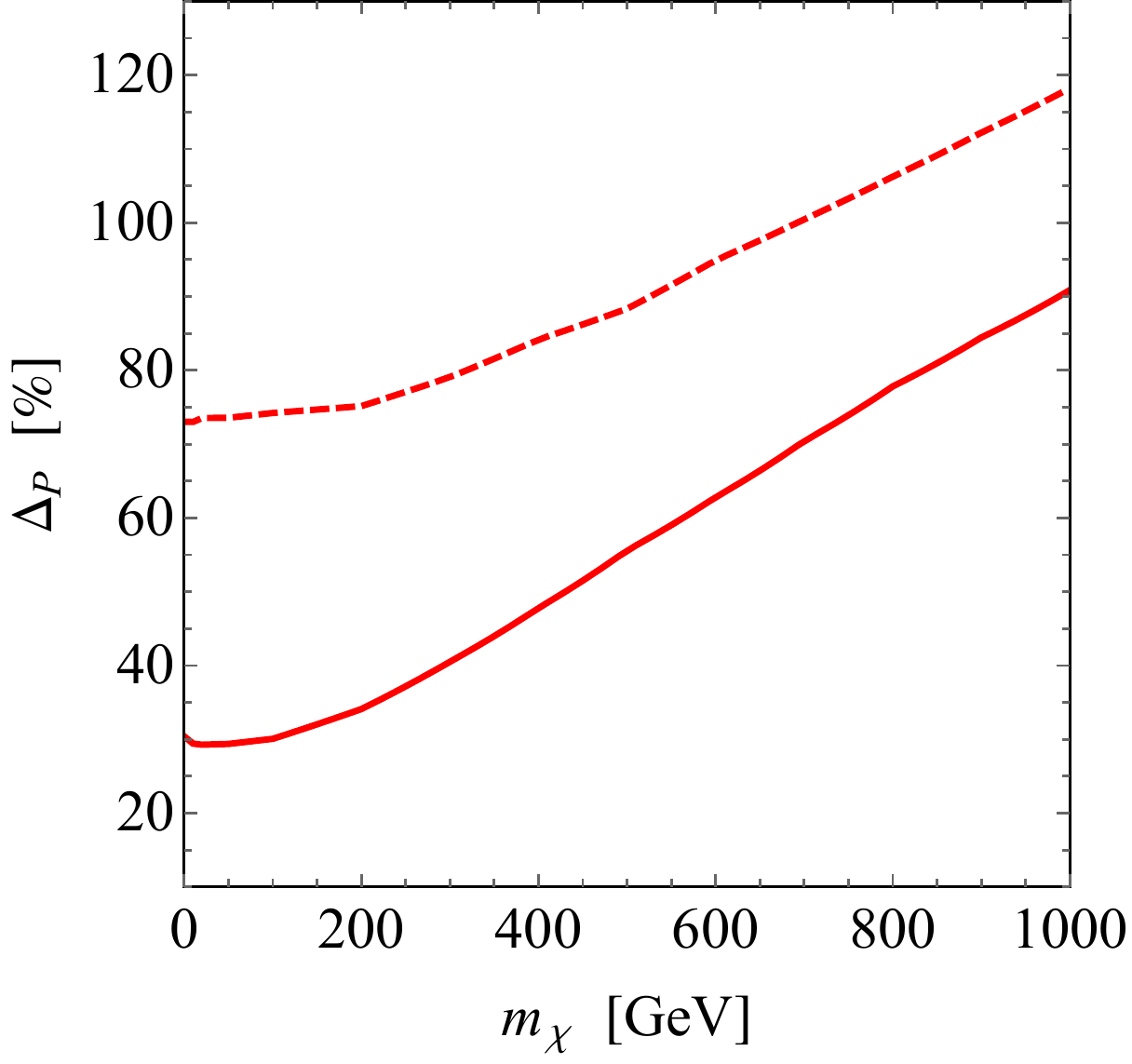} 
\vspace{2mm}
\caption{\label{fig:EFTcomparison} The quantities $\Delta_S$ (left panel) and $\Delta_P$ (right panel) as a function of the DM mass. The solid lines indicate the results that apply in the case of the $8 \, {\rm TeV}$ mono-jet searches, while the dashed curves correspond to our future projections based on $25 \, {\rm fb^{-1}}$ of $14 \, {\rm TeV}$ data. }
\end{center}
\end{figure}

The mono-jet limits on  $\Lambda_{S,P}$  have been derived in the last two sections by employing the exact results for the one-loop $pp \to \slashed{E}_T + j$ amplitudes, but integrating out the mediators that induce the DM-SM interactions. In the following, we compare these bounds to those  that one obtains in the limit of infinite top-quark mass. We call the latter limits $\left ( \Lambda_{S,P} \right )_{m_t \to \infty}$  and define 
\beq \label{eq:DeltaSP}
\Delta_{S,P} = \frac{\left ( \Lambda_{S,P} \right )_{m_t \to \infty}}{\Lambda_{S,P}} - 1\,,
\eeq
which is a measure of the (in)accuracy of the heavy top-quark approximation.  Note that the quantities $\Delta_{S,P}$ depend sensitively on the $\slashed{E}_T$ and $p_{T,j}$ cuts imposed in the experimental analysis. The two panels in Figure~\ref{fig:EFTcomparison} show our results for the quantities $\Delta_{S}$~(left) and  $\Delta_{P}$~(right). As indicated by the solid curves, for the existing mono-jet searches we find $\Delta_S \in [25, 80]\%$ ($\Delta_P \in [30, 90]\%$). Recalling that the ratio of the $\slashed{E}_T + j$ cross sections scales as $\big (1+\Delta_{S,P} \big )^6$, it follows that the $m_t \to \infty$ limit overestimates the exact results by a factor  4  (5) for small DM mass and that the quality of the approximation rapidly degrades  with $m_\chi$, resulting in errors of up to a factor of 32 (48) for the operator $O_S^t$ ($O_P^t$). This clearly shows  that in order to infer faithful bounds on the DM top-quark contact operators (\ref{eq:OSP}), one has to calculate the mono-jet cross section keeping the full top-quark mass dependence \cite{Haisch:2012kf, Fox:2012ru, Haisch:2013fla, Buckley:2014fba, Harris:2014hgav1}. Notice that at the $14 \, {\rm TeV}$ LHC the $m_t \to \infty$ limit is an even  worse approximation than in the LHC run-1 environment. Numerically, we obtain $\Delta_S \in [65, 105]\%$ ($\Delta_P \in [70, 120]\%$), which translates into factors of 19 to 72 (27 to 108) at the level of cross sections. The observed differences between the $8 \, {\rm TeV}$ and the $14 \, {\rm TeV}$ results are easy to understand by remembering  that at higher energies the imposed $\slashed{E}_T$ and $p_{T,j}$ cuts have to be harsher~$\big($see~(\ref{eq:monojetpresent}) and~(\ref{eq:monojetfuture})$\big)$ to differentiate  signal from background. High-energetic initial-state and/or final-state particles are however able to resolve the structure of the top-quark loops that generate the $\slashed{E}_T + j$ signal, so that removing the top quark as an active degree of freedom becomes less and less justified the more stringent the $\slashed{E}_T$ and $p_{T,j}$ selection requirements are. 

The infinite top-quark mass limit is also expected to fail badly at the NLO level.  This means that taking a $K$ factor obtained from Higgs plus jet production to upgrade the LOPS mono-jet cross sections to the NLO level (as done in \cite{Buckley:2014fba}) is hard to defend from a theoretical point of view. A further complication in estimating the size of NLO contributions to $\sigma_{\rm fid} (pp \to \slashed{E}_T + j)$ arises from the fact that in the LHC mono-jet analyses a jet veto is imposed. Such a jet veto tends to decrease the importance of  fixed-order NLO corrections. For instance, in the case of the operator $O_G$ it was found in  \cite{Haisch:2013ata}  that the $K$ factor is reduced from $K \simeq 1.5$ at fixed order to $K \simeq 1.1$ after including PS and hadronisation effects. Although one naively would expect to find a reduction of similar size also in the case considered here, we believe that in order to make a definite statement about the size of  NLO corrections to the $\slashed{E}_T  + j$ signal associated to  the operators $O_{S,P}^t$, an exact ${\cal O} (\alpha_s^4)$  calculation of the top-quark loop-induced mono-jet cross section is unavoidable.

\subsection{Comparison of present DM constraints}
\label{sec:fullpresent}

Below we present the current  LHC exclusion limits on the parameters entering the simplified model (\ref{eq:LSP}), comparing them to those stemming from the first LUX  results~\cite{Akerib:2013tjd} on the SI cross section, the latest Fermi-LAT bounds \cite{Ackermann:2013yva} on the velocity-averaged total DM annihilation cross section and the requirement  not to overclose the Universe. Our exact collider bounds will furthermore be contrasted with the constraints that derive by employing the EFT framework~(\ref{eq:OSP}).

\subsubsection*{Limits from mono-jet searches}
\label{sec:fullpresentmonojet}

\begin{figure}[!t]
\begin{center}
\includegraphics[height=0.45\textwidth]{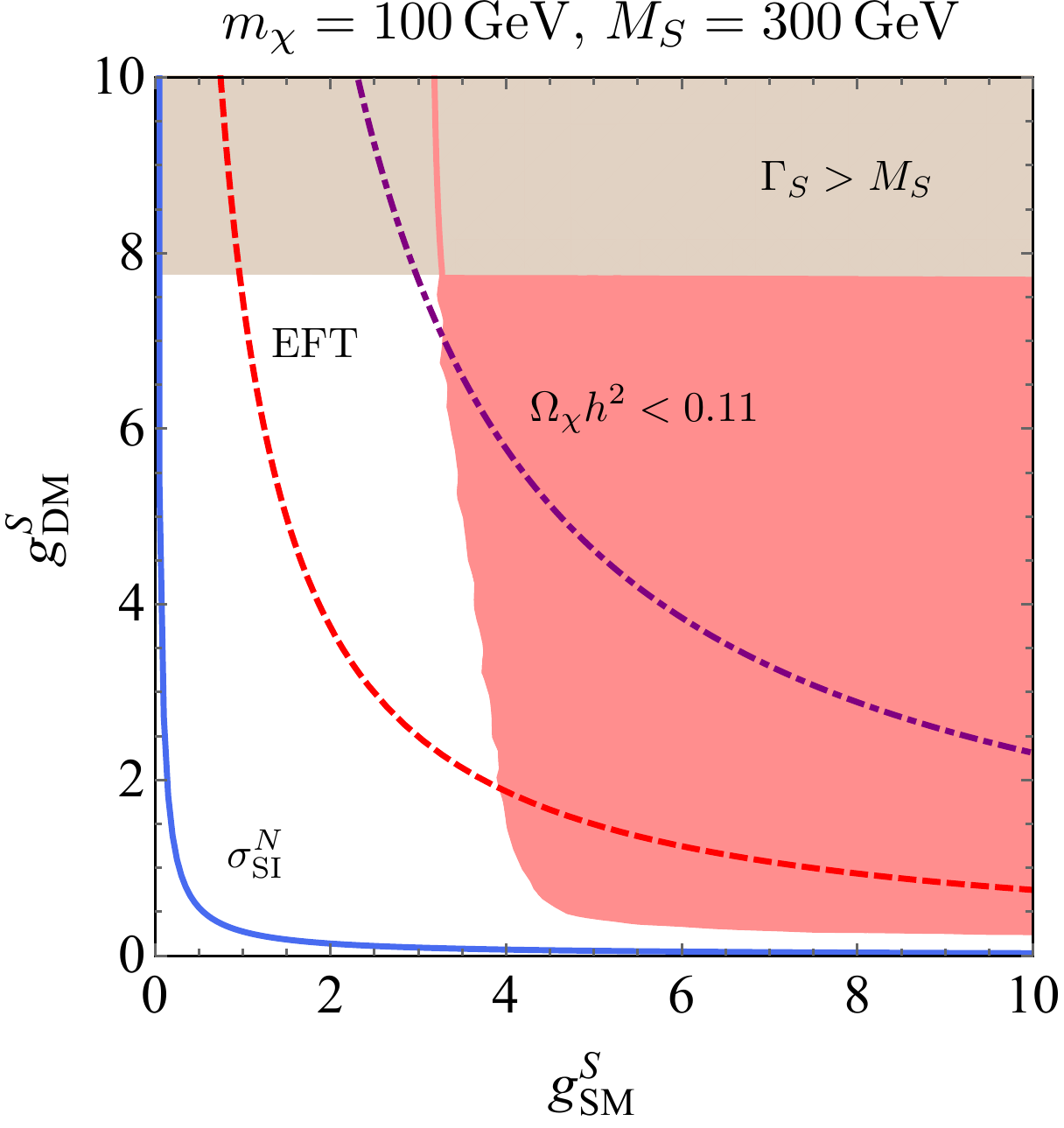} 
\qquad 
\includegraphics[height=0.45\textwidth]{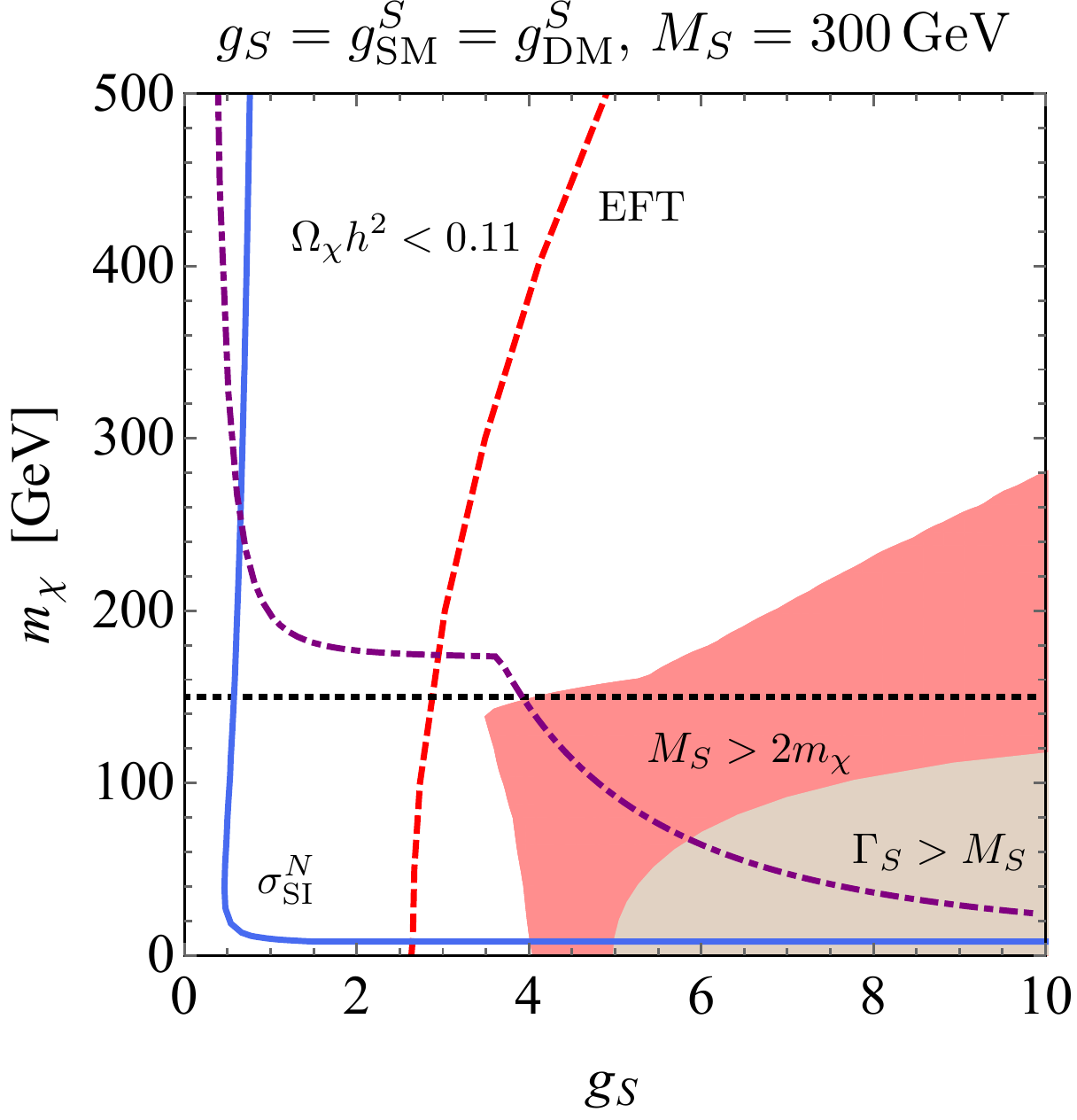} 

\hspace{2mm}

\includegraphics[height=0.45\textwidth]{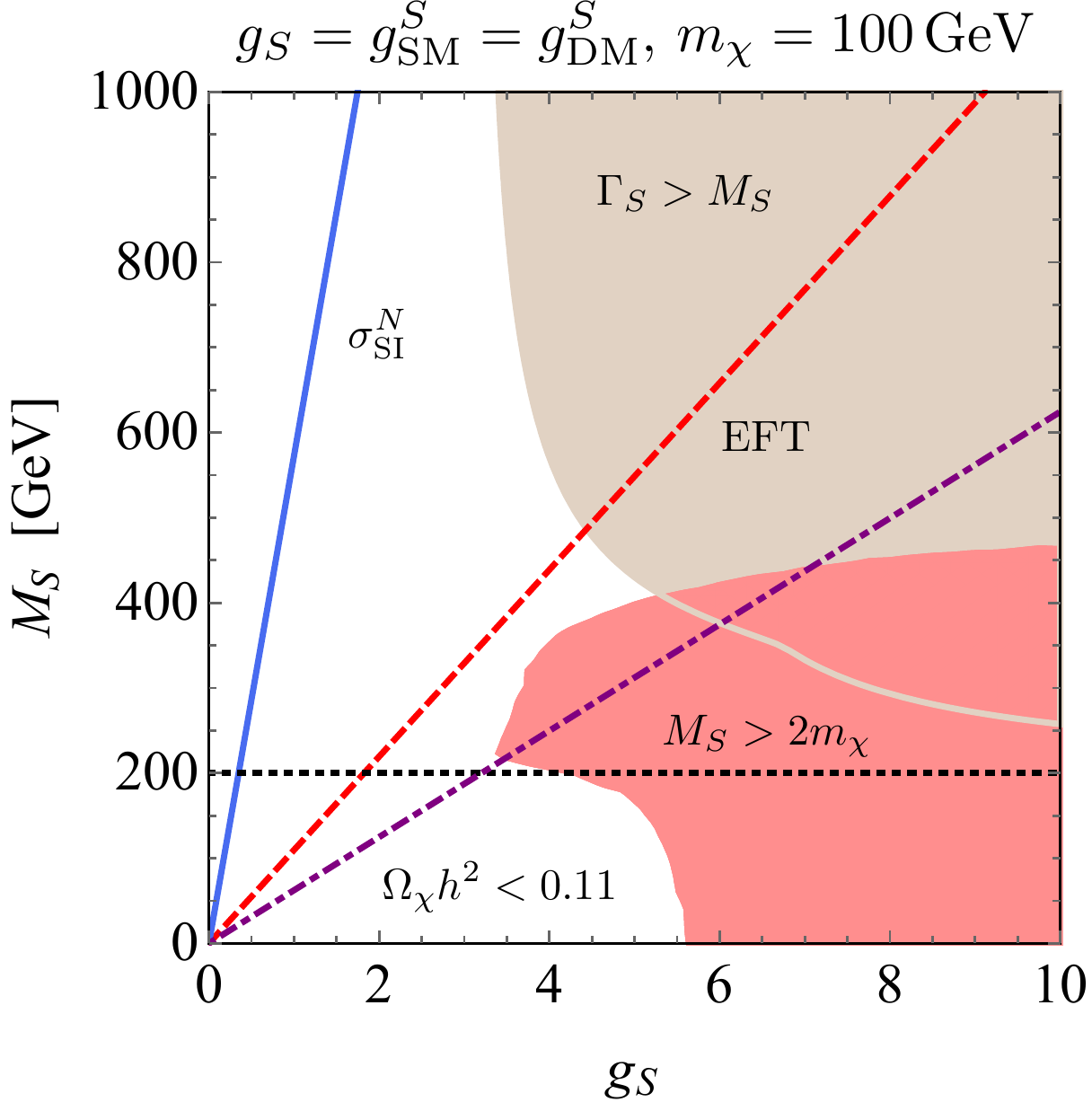} 
\qquad 
\includegraphics[height=0.45\textwidth]{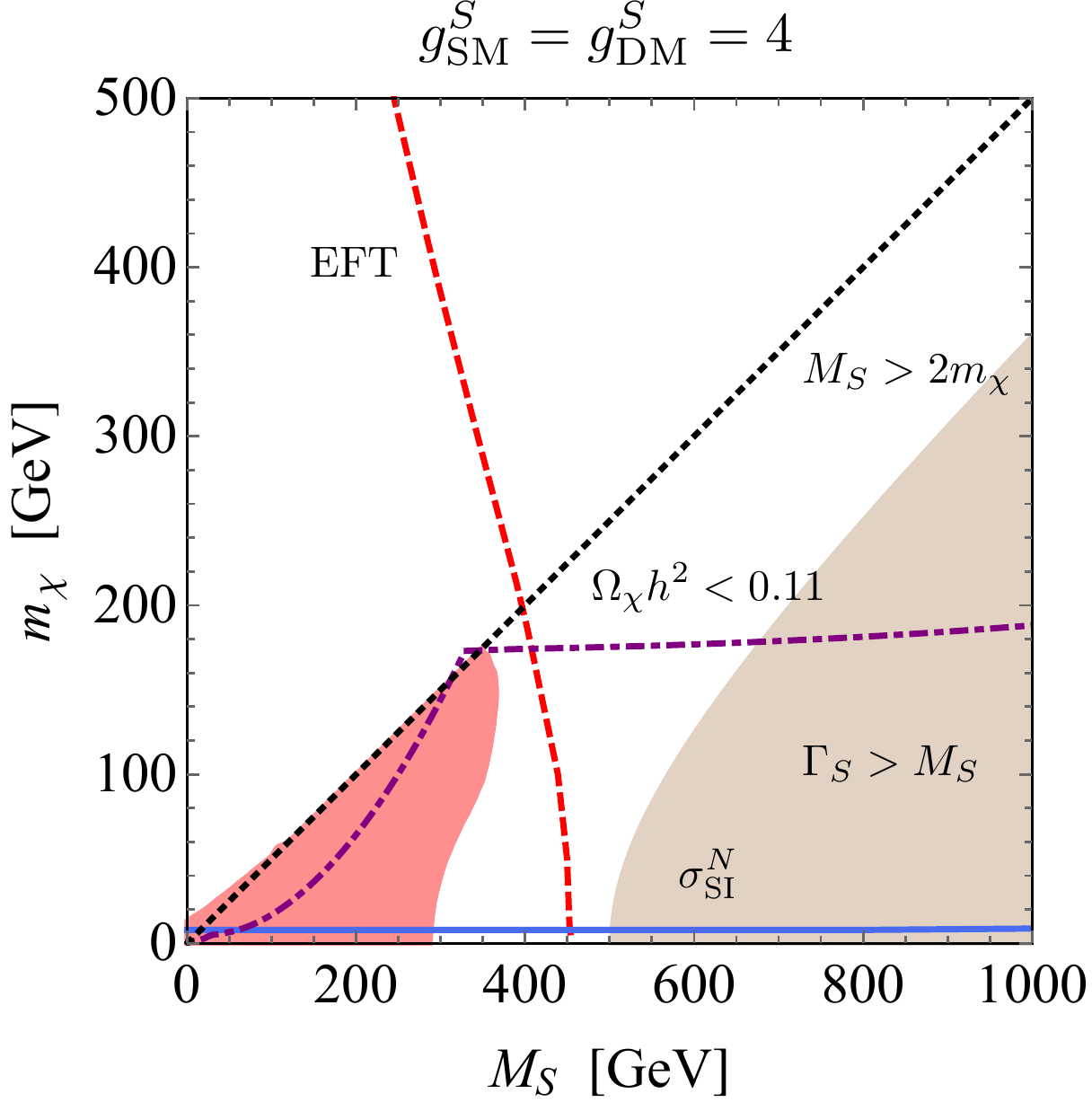} 
\vspace{0mm}
\caption{\label{fig:Splanespresent} Present mono-jet exclusion regions at $95\% \, {\rm CL}$ (red contours) for scalar mediators. In the $g_{\rm SM}^S \hspace{0.25mm} $--$\hspace{0.25mm} g_{\rm DM}^S$ plane (upper left panel) the values $m_\chi = 100 \, {\rm GeV}$ and $M_S = 300 \,{\rm GeV}$ have been employed, while in the $g_S \hspace{0.25mm} $--$\hspace{0.5mm}  m_\chi$  plane (upper right panel) we have identified $g_S = g_{\rm SM}^S = g_{\rm DM}^S$ and fixed the scalar mediator mass to $M_S = 300 \, {\rm GeV}$. The results in the $g_S \hspace{0.25mm} $--$\hspace{0.25mm}  M_S$  plane~(lower left panel) use the same identification and a DM mass of $m_\chi =100 \, {\rm GeV}$, whereas in the $M_S \hspace{0.25mm}$--$\hspace{0.25mm} m_\chi$ plane~(lower right panel) the couplings have been set to $g_{\rm SM}^S = g_{\rm DM}^S =4$. For comparison the regions with $\Gamma_S > M_S$~(brown contours), the current LUX 90\% CL constraint on $\sigma_{\rm SI}^N$ (solid blue curves), the parameter spaces with $\Omega_\chi h^2 < 0.11$ (dot-dashed purple curves), the EFT limits (dashed red curves) and the regions with $M_S > 2 m_\chi$ (dotted black lines) have been indicated.}
\end{center}
\end{figure}

The different constraints on the four parameters $g_{\rm DM}^S$, $g_{\rm SM}^S$, $m_\chi$ and $M_S$ that characterise our simplified scalar mediator models  are summarised in the panels of Figure~\ref{fig:Splanespresent}. Here and below the widths of the mediators are calculated using  (\ref{eq:partial}). In physical terms this means that we assume that the total decay widths $\Gamma_{S,P}$ are minimal. The shown limits are obtained from (\ref{eq:monofidpresent}) and  take into account theoretical uncertainties due to scale variations. To be conservative, we include these uncertainties by setting the bounds  using  the signal cross sections calculated for $\xi = 2$ (see Section~\ref{sec:MC}). 

Turning our attention to the $g_{\rm SM}^S \hspace{0.25mm} $--$\hspace{0.25mm} g_{\rm DM}^S$ plane~(upper left panel), we see that  for $m_\chi = 100 \, {\rm GeV}$ and $M_S = 300 \, {\rm GeV}$, the present lower limit (\ref{eq:monofidpresent}) on the mono-jet cross section allows to probe only simplified models with $g_{\rm SM}^S \gtrsim 3$ and $g_{\rm DM}^S \gtrsim 0.2$ (red contour). This finding is in line with the observation made recently in \cite{Harris:2014hgav1} that current $\slashed{E}_T + j$ searches are not sensitive to weakly-coupled realisations of (\ref{eq:LSP}). Although a direct comparison is difficult, our conclusions also seem to agree with the results presented in~\cite{Buckley:2014fba}. We have explicitly verified that even for light scalar mediators with $M_S < 100 \, {\rm GeV}$ and $m_\chi < 50 \, {\rm GeV}$, values of $g_{{\rm SM}, {\rm DM}}^S <1$ remain inaccessible, if the mediator width $\Gamma_S$ is calculated using the formulas (\ref{eq:partial}). Our finding that existing mono-jet searches are only sensitive to scenarios with couplings $g_{\rm SM}^S$ of  order of a few is hence robust against variations of the remaining parameters.  

The exact $95\% \, {\rm CL}$ exclusion region should be contrasted with the limits that follow from an EFT interpretation (dashed red curve) of the $\slashed{E}_T + j$ searches. We observe that for the considered values of $m_\chi$ and $M_S$, the EFT bounds are  too strong (weak) for $g_{\rm DM}^S \gtrsim 2$ ($g_{\rm DM}^S \lesssim 2$). This feature is easy to understand by noticing that for our choice of parameters one has $M_S < 2 m_t$, which implies that  $\Gamma_S \simeq \Gamma (S \to \bar \chi \chi) \propto \left ( g_{\rm DM}^S \right)^2$. The width of the scalar mediator thus grows quadratically with $g_{\rm DM}^S$ and for $g_{\rm DM}^S \gtrsim 6$ one ends up in the unphysical situation where $\Gamma_S > M_S$ (brown contour).  For a broad (narrow) resonance it is however known (see for instance~\cite{Fox:2011pm,Fox:2012ru,Buchmueller:2014yoa} for the case of vector and axial-vector mediators) that EFT cross sections tend to overestimate (underestimate) the exact results.  This is a general shortcoming of the EFT framework that can only be overcome by calculating $\slashed{E}_T$ signals in a simplified model such as (\ref{eq:LSP}). 

\begin{figure}[!t]
\begin{center}
\includegraphics[height=0.45\textwidth]{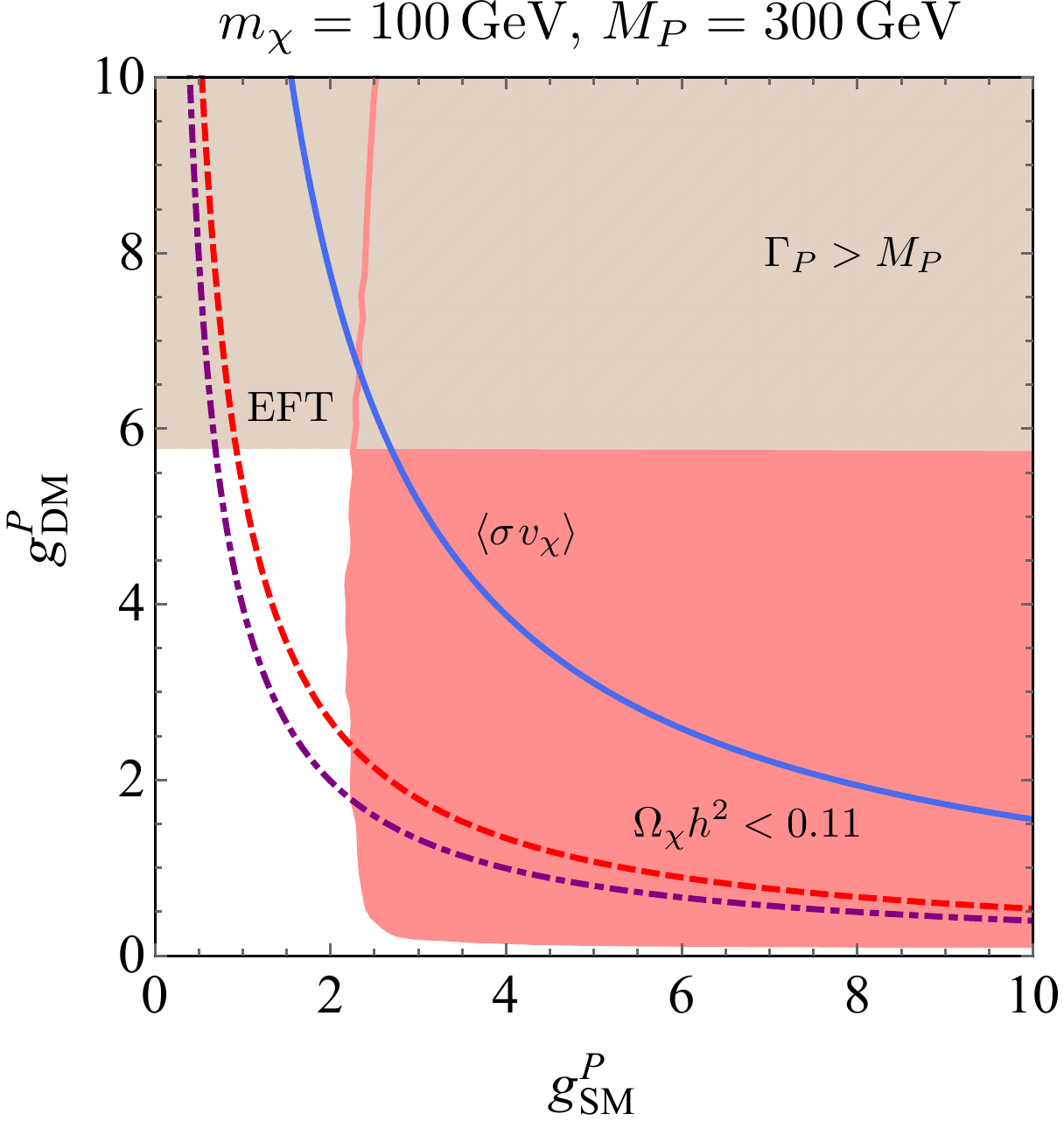} 
\qquad 
\includegraphics[height=0.45\textwidth]{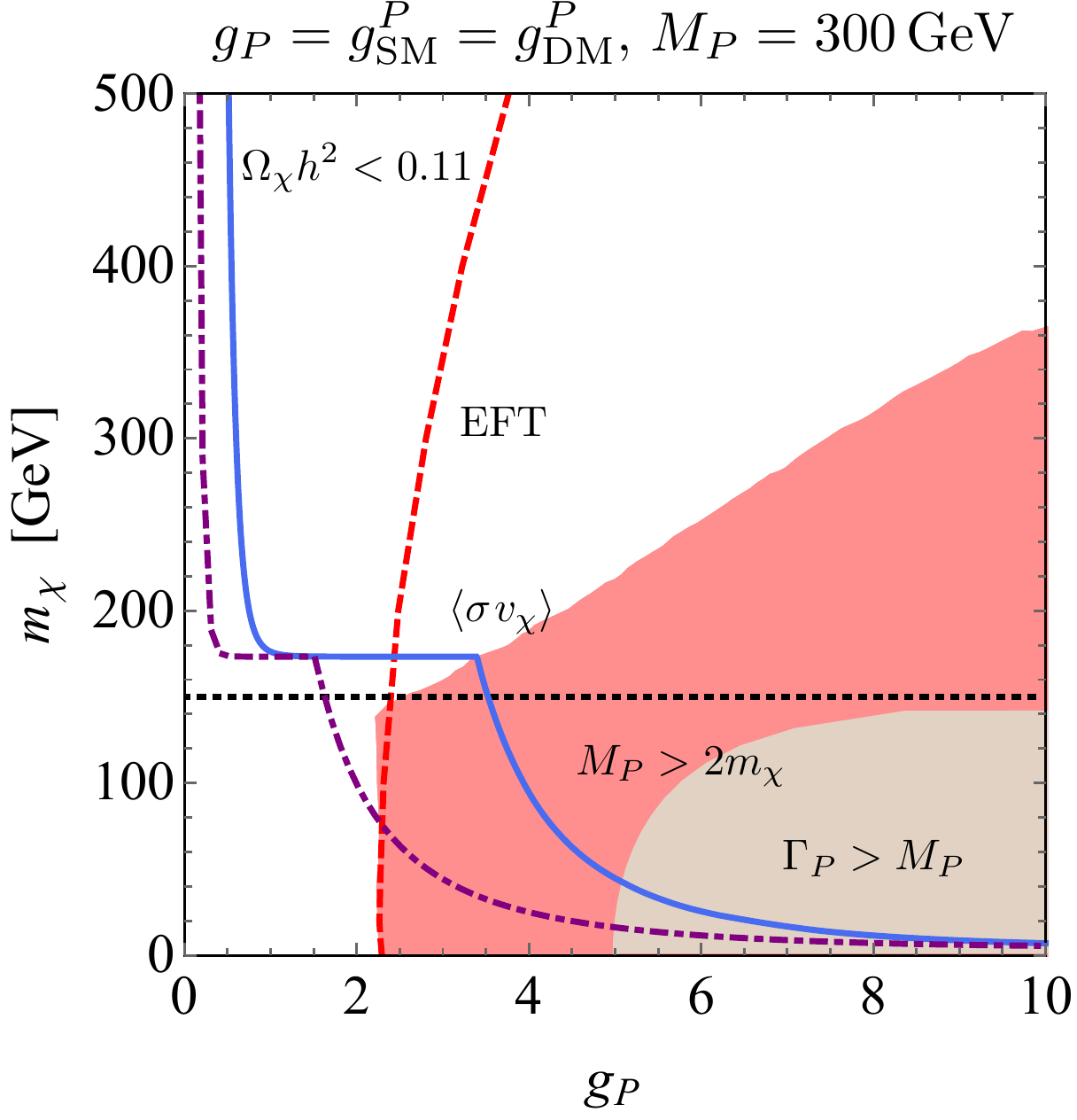} 

\hspace{2mm}

\includegraphics[height=0.45\textwidth]{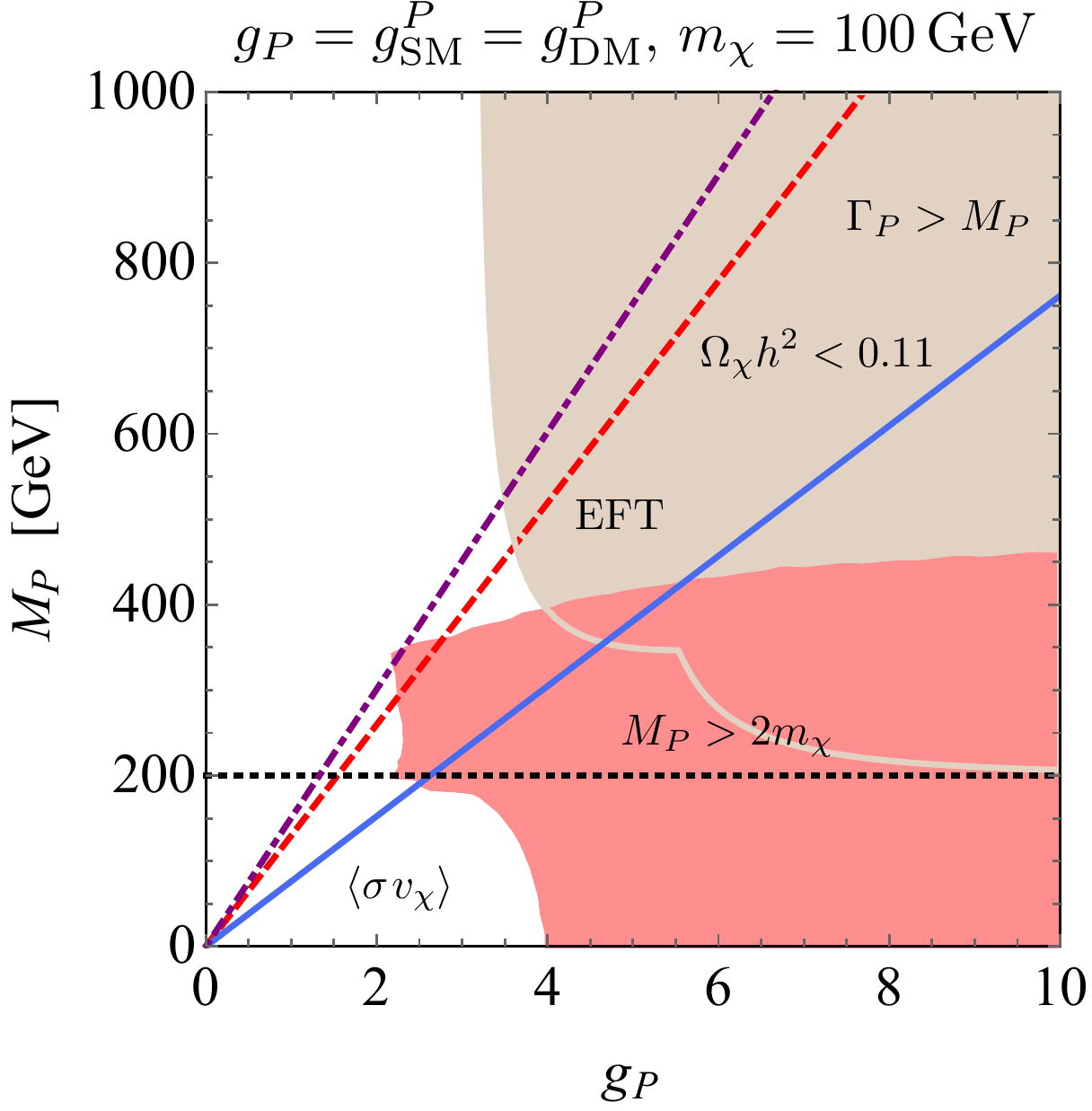} 
\qquad 
\includegraphics[height=0.45\textwidth]{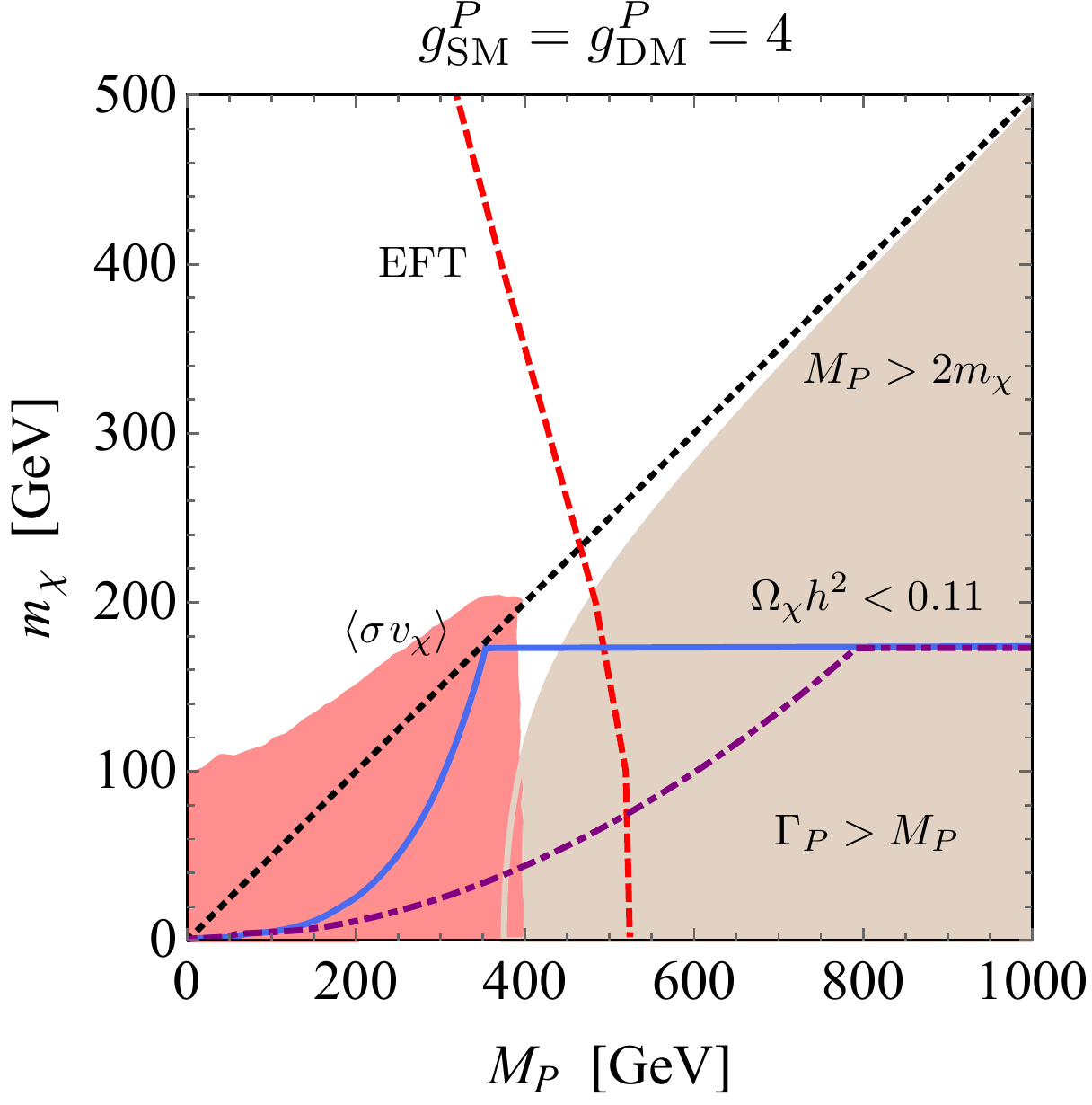} 
\vspace{0mm}
\caption{\label{fig:Pplanespresent}  Present mono-jet exclusion regions at $95\% \, {\rm CL}$ for pseudo-scalar mediators. The latest Fermi-LAT 95\% CL bound on the total velocity-averaged DM annihilation cross section $\left \langle  \sigma \hspace{0.25mm} v_\chi \right \rangle$ is indicated by the solid blue curves. Apart from this the same colour coding and choice of parameters as in Figure~\ref{fig:Splanespresent} is adopted.}
\end{center}
\end{figure}

For comparison we also show in the $g_{\rm SM}^S \hspace{0.25mm} $--$\hspace{0.25mm} g_{\rm DM}^S$ plane  the  restriction on $\sigma_{\rm SI}^S$ provided by LUX (solid blue curve) and the DM relic density (dot-dashed purple curve). Since in the case of scalar exchange the elastic DM-nucleon scattering is SI and unsuppressed~$\big($cf.~(\ref{eq:sigmaSIN})$\big)$, the limits from the existing direct detection experiments are significantly more stringent than the collider bounds, and essentially exclude the entire $g_{{\rm SM}, {\rm DM}}^S$ parameter space for $M_S = 300 \, {\rm GeV}$. Notice that the  constraints arising from the limits on $\sigma_{\rm SI}^S$ can in principle be evaded by assuming that $\chi$ is not stable on cosmological time scales, but lives long enough to escape the ATLAS and CMS detectors. We add that the limits from direct detection are plagued by systematic errors due to the uncertain local DM density and velocity distribution, which play no role in the case of  the collider bounds. It is also evident that compared to the exact LHC exclusion, the requirement not to have a too high DM relic density, i.e.~$\Omega_\chi h^2 < 0.11$, further pushes $g_{{\rm SM}, {\rm DM}}^S$ to larger values. We add that the limits following from the relic abundance calculation are  more model dependent than the remaining bounds, because they depend strongly on the full particle content and all the interactions of the underlying theory. For instance,  opening up additional DM annihilation channels will generically have a more visible impact on $\Omega_\chi h^2$ than on $\sigma  ( pp \to \slashed{E}_T + j )$ and $\sigma_{\rm SI}^N$. These loopholes  should be kept clearly in mind when interpreting bounds associated to the thermal DM relic density. 

\begin{figure}[!t]
\begin{center}
\includegraphics[height=0.45\textwidth]{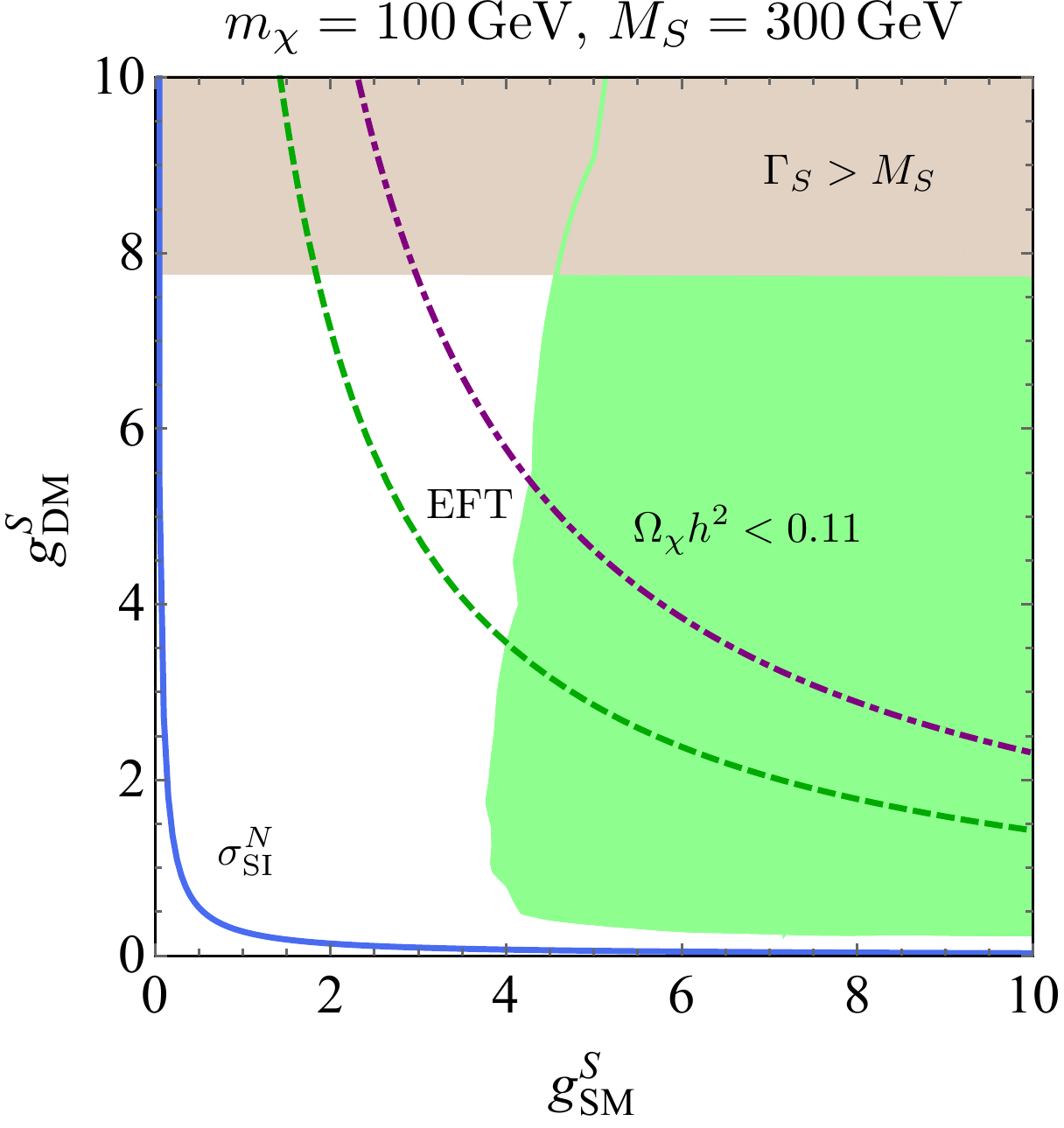} 
\qquad 
\includegraphics[height=0.45\textwidth]{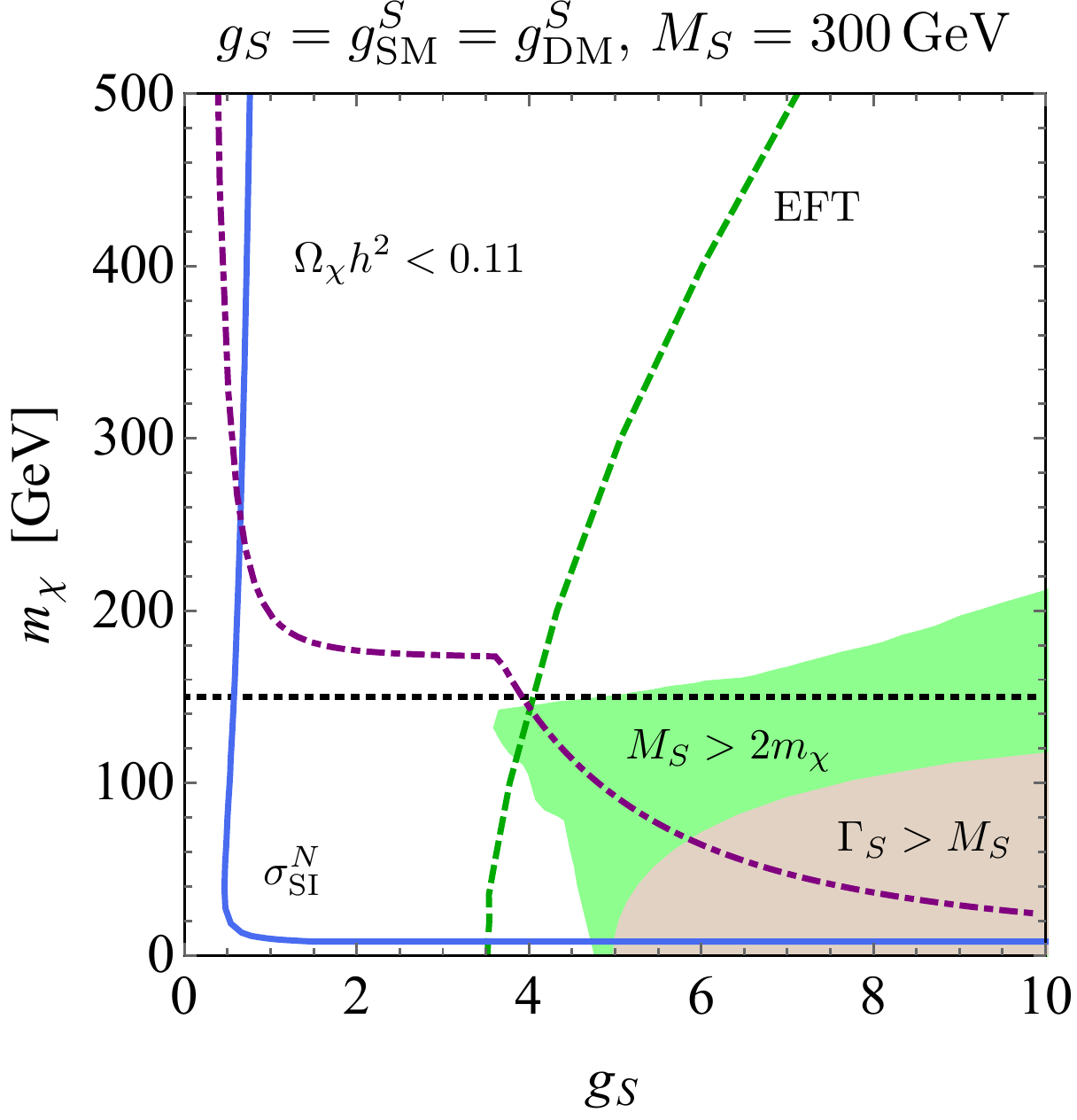} 

\hspace{2mm}

\includegraphics[height=0.45\textwidth]{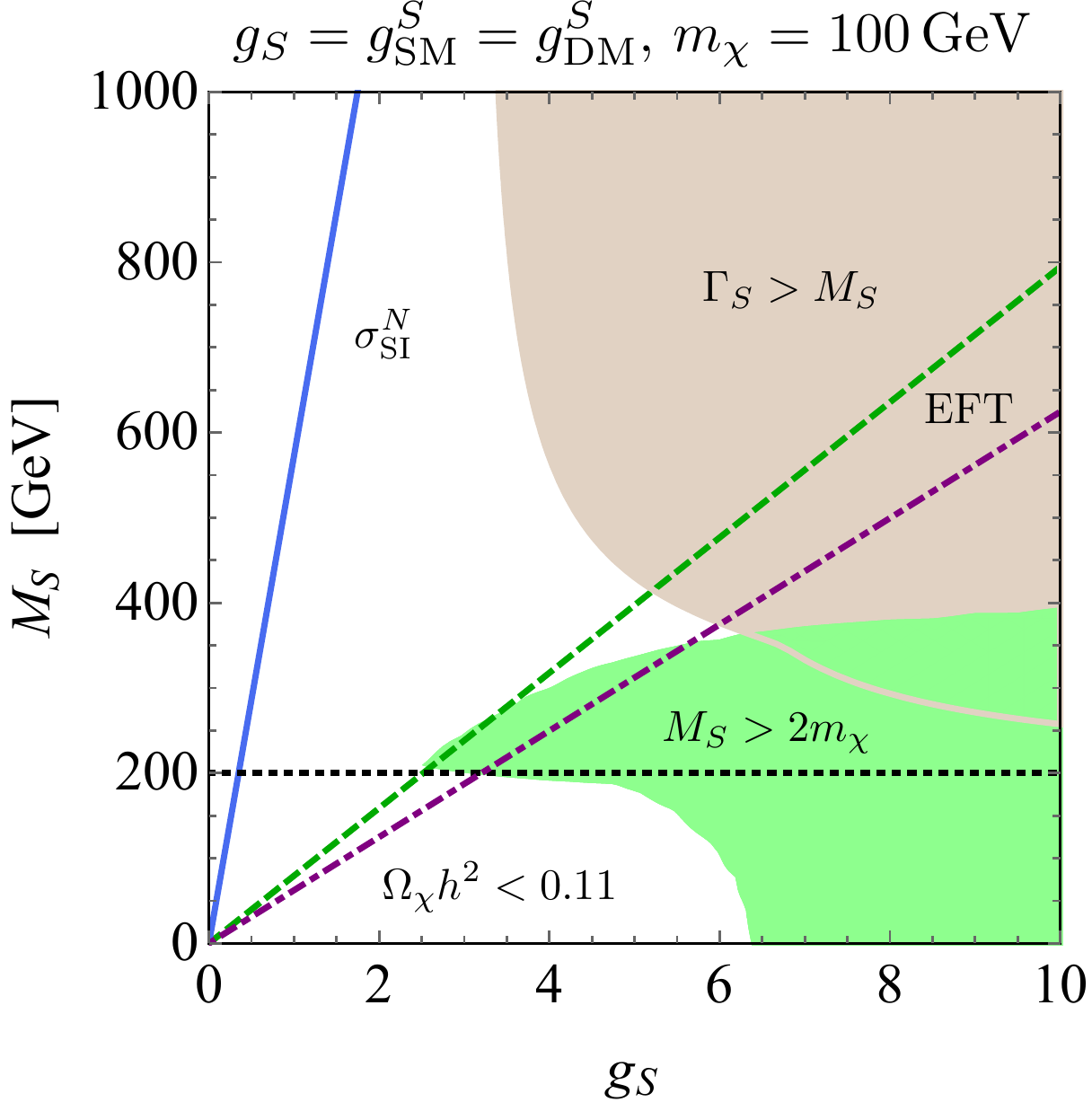} 
\qquad 
\includegraphics[height=0.45\textwidth]{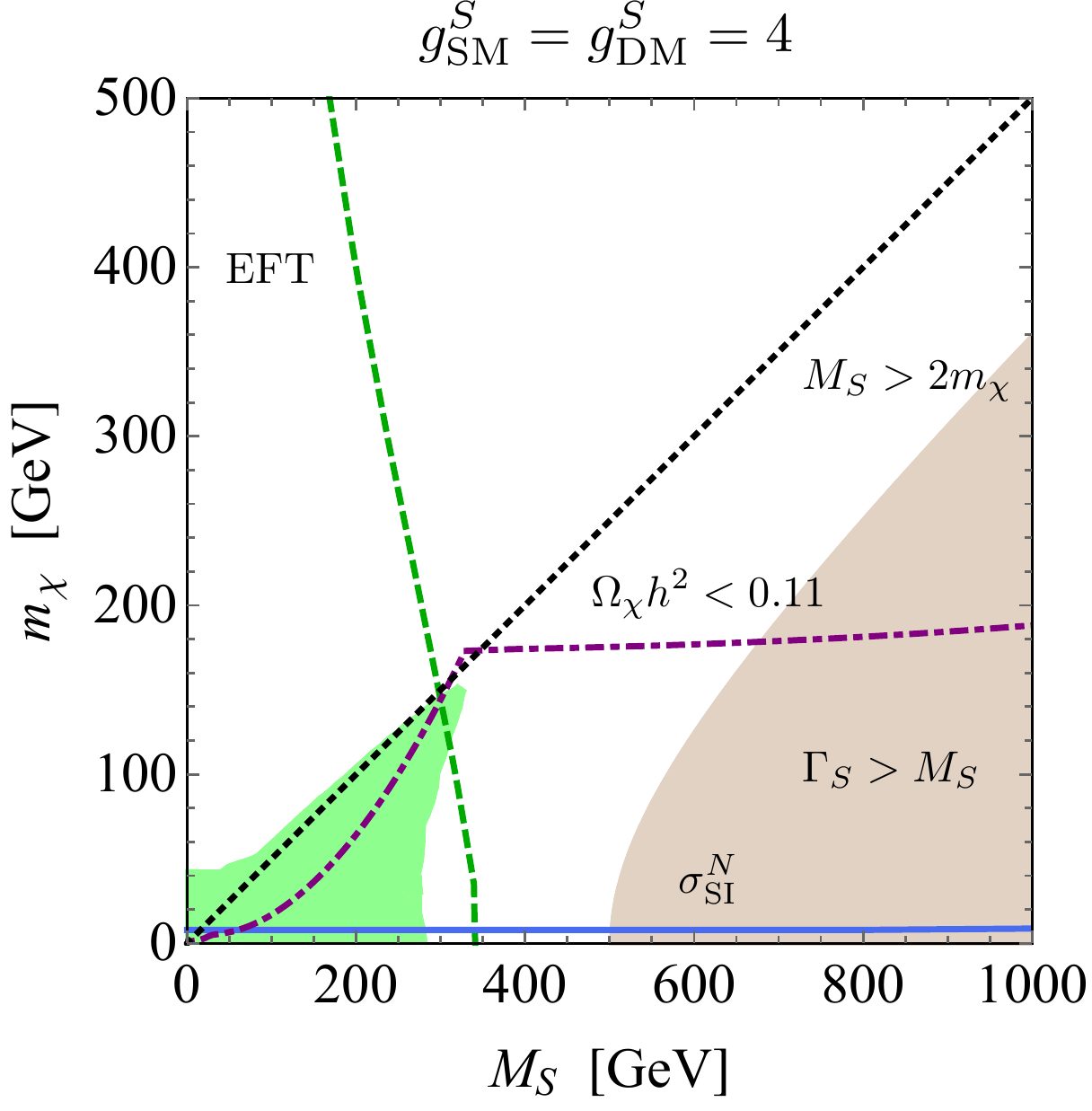} 
\vspace{0mm}
\caption{\label{fig:SplanespresentttMET} Exclusion regions at $95\% \, {\rm CL}$ for scalar mediators following from the present $\slashed{E}_T + \bar t t$ searches in the single-lepton channel (green contours). In the $g_{\rm SM}^S \hspace{0.25mm} $--$\hspace{0.25mm} g_{\rm DM}^S$ plane (upper left panel) the values $m_\chi = 100 \, {\rm GeV}$ and $M_S = 300 \,{\rm GeV}$ have been used, while in the $g_S \hspace{0.25mm} $--$\hspace{0.5mm}  m_\chi$  plane (upper right panel) we have set $g_S = g_{\rm SM}^S = g_{\rm DM}^S$ and $M_S = 300 \, {\rm GeV}$. The results in the $g_S \hspace{0.25mm} $--$\hspace{0.25mm}  M_S$  plane~(lower left panel) use the same couplings and  $m_\chi =100 \, {\rm GeV}$, whereas in the $M_S \hspace{0.25mm}$--$\hspace{0.25mm} m_\chi$ plane~(lower right panel) the couplings have been fixed to $g_{\rm SM}^S = g_{\rm DM}^S =4$. The regions with $\Gamma_S > M_S$~(brown contours), the current LUX constraint on $\sigma_{\rm SI}^N$ (solid blue curves), the parameter spaces with $\Omega_\chi h^2 < 0.11$~(dot-dashed purple curves), the EFT limits~(dashed green curves) and the regions with $M_S > 2 m_\chi$~(dotted black lines) are also shown.}
\end{center}
\end{figure}

Further insights into the limitations of the EFT description of the mono-jet signal can be gained by examining the predictions in the $g_S \hspace{0.25mm} $--$\hspace{0.5mm}  m_\chi$  (upper right panel),  $g_S \hspace{0.25mm} $--$\hspace{0.25mm}  M_S$~(lower left panel) and the $M_S \hspace{0.25mm}$--$\hspace{0.25mm} m_\chi$ planes~(lower right panel). A simple criterion that has been proposed (cf.~\cite{Bai:2010hh,Goodman:2010ku,Fox:2011pm}) to assess the validity of the EFT approach is to demand that $M_S > 2 m_\chi$. To show how this requirement restricts the domain of the EFT limits we have included it into the plots (dotted black lines).  From the $g_S \hspace{0.25mm} $--$\hspace{0.5mm}  m_\chi$ ($g_S \hspace{0.25mm} $--$\hspace{0.25mm}  M_S$) plane, we see that in the region $m_\chi  > 150 \, {\rm GeV}$ ($M_S < 200 \, {\rm GeV}$) off-shell production of DM pairs is numerically important, and as a result the exact exclusion extends far into the parameter region with $M_S < 2 m_\chi$.  In the case of the  $M_S \hspace{0.25mm}$--$\hspace{0.25mm} m_\chi$  plane, on the other hand, one observes that combining the EFT bound with the criterion $M_S > 2 m_\chi$ singles out a slice in parameter space that at least qualitatively resembles the real exclusion contour. This shows that although simple criteria to gauge the applicability of the EFT cannot be used to do precision physics, they are still useful in the sense that they can serve as a sanity check of the calculation in the full theory. We finally remark that  for the same   choice of parameterisation of $g_{\rm DM}^S$ and identical input parameters the shapes and locations of our exact mono-jet exclusion contours in the $M_S \hspace{0.25mm}$--$\hspace{0.25mm} m_\chi$ plane do not resemble the results of \cite{Harris:2014hgav1}.  

\begin{figure}[!t]
\begin{center}
\includegraphics[height=0.45\textwidth]{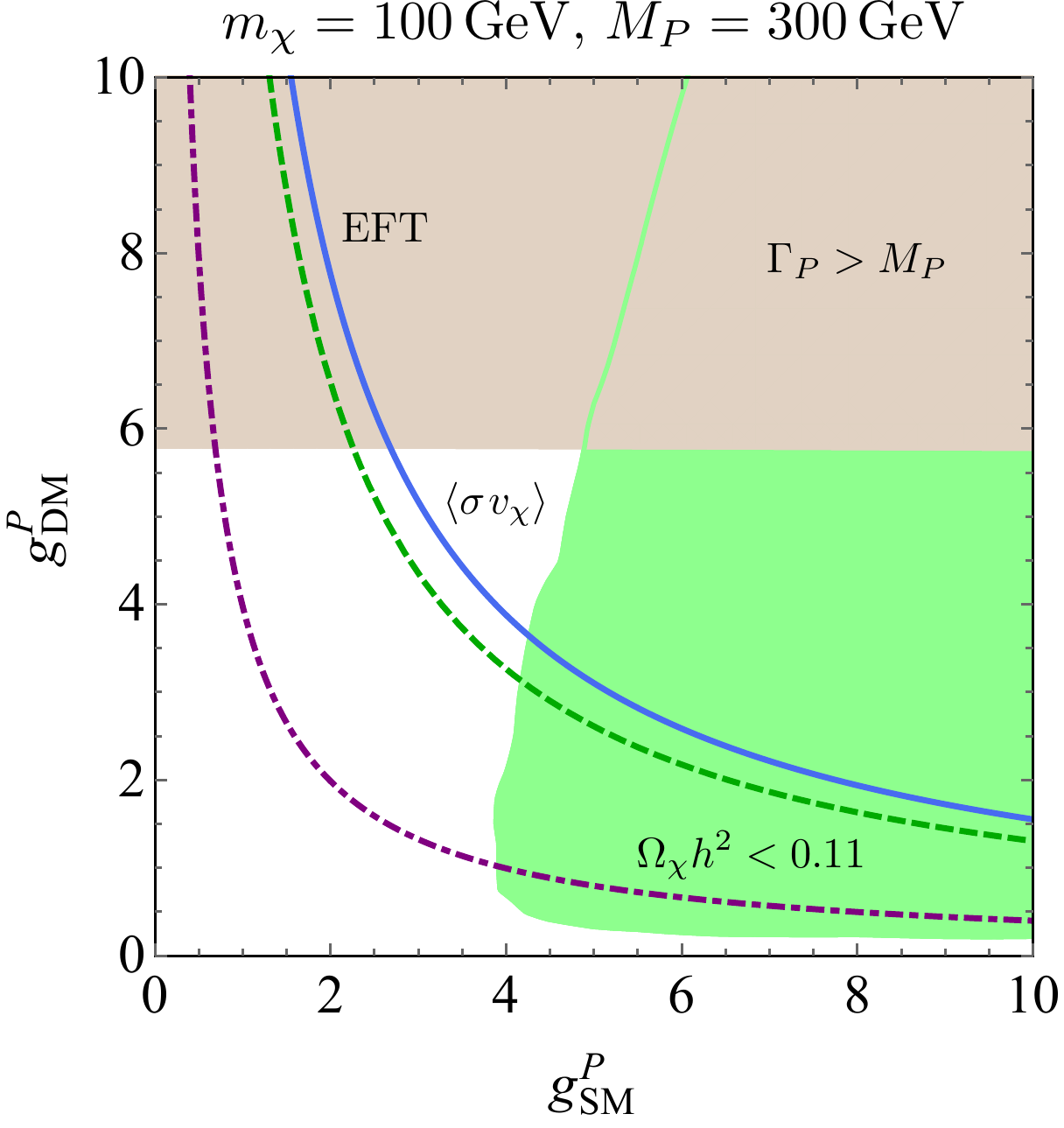} 
\qquad 
\includegraphics[height=0.45\textwidth]{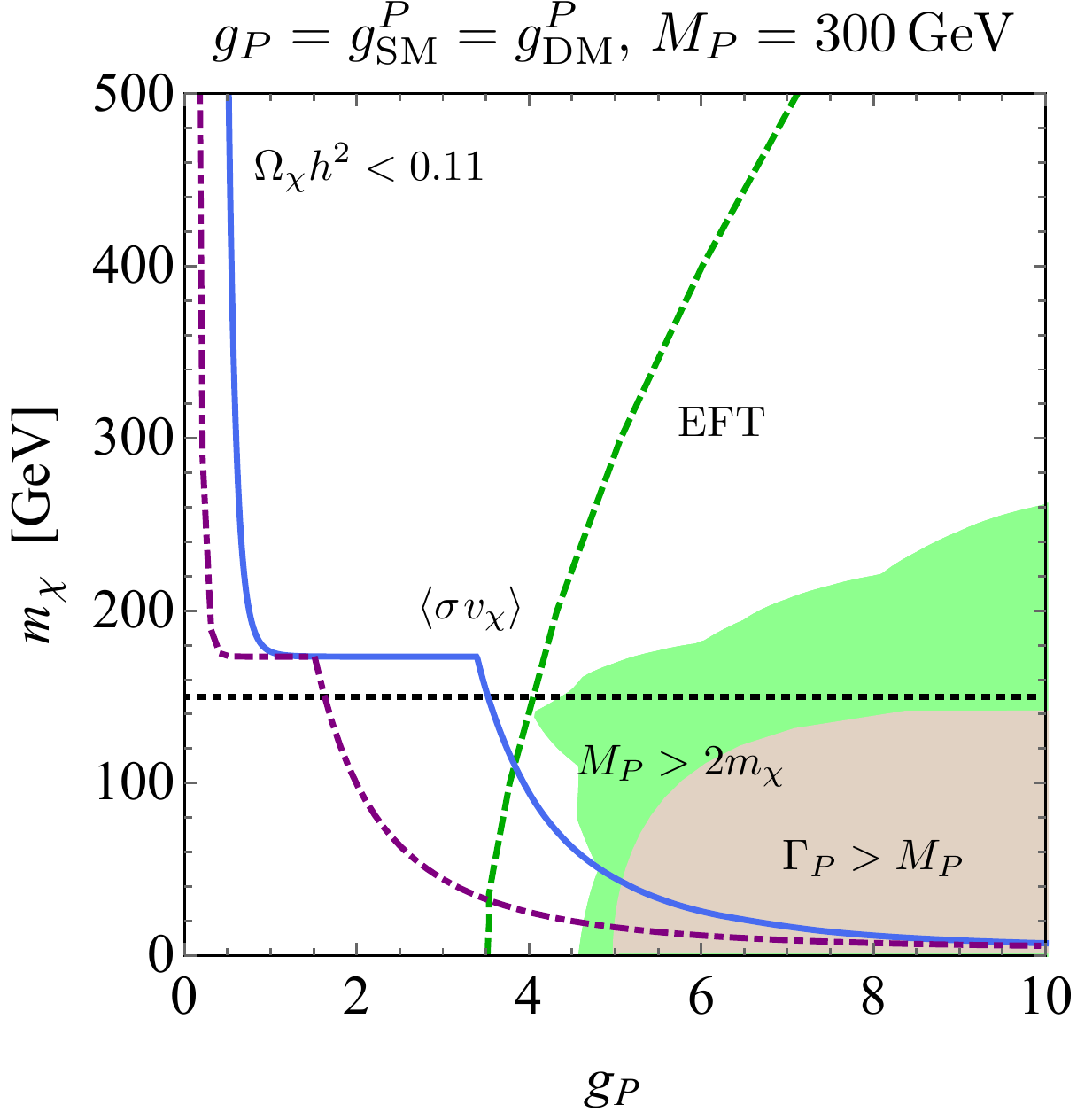} 

\hspace{2mm}

\includegraphics[height=0.45\textwidth]{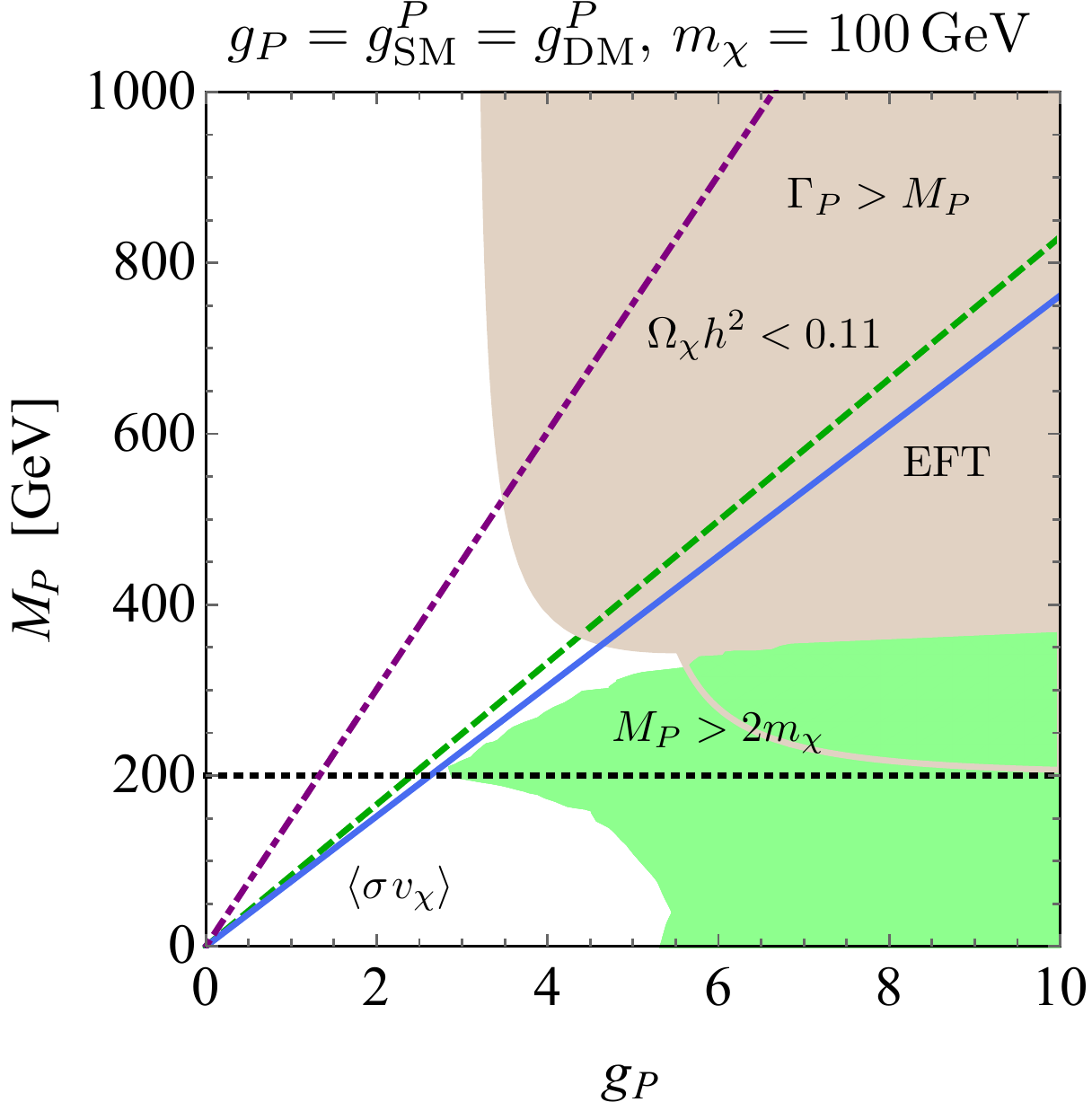} 
\qquad 
\includegraphics[height=0.45\textwidth]{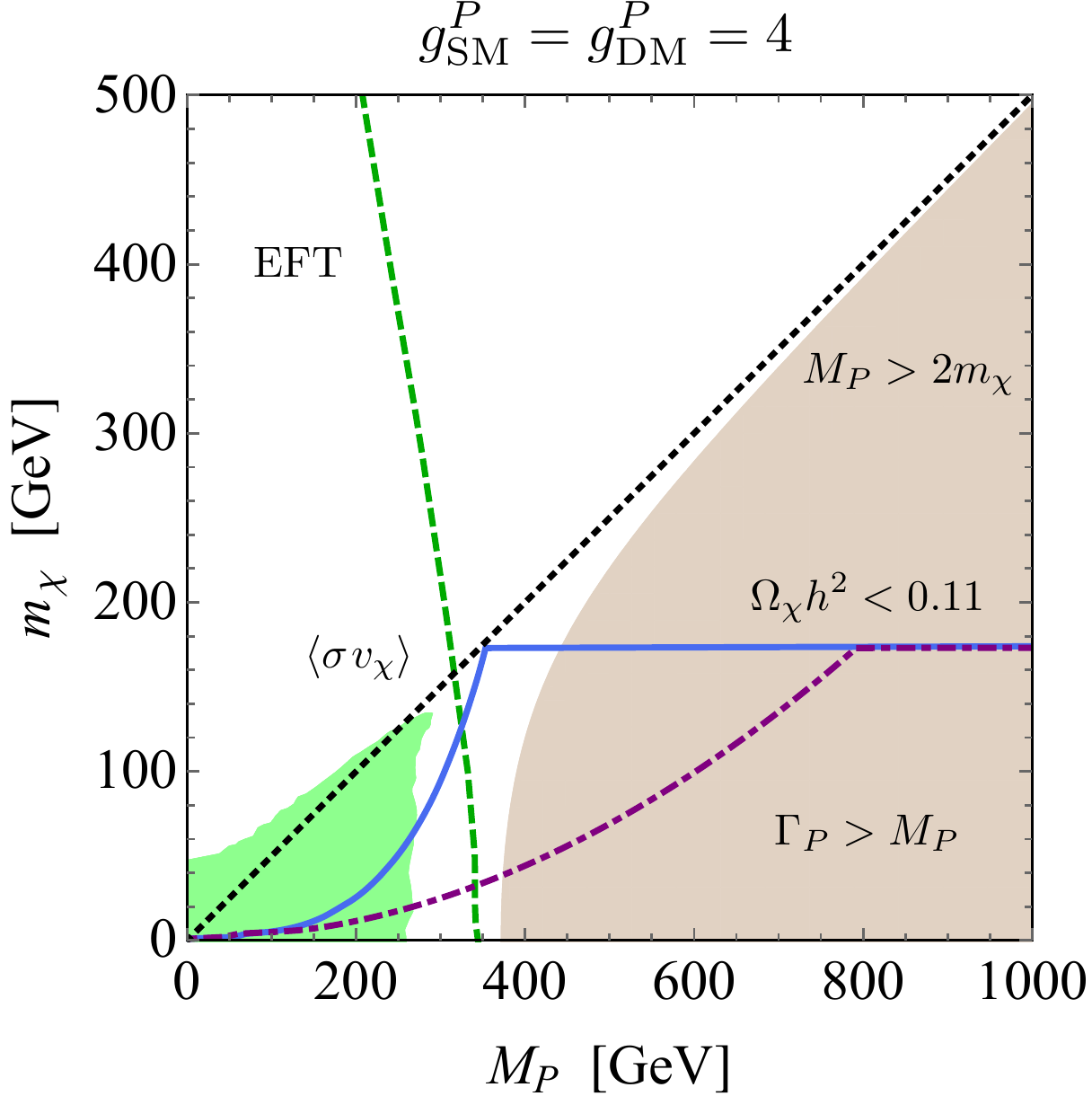} 
\vspace{0mm}
\caption{\label{fig:PplanespresentttMET} Exclusion regions at $95\% \, {\rm CL}$ for pseudo-scalar mediators following from the current $\slashed{E}_T + \bar t t$ searches in the single-lepton final state. The present Fermi-LAT 95\% CL limit on the total velocity-averaged DM annihilation cross section $\left \langle  \sigma \hspace{0.25mm} v_\chi \right \rangle$ is indicated by the solid blue curves. The colour coding and choice of parameters  otherwise resembles the one used in Figure~\ref{fig:SplanespresentttMET}.}
\end{center}
\end{figure}

We now repeat the above exercise for the case of pseudo-scalar $s$-channel exchange. The corresponding results are shown in the panels of Figure~\ref{fig:Pplanespresent}. One observes that the shapes of the exact exclusions (red contours) for pseudo-scalar exchange are quite similar to those found in the scalar case, but that the disfavoured regions are larger in all panels. The reason for the latter feature is twofold. In the regions $M_P > 2 m_\chi$ (dotted black lines) it is a result of (\ref{eq:OGG}), which implies that for the same model parameters the mono-jet cross sections associated to pseudo-scalar exchange are larger than those for scalars by a factor of roughly $2$. For $M_P < 2 m_\chi$, one instead has to take into account that the squared matrix element of $P \to \bar \chi \chi$ scales as $m_{\chi \chi}^2$, whereas for the $S \to \bar \chi \chi$ channel one obtains  $m_{\chi \chi}^2 - 4 m_\chi^2$. The effects of off-shell DM pair production is thus more pronounced if the mediator couples to $\bar \chi \gamma_5 \chi$ rather than to $\bar \chi \chi$. Our results for the mono-jet exclusion regions are again in qualitative (though not  quantitative) agreement with \cite{Harris:2014hgav1}. Two other visible differences are that the regions~(brown contours) where the mediator width exceeds its mass, i.e.~$\Gamma_P > M_P$,  cover more of the shown parameter space, and that direct detection does not provide relevant constraints, because the DM-nucleon scattering cross section is spin-dependent and momentum-suppressed for pseudo-scalar interactions. The leading non-collider constraints arise therefore from the Fermi-LAT measurements  of the $\gamma$-ray flux of dwarf spheroidal satellite galaxies of the Milky Way (solid blue curves). In fact, one observes that combining the LHC constraints with those stemming from $\left \langle  \sigma \hspace{0.25mm} v_\chi \right \rangle$, $\Omega_\chi h^2$ and $\Gamma_P > M_P$ restricts the allowed parameter space visibly (white regions between the solid blue and dot-dashed purple curves). Notice however that the bounds that follow from determinations of the velocity-averaged total DM annihilation cross sections are model dependent. For instance, they can be  weakened significantly in the region $m_\chi < m_t$, if the pseudo-scalar mediator does not couple to down-type quarks, i.e.~by choosing $g^{P}_d = 0$ and $g^{P}_u \neq 0$~$\big($see remark below~(\ref{eq:LSP})$\big)$. Such a choice is compatible with the MFV hypothesis and will not affect the mono-jet limits that result from top-quark loops. 

\subsubsection*{\boldmath Limits from single-lepton $\slashed{E}_T + \bar t t$ channel}
\label{sec:fullpresentttMET}

The restrictions on the simplified scalar and pseudo-scalar mediator models that follow from the bound (\ref{eq:singleleptonpresent}) on the fiducial cross section of $pp \to \slashed{E}_T +  jbl$ are shown in Figure~\ref{fig:SplanespresentttMET} and~\ref{fig:PplanespresentttMET}, respectively. One first observes that the constraints on the four-dimensional parameter space are in the case of the $\slashed{E}_T + \bar tt$ signal very similar for scalar and pseudo-scalar interactions. This is a consequence of the observation made earlier that $\sigma (pp \to S \to \slashed{E}_T + \bar tt) \simeq \sigma (pp \to P \to \slashed{E}_T + \bar tt)$ if $M_S = M_P$, $g^S_{\rm DM}=g^P_{\rm DM}$, $g^S_{\rm SM}=g^P_{\rm SM}$ and~$m_\chi$ is sufficiently light.  From the different panels one can  furthermore see that for scalar interactions the single-lepton $\slashed{E}_T + \bar tt$ channel can provide stronger constraints than the $\slashed{E}_T + j$ searches in regions of parameter space with dominant off-shell production, while for pseudo-scalar mediators this is  not the case. Given the discussion of the EFT limits  in Section~\ref{sec:EFTpresent},  this is an unexpected finding. Qualitatively our observation can be explained  as follows.  First, in the case of the mono-jet signal there is a strong form-factor suppression at work that  originates from the momentum dependence of the top-quark loop amplitudes. As a consequence, the EFT limits typically tend to be too strong when compared to the exact exclusions (see~e.g.~the $M_S \hspace{0.25mm}$--$\hspace{0.25mm} m_\chi$ plane in Figure~\ref{fig:Splanespresent}).  In the case of the  single-lepton $\slashed{E}_T + \bar tt$ channel, on the other hand, one can also find parameter regions where the opposite behaviour is observed (see~e.g.~$M_S \hspace{0.25mm}$--$\hspace{0.25mm} m_\chi$ plane in Figure~\ref{fig:SplanespresentttMET}). This example shows clearly that depending on the dynamic of the considered $\slashed{E}_T$ process an EFT description can lead to both too aggressive and too conservative bounds. In order to determine  the exact exclusions a calculation in a simplified DM model is therefore mandatory.

\begin{figure}[!t]
\begin{center}
\includegraphics[height=0.45\textwidth]{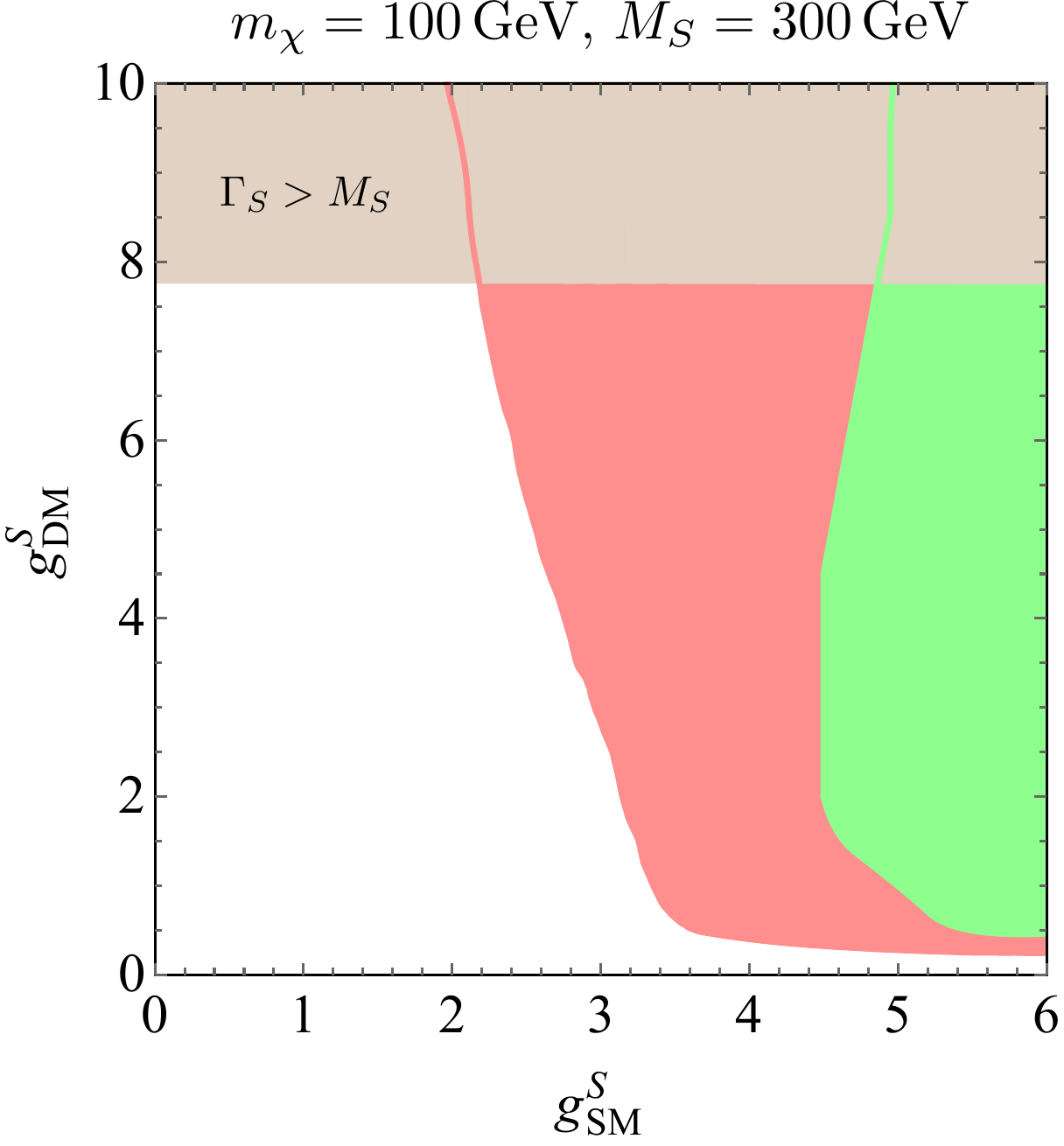} 
\qquad 
\includegraphics[height=0.45\textwidth]{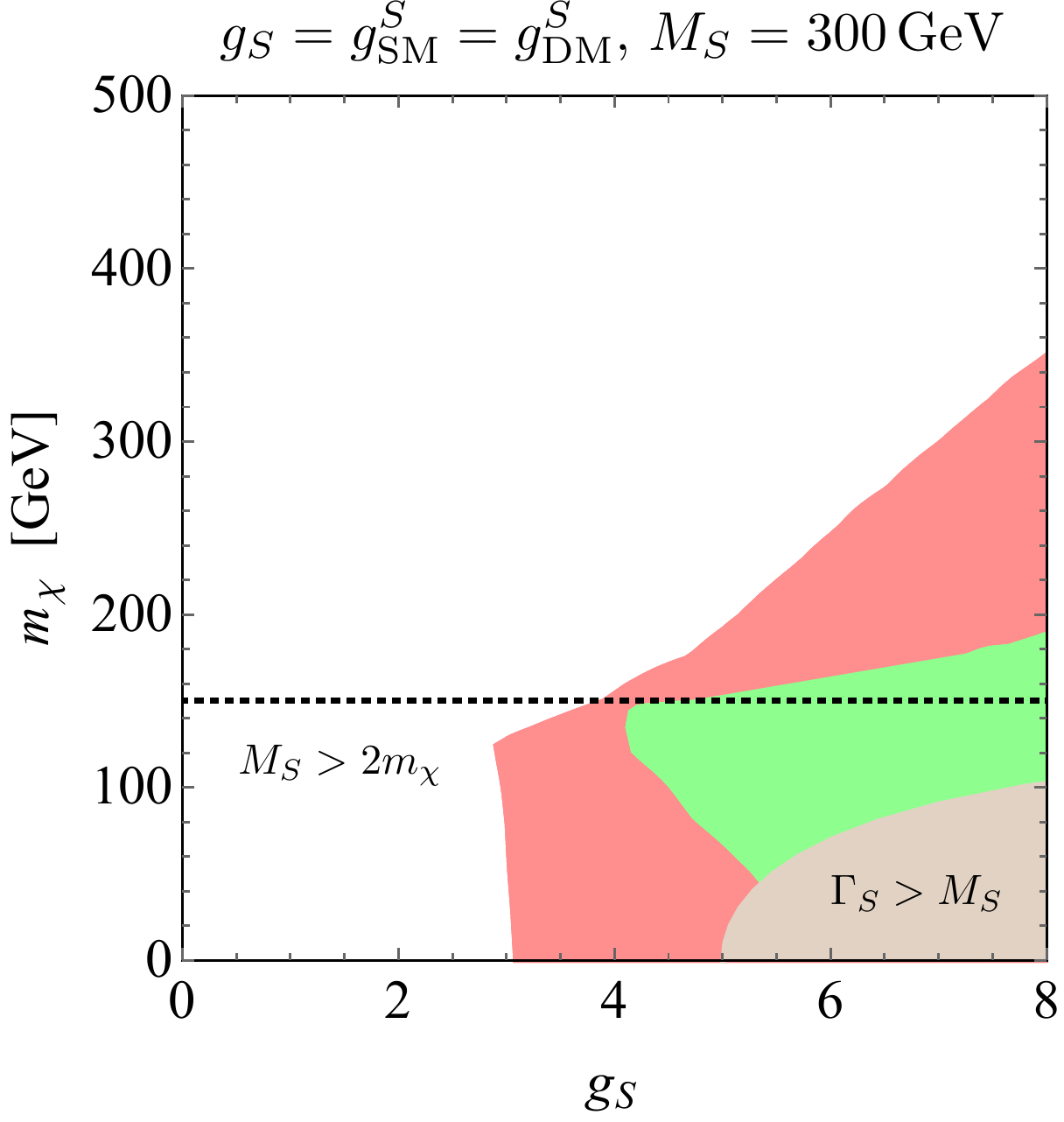} 

\hspace{2mm}

\includegraphics[height=0.45\textwidth]{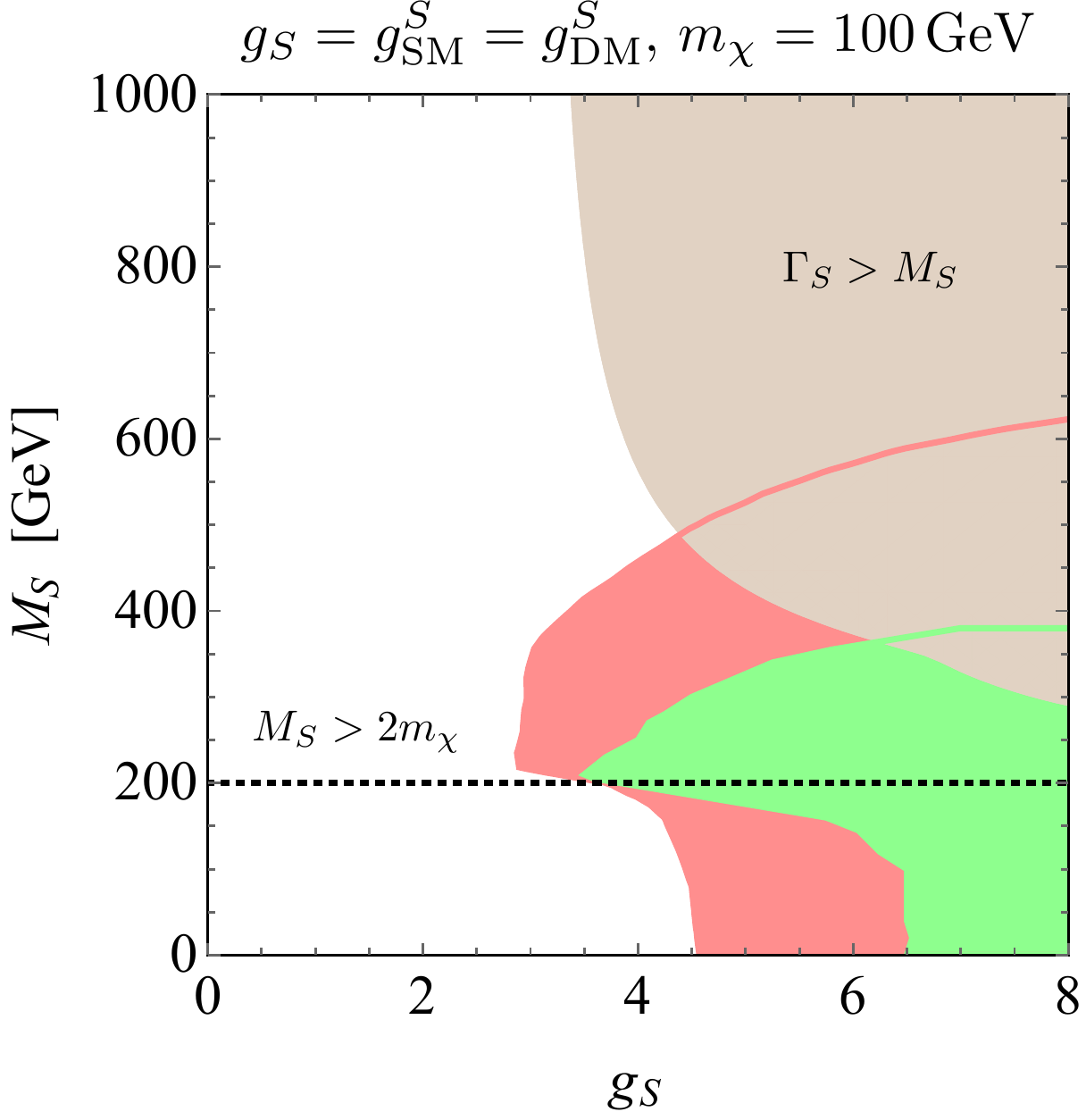} 
\qquad 
\includegraphics[height=0.45\textwidth]{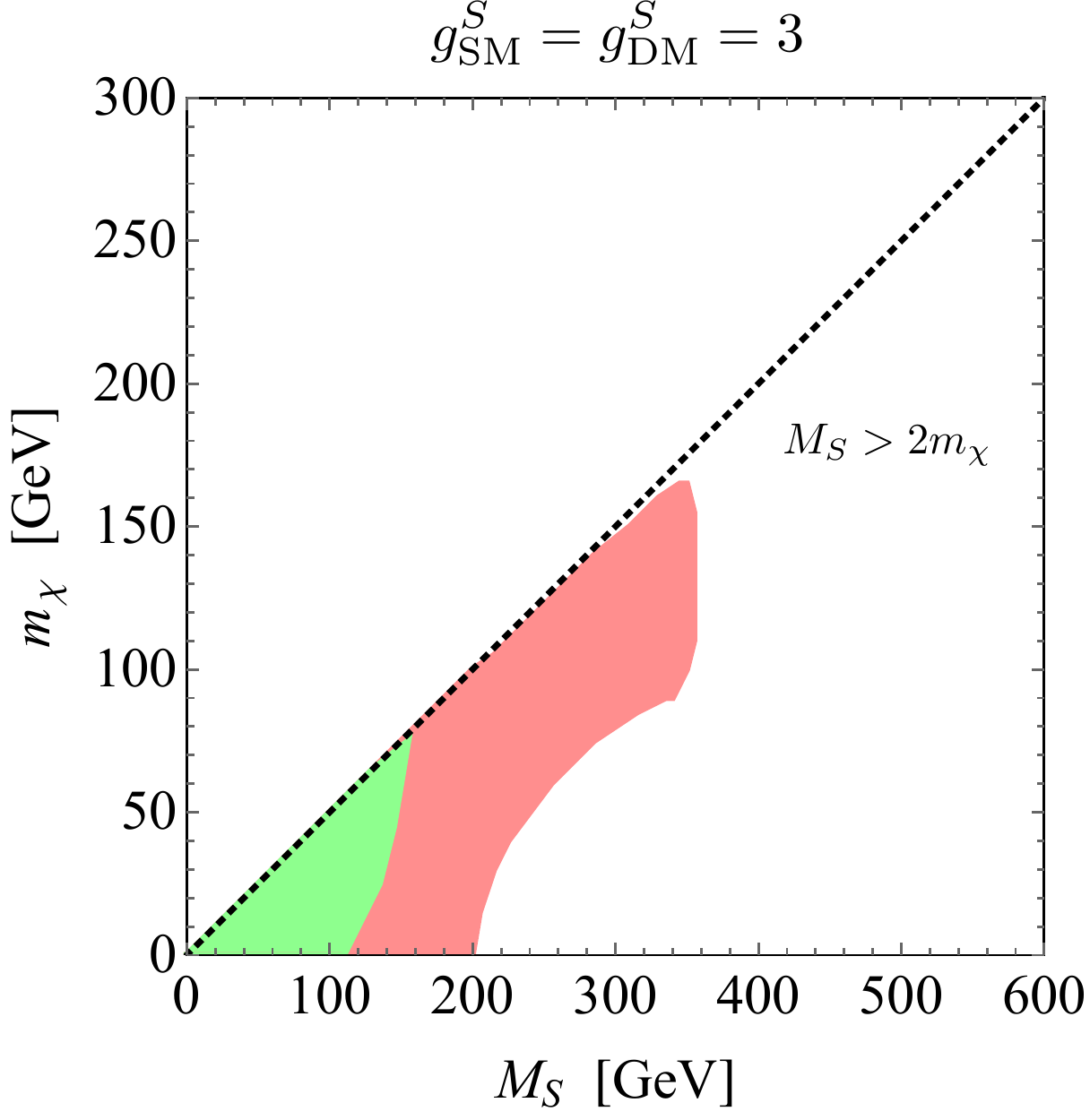} 
\vspace{0mm}
\caption{\label{fig:Sfuture} Exclusion contours at $95\%$~CL for scalar mediators following from hypothetical measurements of a $\slashed{E}_T + j$ signal (red regions) and studies of the $\slashed{E}_T + t \bar t $ single-lepton channel (green regions). The shown predictions assume $25 \, {\rm fb}^{-1}$ of $14 \, {\rm TeV}$ LHC data. In the $g_{\rm SM}^S \hspace{0.25mm} $--$\hspace{0.25mm} g_{\rm DM}^S$ plane (upper left panel), we employed $m_\chi = 100 \, {\rm GeV}$ and $M_S = 300 \,{\rm GeV}$, while in the $g_S \hspace{0.25mm} $--$\hspace{0.5mm}  m_\chi$  plane (upper right panel) we have set $g_S = g_{\rm SM}^S = g_{\rm DM}^S$ and $M_S = 300 \, {\rm GeV}$. The results in the $g_S \hspace{0.25mm} $--$\hspace{0.25mm}  M_S$  plane~(lower left panel) correspond to the same couplings and  $m_\chi =100 \, {\rm GeV}$, while in the $M_S \hspace{0.25mm}$--$\hspace{0.25mm} m_\chi$ plane~(lower right panel) the couplings have been fixed to $g_{\rm SM}^S = g_{\rm DM}^S =3$. The regions with $\Gamma_S > M_S$~(brown contours) and the regions with $M_S > 2 m_\chi$ (dotted black lines) have been indicated for comparison.}
\end{center}
\end{figure}

Another interesting characteristic of  Figures~\ref{fig:Splanespresent} to~\ref{fig:PplanespresentttMET} is that even so the areas of the excluded regions differ  their shapes are quite similar, if one considers the same parameter plane. This feature can be understood by realising that in their present form the $\slashed{E}_T + j$ and $\slashed{E}_T + \bar t t$ searches are simple cut-and-count experiments that measure the total number of  events in the tails of distributions such as the $\slashed{E}_T$ spectrum. Yet, the size of these tails depends to first approximation only on  the overall production rate, but is rather insensitive to the precise form of the DM-SM interactions that lead to a given final state. This implies that while the existing $\slashed{E}_T$ searches are well suited to bound/discover DM, they are unlike to provide enough information to determine further DM properties. For instance,  with the existing $\slashed{E}_T + \bar tt$  searches it is impossible to distinguish a $\slashed{E}_T$ signal arising from $S \to \bar \chi \chi$ or $P \to \bar \chi \chi$. Some of these limitations can however be overcome by studying two-particle (or multi-particle) correlations in processes involving $\slashed{E}_T$ \cite{Haisch:2013fla,Cotta:2012nj,Crivellin:2015wva}. In the case of $\slashed{E}_T + \bar tt$ production, a sensitive probe of the Lorentz structure of the DM-SM interactions is provided by  the pseudo-rapidity difference  $\Delta \eta_{b_1 b_2}$ ($\Delta \eta_{l^+ l^-}$) between the two $b$-jets  (charged leptons) that result from the top-quark decays \cite{UH}. Like in the case of Higgs physics~(see~e.g.~\cite{Ellis:2013yxa,Demartin:2014fia}) studies of the correlations of the top-quark decay products in associated production also offer in the context of a $\slashed{E}_T + \bar tt$ signal unique opportunities to probe the DM mediator  top-quark interactions. Any dedicated effort at  LHC run-2 in this direction is thus more than welcome.  

\subsection{Future sensitivities of DM searches}
\label{sec:fullfuture}

\begin{figure}[!t]
\begin{center}
\includegraphics[height=0.45\textwidth]{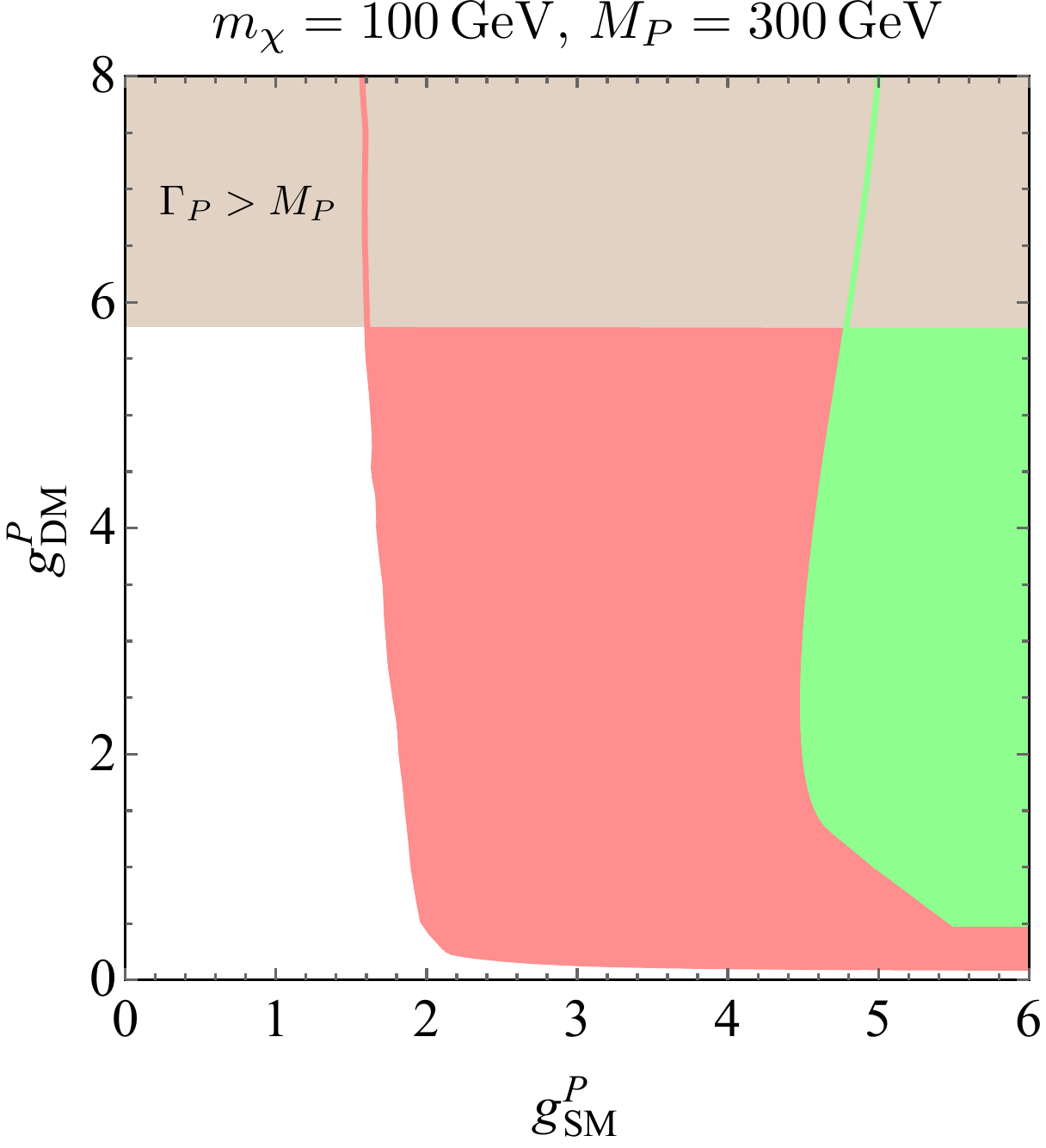} 
\qquad 
\includegraphics[height=0.45\textwidth]{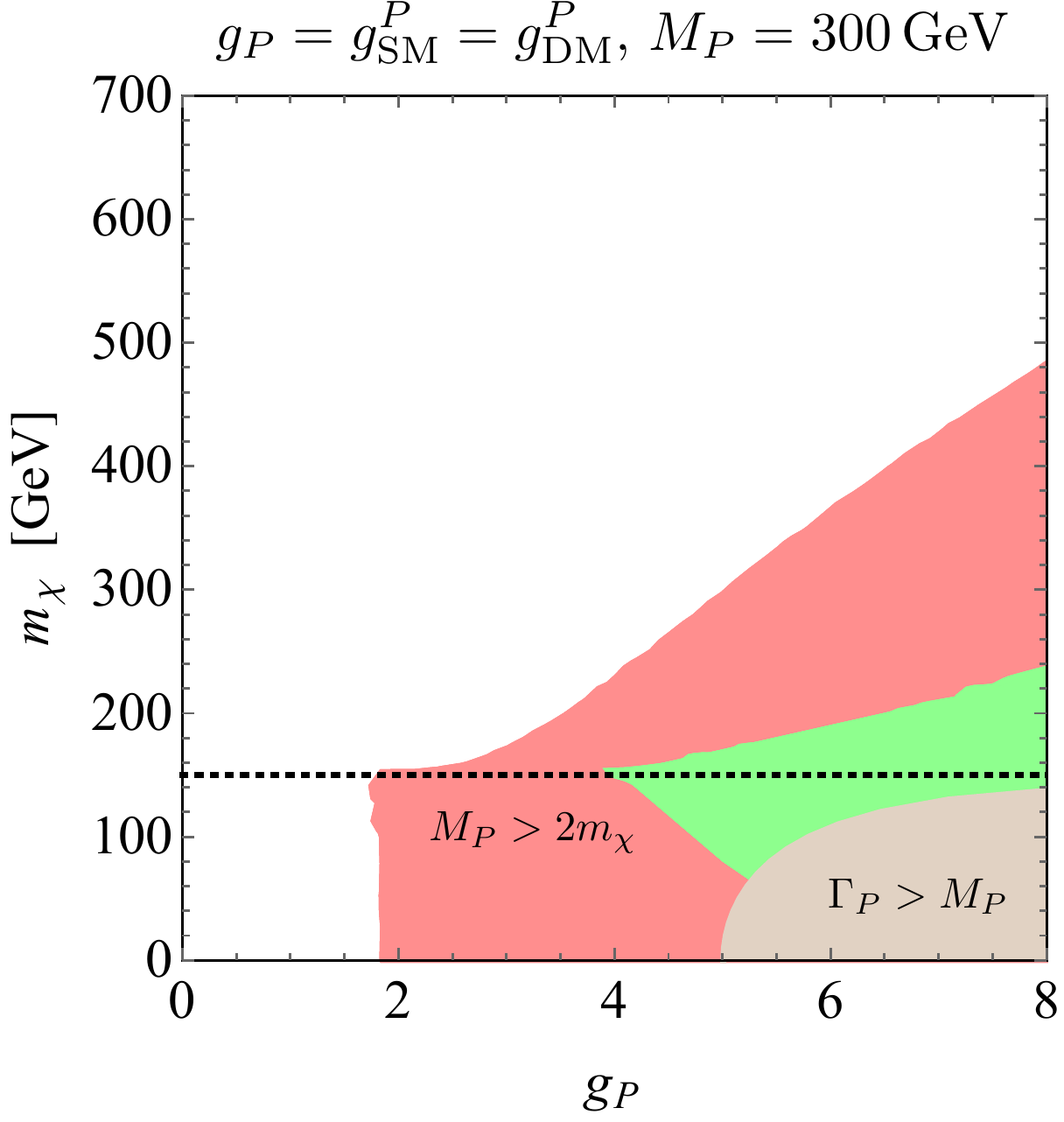} 

\hspace{2mm}

\includegraphics[height=0.45\textwidth]{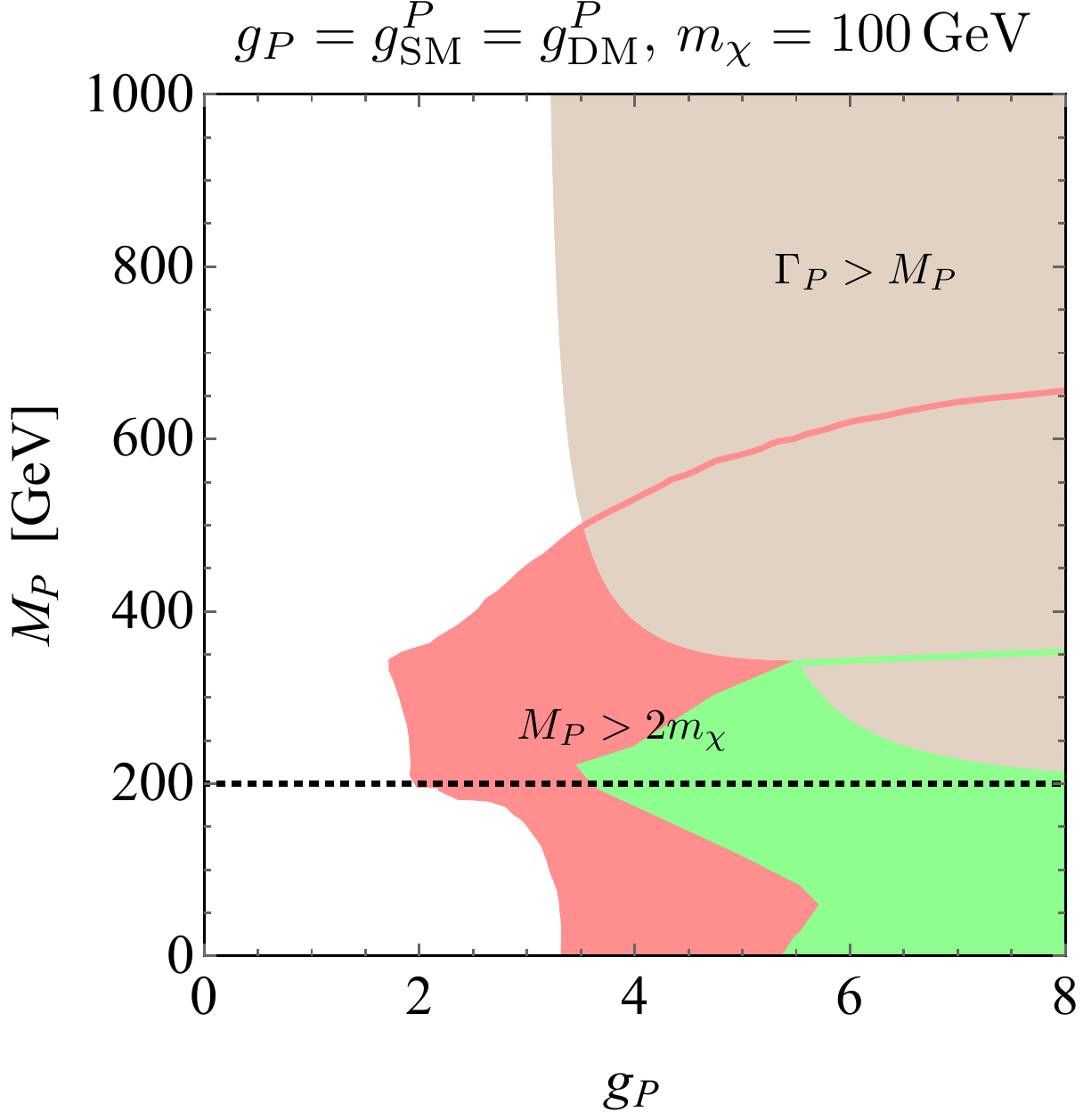} 
\qquad 
\includegraphics[height=0.45\textwidth]{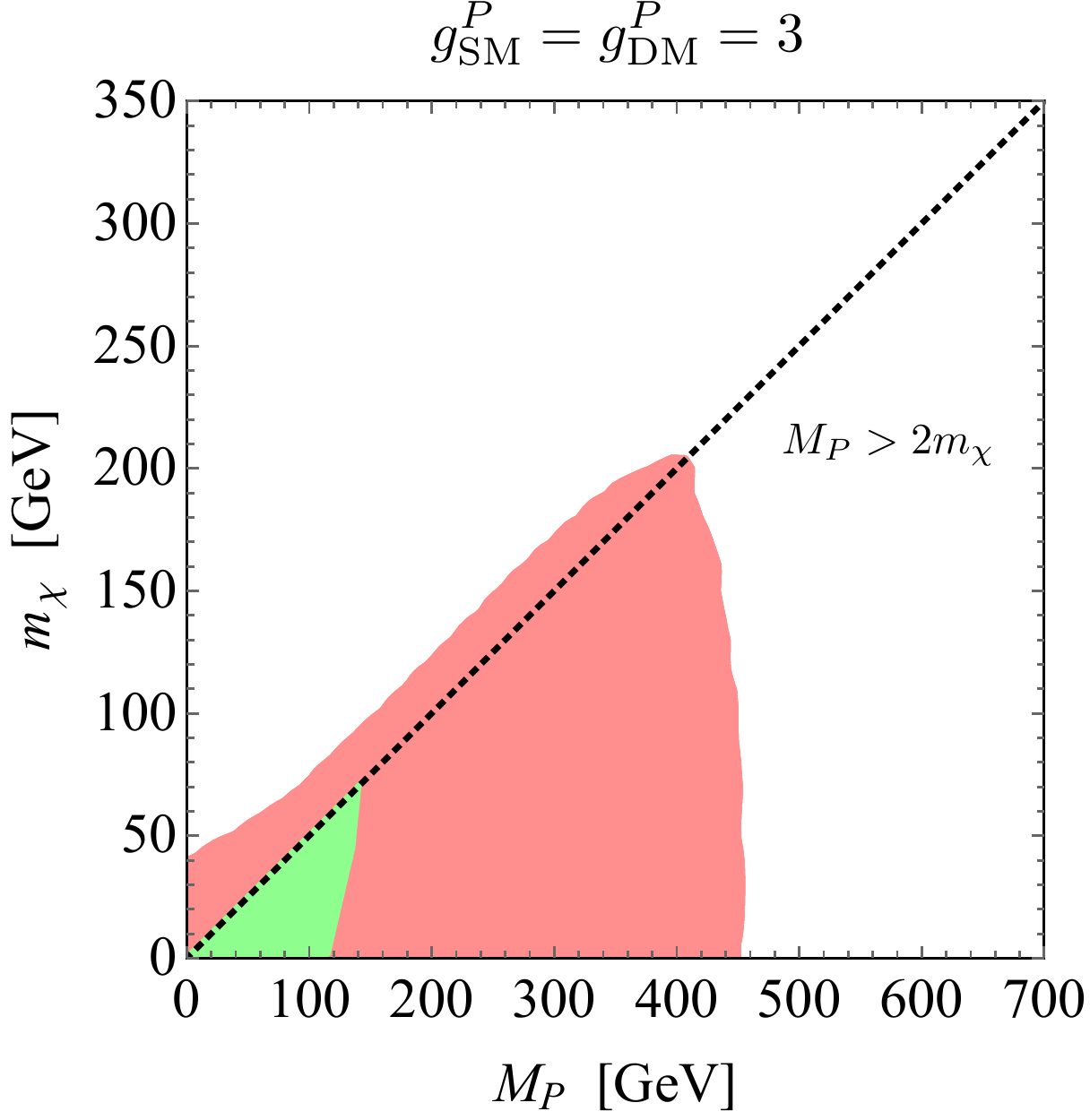} 
\vspace{0mm}
\caption{\label{fig:Pfuture} Exclusion contours at $95\%$~CL for pseudo-scalar mediators following from hypothetical measurements of a $\slashed{E}_T + j$ signal (red regions) and studies of the $\slashed{E}_T + t \bar t $ single-lepton channel (green regions). The shown predictions assume $25 \, {\rm fb}^{-1}$ of $14 \, {\rm TeV}$ LHC data. The colour coding and choice of parameters   is identical to the one used in Figure~\ref{fig:Sfuture}.}
\end{center}
\end{figure}

We finally study how the bounds on the parameter space of the simplified scalar and pseudo-scalar models may improve at future LHC runs. As a baseline for our analysis, we  consider~$25 \, {\rm fb}^{-1}$ of integrated luminosity collected at $14 \, {\rm TeV}$, which corresponds to around one year of data taking. The panels in Figure~\ref{fig:Sfuture}~and~\ref{fig:Pfuture} show our results for $s$-channel scalar and pseudo-scalar mediators, respectively. In all plots the red  and green contours correspond to the 95\%~CL exclusions obtained from (\ref{eq:fidZnunu}) in the case of the $\slashed{E}_T + j$ signal and  (\ref{eq:singleleptonfuturebackground}) for what concerns  the $\slashed{E}_T + \bar tt$  single-lepton signature. To allow for an easy  comparison between  the present and the future constraints, we have employed in the $g_{\rm SM}^{S,P} \hspace{0.25mm} $--$\hspace{0.25mm} g_{\rm DM}^{S,P}$, $g_{S,P} \hspace{0.25mm} $--$\hspace{0.5mm}  m_\chi$ and the $g_{S,P} \hspace{0.25mm} $--$\hspace{0.25mm}  M_{S,P}$  planes the same choice of parameters that has been previously used in Figures~\ref{fig:Splanespresent}~to~\ref{fig:PplanespresentttMET}. One observes that while the shapes of the contours remain qualitatively the same, the bounds that one might be able to set with upcoming data will improve notable compared to the limits obtained at $8 \, {\rm TeV}$. One however also sees that even with $25 \, {\rm fb}^{-1}$ of~$14 \, {\rm TeV}$ data, only model realisations in which the mediators have masses not too far above the weak scale,~i.e.~$M_{S,P} \ll 1 \, {\rm TeV}$, and couple strong enough to the SM,~i.e.~$g_{\rm SM}^{S,P} > 1$, can be explored. This feature is further illustrated by the exclusion contours in the $M_{S} \hspace{0.25mm}$--$\hspace{0.25mm} m_\chi$ and $M_{P} \hspace{0.25mm}$--$\hspace{0.25mm} m_\chi$ planes, which have been obtained for $g_{\rm SM}^S = g_{\rm DM}^S =3$ and $g_{\rm SM}^P = g_{\rm DM}^P =3$.  Notice finally that in our sensitivity study the constraints from the single-lepton $\slashed{E}_T + \bar t t$ channel are in the entire parameter and theory space weaker than the restrictions that derive from the $\slashed{E}_T + j$ signal. As already mentioned at the end of Section~\ref{sec:EFTfuture}, since for realistic cuts  $\sigma_{\rm fid} (pp \to \slashed{E}_T + \bar t t \, (\to jbl\nu))$ is much smaller than $\sigma_{\rm fid} (pp \to \slashed{E}_T + j)$, the $\slashed{E}_T + \bar t t$ channel will only become competitive to the mono-jet signature at the phase-1 and phase-2 upgrades at~$14 \, {\rm TeV}$. Realising that the  existing $\slashed{E}_T + \bar t t  \, (\to jbl\nu)$ analyses are  all recasts of top-squark searches (e.g.~\cite{Aad:2014vea} relies on \cite{Aad:2014kra}), the LHC reach might even be improved further by trying to optimise these searches to the specific topology of the $\slashed{E}_T + \bar t t$ signature arising in simplified scalar and pseudo-scalar models. This issue deserves additional studies. 
 
\section{Conclusions}
\label{sec:conclusions}

Dedicated searches for DM candidates represent an integral part of the physics programme at the LHC. Given our ignorance of the dynamics that may connect the SM to the dark sector, it is important that these searches are as model independent as possible and sensitive to many different types of DM pair production. One  way to achieve this is to employ an EFT in which the SM couples to DM via contact interactions.  In fact, the EFT approach has proven to be useful in the analysis of LHC data, because it allows to derive stringent bounds on effective DM-SM interactions involving light and heavy quarks, gluons, photons, electroweak gauge bosons and the Higgs that can be easily compared to the limits of direct and indirect DM searches. 

In this article, we have studied $\slashed{E}_T$ signatures  that result from interactions between DM and top quarks. Scalar and pseudo-scalar couplings of this type have been searched for in two ways at the LHC. First, by looking for a $pp \to {\slashed E}_T + j$ signal, where the DM pair is emitted from a top-quark loop, and second by trying to detect the top-quark decay products that arise from the  tree-level  transition $pp \to {\slashed E}_T + \bar t t$. Our EFT analysis shows that the strongest constraints that the $8 \, {\rm TeV}$ LHC can place on both scalar and pseudo-scalar DM top-quark interactions come from mono-jet searches. Let us add that DM top-quark couplings can also be probed at the LHC in associated production of a single top quark and a $W$ boson and $t$-channel single-top production. We plan to return to these complementary search strategies in a forthcoming publication. 

Already at $8 \, {\rm TeV}$ the LHC however operates in a regime where the EFT interpretations of the $\slashed{E}_T$ signals often do not apply, and these limitations  will become more severe at a future energy upgrade to $14 \, {\rm TeV}$. In order to derive faithful bounds on  the DM-SM interactions, one has thus turn to simplified models, where the dynamics of the contact terms is resolved. Motivated by this general observation, we have investigated in detail the phenomenology of  the simplified models that give rise to interactions between DM and top-quark pairs through the $s$-channel exchange of a scalar or pseudo-scalar resonance. Under the assumption of MFV, the parameter space of these simplified models is four dimensional and consists of the two couplings $g_{\rm DM}^{S}$ ($g_{\rm DM}^{P}$)   and $g_{\rm SM}^{S}$ ($g_{\rm SM}^{P}$) as well as the DM and the mediator masses $m_\chi$ and $M_{S}$ ($M_P$).  In our analysis the mediator width $\Gamma_S$~($\Gamma_P$) is not treated as a free parameter, but calculated from the remaining parameters. By scanning the four-dimensional parameter spaces, we observe that while the signal cross sections dependent  strongly on $g_{{\rm DM}, {\rm SM}}^{S,P}$, $m_\chi$ and $M_{S,P}$, the ratio between the LOPS and the LO cross section is, for a given signal region, essentially independent of the choice of model parameters. To give an example, in the case of the $8 \, {\rm TeV}$ ($14 \, {\rm TeV}$) mono-jet search, we find a flat ratio close to  $60 \%$ ($45 \%)$ (see Appendix~\ref{app:effectiveMC} for more details). While we have not made use of this feature in our work, this property can be exploited to  efficiently  generate large MC samples  with our {\tt POWHEG~BOX} implementations. 

One main outcome of our comprehensive analysis of the simplified models is that even with the full LHC run-1 data set theories with couplings $g_{\rm SM}^{S,P}$ below 1 cannot be tested. By increasing the LHC centre-of-mass energy to $14 \, {\rm TeV}$, larger parts of the parameter spaces can be explored, but discoveries are still only possible if the mediators have masses of the order of the weak scale and couple sufficiently strong to top-quark pairs. Particles with such properties contribute indirectly also to other observables such as the total $\bar t t$ cross section, the Peskin-Takeuchi parameters and necessarily change the properties of the Higgs boson, which provides further avenues to search for them. We find that while mono-jet searches typically provide the dominant restrictions, in the case of simplified models with sizeable couplings between the scalar mediator and top quarks,  top-quark pair production in association with $\slashed{E}_T$ can allow to better probe parameter regions with dominant off-shell production. In contrast, $\slashed{E}_T + j$ searches are superior to $\slashed{E}_T + \bar t t$ searches for pseudo-scalar $s$-channel exchange in the entire parameter space. This observation shows the complementarity of these two different channels and underpins the importance of performing dedicated searches for $pp \to \slashed{E}_T + \bar t t$ (and likewise $pp \to \slashed{E}_T + \bar b b$) in the upcoming LHC runs. Our exact calculations furthermore allow us to determine under which circumstances an EFT interpretation of LHC bounds is possible. While it turns out that simple criteria like $M_{S,P} > 2 m_\chi$ combined with the EFT bounds typically fail to reproduce the exact exclusions, they still provide enough information to get a first idea about the sensitivity of different $\slashed{E}_T$ channels, which makes the EFT framework a particularly useful tool when designing new search strategies. We finally compared our exact LHC results to the bounds obtained from direct and indirect detection experiments as well as the constraints arising from the thermal DM relic abundance and discussed some of the caveats of such a comparison.

With the start of LHC run-2, collider searches for $\slashed{E}_T$ signatures are soon to explore new territory, and the large statistics expected at the phase-1 and phase-2 upgrades at $14 \, {\rm TeV}$  have the potential to revolutionise our understanding of  DM. New theoretical developments that allow for a better description of both signals and backgrounds  have to go along with the experimental advances  in order to exploit the full physics potential of the LHC. Studies based on the simplified DM models we have discussed here may play a key  role  in this effort. 

\section*{Note added}

After communications with us, the authors of \cite{Harris:2014hgav1} revised their mono-jet analysis. They found that in their original study they  forgot to implement the jet veto, which is supposed to reject events  if they contain more than two jets. In the scalar and pseudo-scalar case this omission results in fiducial cross sections that are too large by at least a factor of~3. After correcting this mistake the results  \cite{Harris:2014hgav2} now seem to be in fair agreement with our findings. Still at close inspections quantitative differences can be observed. For instance, the fact that at $8 \, {\rm TeV}$ the exclusion contours in the $M_{S,P} \hspace{0.25mm}$--$\hspace{0.25mm} m_\chi$ planes as shown in  \cite{Harris:2014hgav2} do not extend up to $M_{S,P} = 2 m_\chi$ calls in our opinion for an explanation. 

\acknowledgments We are grateful to  Chris~McCabe and Felix~Kahlhoefer for useful general discussions and thank Phil~Harris for helpful information concerning  $\slashed{E}_T + \bar t t$ searches. We also thank David~Salek for useful comments on the manuscript and Chris~McCabe for pointing out that the value of the nucleon form factor $f_N$ that we used in version 2 of this work was incorrect.  UH acknowledges the warm hospitality and support of the CERN theory division.

\begin{appendix}

\section{Generation of mono-jet samples on a budget}
\label{app:effectiveMC}

\begin{figure}[!t]
\begin{center}
\includegraphics[height=0.385\textwidth]{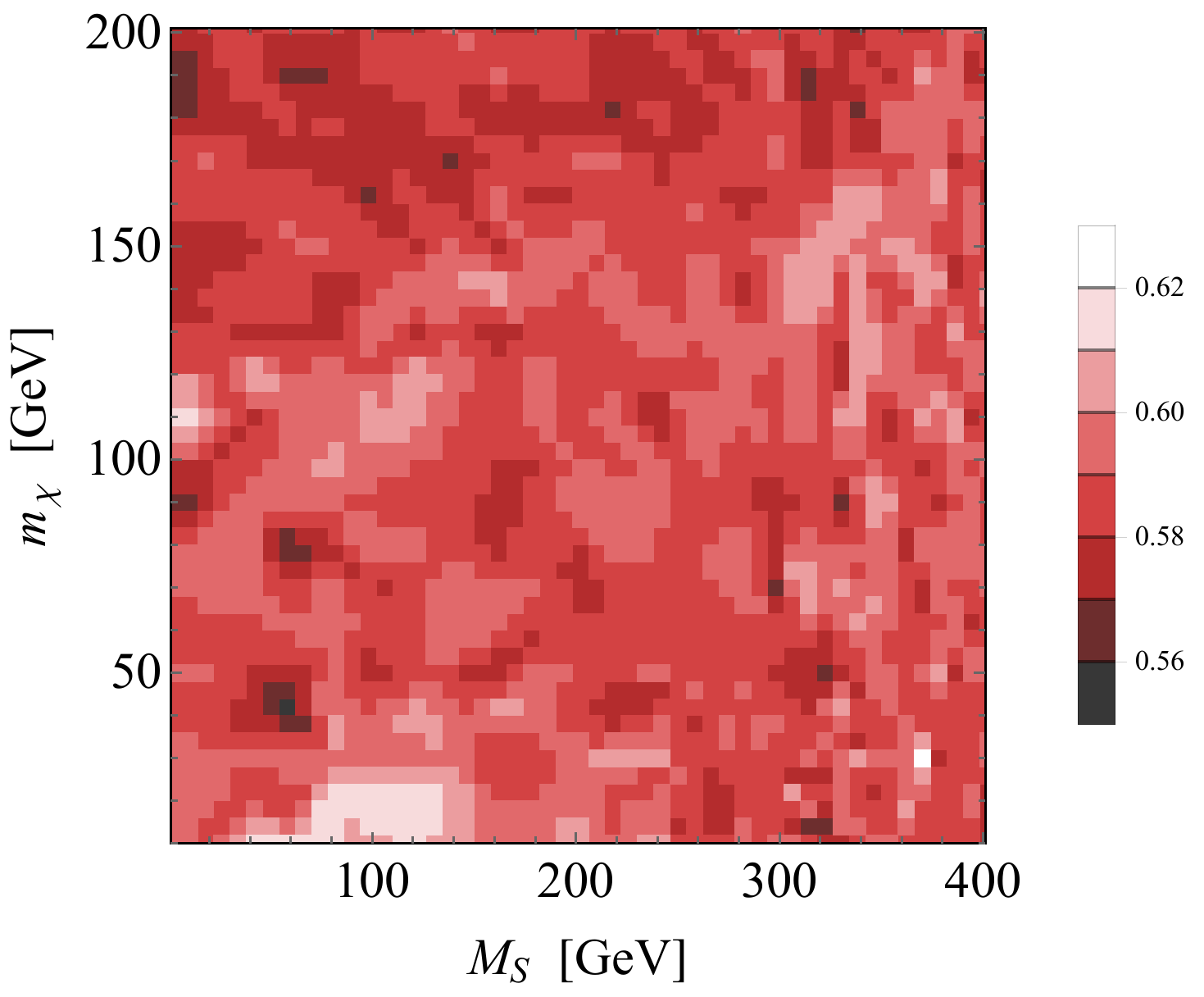}  \qquad 
\includegraphics[height=0.385\textwidth]{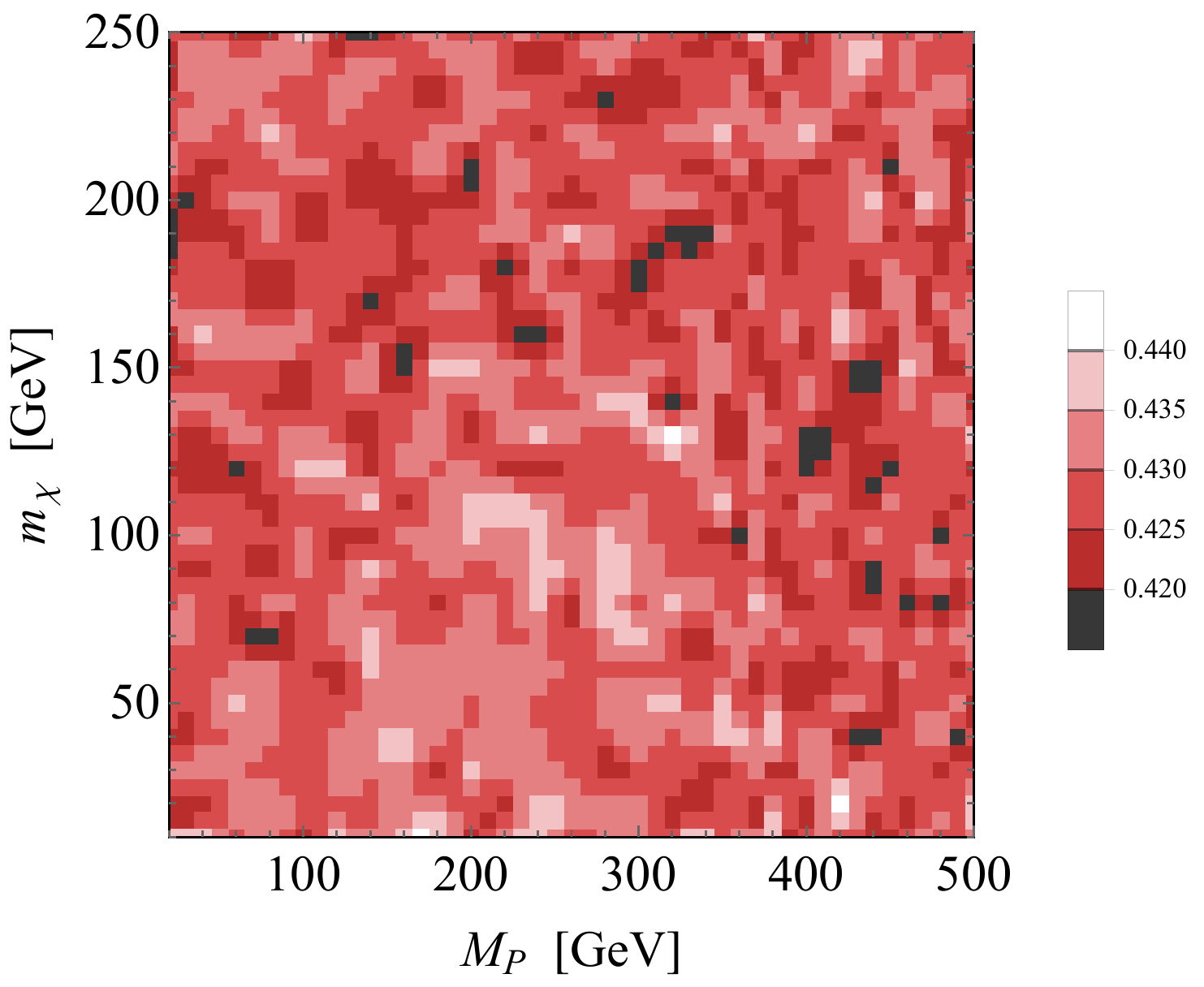} 
\vspace{-2mm}
\caption{\label{fig:efficiencies} Contour plot of the ratio of the LOPS and LO cross sections for mono-jet production as a function of the mediator mass $M_S$ ($M_P$) and the DM mass $m_\chi$. The left (right) panel shows the results for a scalar (pseudo-scalar) mediator at $8 \, {\rm TeV}$ ($14 \, {\rm TeV}$).}
\end{center}
\end{figure}

The studies of the simplified DM models performed in this work require to scan the four-dimensional parameter spaces spanned by $g_{{\rm DM}}^{S,P}$, $g_{{\rm SM}}^{S,P}$, $m_\chi$ and $M_{S,P}$. If the parameter space is sampled brute force, i.e.~by calculating the LOPS  fiducial cross sections for each parameter point, this is a time-consuming task that can be done in a finite amount of time only on a computer farm. It is therefore worthwhile to ask if accurate results for the fiducial cross sections can be obtained without running a full MC chain including a PS  and a detector simulation. In fact,  it is possible to achieve precise results in an efficient way, if one makes use of the observation that the ratio of the LOPS and the LO results for the fiducial cross sections (which is a measure of the acceptance) of simple $\slashed{E}_T$ signatures are to very good approximation independent of the specific choice of model parameters. This feature is illustrated in the two panels of Figure~\ref{fig:efficiencies} for the case of a mono-jet signal arising from our simplified scalar and pseudo-scalar models. On the left-hand (right-hand) side we show  the ratio of the LO and the LOPS cross sections in the $M_S\hspace{0.25mm} $--$\hspace{0.25mm} m_\chi$ ($M_P\hspace{0.25mm} $--$\hspace{0.25mm} m_\chi$) for $\slashed{E}_T + j$ production at the $8 \, {\rm TeV}$ ($14 \, {\rm TeV}$). The signal regions are specified in (\ref{eq:monojetpresent}) and (\ref{eq:monojetfuture}), respectively. Both panels demonstrate clearly that the ratio between the  LOPS and the LO cross sections is essentially flat in parameter space and only depends on the experimental environment,~i.e.~the centre-of-mass energy of the collisions and the event selection requirements that define the fiducial signal regions. Numerically, we find that for the considered searches and the imposed generation cuts the cross section ratios lie in the range $[0.55,0.62]$ and $[0.40,0.46]$, which corresponds to a relative error of $\pm 6\%$ and $\pm 7\%$.  Compared to the scale ambiguities that amount to around $\pm 40\%$ in the case of mono-jet searches the latter uncertainties are hence subleading and can be neglected to first approximation. We have also verified that the above observation applies to the case of a mono-jet signature resulting from vector and axial-vector mediator $s$-channel exchange, which can be simulated at NLO and NLOPS level using the {\tt POWHEG~BOX} implementation presented in \cite{Haisch:2013ata}. Notice that these findings strongly suggest that in the mono-jet  case, the signal acceptance is a rather flat function in both parameter and theory space. 

From the above discussion it should be clear that to achieve accurate results it is sufficient to calculate the ratio between the fiducial cross sections before and after including a PS and a detector simulation for a few parameter points only. The determined ratio can then be used to promote the fixed-order results to the true fiducial cross sections. If this is done the problem of calculating the LOPS (or NLOPS) results boils down to  generating the LO (or NLO) fixed-order  fiducial cross sections efficiently. In the case of the mono-jet signal this can  be done by generating Born-level configurations with a suitable cut on the minimal $p_{T,j_1}$ of the leading jet. Rather than imposing a fixed cut and generating unweighted events  it turns out  to be advantages \cite{Haisch:2013ata,Alioli:2010qp} to generate weighted events of relative weight  $w = (p_{T, \bar \chi \chi}^2 + p_{T, \rm sup}^2)/p_{T, \bar \chi \chi}^2$. Here $p_{T,\bar \chi \chi}$ is the transverse momentum of the DM pair and $p_{T, \rm sup}$ is a parameter that suppresses the generation of phase space points with small $p_{T,\bar \chi \chi}$. In the case of mono-jet searches one has to choose $p_{T, \rm sup}$ sufficiently below the~$\slashed{E}_T$ restriction imposed in the experimental analysis. Using this method to populate the phase space will lead to a rather uniform distribution of events in the entire $p_{T, \bar \chi \chi}$ range, but the LOPS~(or NLOPS) accuracy is recovered, because the few events at low $p_{T, \bar \chi \chi}$ will have a large relative weight $w$. In this way, the correct cross section is reproduced.  To allow for maximal flexibility both methods of sampling the phase space have been implemented in our {\tt POWHEG~BOX} add-on that can simulate the $\slashed{E}_T + j$ signal arising from top-quark loops.

\end{appendix}


\begin{thebibliography}{99}

\bibitem{Askew:2014kqa} 
  A.~Askew, S.~Chauhan, B.~Penning, W.~Shepherd and M.~Tripathi,
  Int.\ J.\ Mod.\ Phys.\ A {\bf 29}, 1430041 (2014)
  [arXiv:1406.5662 [hep-ph]].
  
\bibitem{Bai:2010hh} 
  Y.~Bai, P.~J.~Fox and R.~Harnik,
  JHEP {\bf 1012}, 048 (2010)
  [arXiv:1005.3797 [hep-ph]].
  
\bibitem{Fox:2011pm} 
  P.~J.~Fox, R.~Harnik, J.~Kopp and Y.~Tsai,
  Phys.\ Rev.\ D {\bf 85}, 056011 (2012)
  [arXiv:1109.4398 [hep-ph]].
  
\bibitem{D'Ambrosio:2002ex} 
  G.~D'Ambrosio, G.~F.~Giudice, G.~Isidori and A.~Strumia,
  Nucl.\ Phys.\ B {\bf 645}, 155 (2002)
  [hep-ph/0207036].
  
\bibitem{Abdallah:2014hon} 
  J.~Abdallah, A.~Ashkenazi, A.~Boveia, G.~Busoni, A.~De Simone, C.~Doglioni, A.~Efrati and E.~Etzion {\it et al.},
  arXiv:1409.2893 [hep-ph].
  
\bibitem{Haisch:2012kf} 
  U.~Haisch, F.~Kahlhoefer and J.~Unwin,
  JHEP {\bf 1307}, 125 (2013)
  [arXiv:1208.4605 [hep-ph]].
  
\bibitem{Cheung:2010zf} 
  K.~Cheung, K.~Mawatari, E.~Senaha, P.~Y.~Tseng and T.~C.~Yuan,
  JHEP {\bf 1010}, 081 (2010)
  [arXiv:1009.0618 [hep-ph]].
  
\bibitem{Lin:2013sca} 
  T.~Lin, E.~W.~Kolb and L.~T.~Wang,
  Phys.\ Rev.\ D {\bf 88},  063510 (2013)
  [arXiv:1303.6638 [hep-ph]].
  
\bibitem{Fox:2012ru} 
  P.~J.~Fox and C.~Williams,
  Phys.\ Rev.\ D {\bf 87},  054030 (2013)
  [arXiv:1211.6390 [hep-ph]].
  
\bibitem{Haisch:2013fla} 
  U.~Haisch, A.~Hibbs and E.~Re,
  Phys.\ Rev.\ D {\bf 89}, 034009 (2014)
  [arXiv:1311.7131 [hep-ph]].
  
\bibitem{Buckley:2014fba} 
  M.~R.~Buckley, D.~Feld and D.~Goncalves,
  Phys.\ Rev.\ D {\bf 91}, no. 1, 015017 (2015)
  [arXiv:1410.6497v2 [hep-ph]].
  
\bibitem{Harris:2014hgav1} 
  P.~Harris, V.~V.~Khoze, M.~Spannowsky and C.~Williams,
  arXiv:1411.0535v1 [hep-ph].
  
\bibitem{Artoni:2013zba} 
  G.~Artoni, T.~Lin, B.~Penning, G.~Sciolla and A.~Venturini,
  arXiv:1307.7834 [hep-ex].
  
\bibitem{CMS:2014mxa} 
  CMS Collaboration,
  \href{http://cds.cern.ch/record/1697173/files/B2G-13-004-pas.pdf}{http://cds.cern.ch/record/1697173/files/B2G-13-004-pas.pdf}
  
\bibitem{CMS:2014pvf} 
  CMS Collaboration,
  \href{http://cds.cern.ch/record/1749153/files/B2G-14-004-pas.pdf}{http://cds.cern.ch/record/1749153/files/B2G-14-004-pas.pdf}
  
\bibitem{Aad:2014vea} 
  G.~Aad {\it et al.}  [ATLAS Collaboration],
  Eur.\ Phys.\ J.\ C {\bf 75}, no. 2, 92 (2015)
  [arXiv:1410.4031 [hep-ex]].
  
\bibitem{Batell:2011tc} 
  B.~Batell, J.~Pradler and M.~Spannowsky,
  JHEP {\bf 1108}, 038 (2011)
  [arXiv:1105.1781 [hep-ph]].
  
\bibitem{UH}   
 U.~Haisch, talk at DM @ LHC 2014,  \href{http://indico.cern.ch/event/312657/session/1/contribution/40/material/slides/0.pdf}{http://indico.cern.ch/event/312657/session/1/ contribution/40/material/slides/0.pdf}
  
\bibitem{Czakon:2013goa} 
  M.~Czakon, P.~Fiedler and A.~Mitov,
  Phys.\ Rev.\ Lett.\  {\bf 110}, 252004 (2013)
  [arXiv:1303.6254 [hep-ph]].

\bibitem{Dolan:2014ska} 
  M.~J.~Dolan, C.~McCabe, F.~Kahlhoefer and K.~Schmidt-Hoberg,
  JHEP {\bf 1503}, 171 (2015)
  [arXiv:1412.5174 [hep-ph]].
   
\bibitem{Shifman:1978zn} 
  M.~A.~Shifman, A.~I.~Vainshtein and V.~I.~Zakharov,
  Phys.\ Lett.\ B {\bf 78}, 443 (1978).
  
 \bibitem{Haisch:2013ata} 
  U.~Haisch, F.~Kahlhoefer and E.~Re,
  JHEP {\bf 1312}, 007 (2013)
  [arXiv:1310.4491 [hep-ph]].
  
\bibitem{Crivellin:2013ipa} 
  A.~Crivellin, M.~Hoferichter and M.~Procura,
  Phys.\ Rev.\ D {\bf 89}, 054021 (2014)
  [arXiv:1312.4951 [hep-ph]].
  
\bibitem{Hinshaw:2012aka} 
  G.~Hinshaw {\it et al.}  [WMAP Collaboration],
  Astrophys.\ J.\ Suppl.\  {\bf 208}, 19 (2013)
  [arXiv:1212.5226 [astro-ph.CO]].
  
\bibitem{Kolb:1990vq} 
  E.~W.~Kolb and M.~S.~Turner,
  Front.\ Phys.\  {\bf 69}, 1 (1990).
  
\bibitem{Alioli:2010xd} 
  S.~Alioli, P.~Nason, C.~Oleari and E.~Re,
  JHEP {\bf 1006}, 043 (2010)
  [arXiv:1002.2581 [hep-ph]].
  
\bibitem{MCFM} 
J.~Campbell, R.~K.~Ellis and C.~Williams, \href{http://mcfm.fnal.gov} {http://mcfm.fnal.gov}   
  
\bibitem{Ellis:1987xu} 
  R.~K.~Ellis, I.~Hinchliffe, M.~Soldate and J.~J.~van der Bij,
  Nucl.\ Phys.\ B {\bf 297}, 221 (1988).
  
\bibitem{Spira:1995rr} 
  M.~Spira, A.~Djouadi, D.~Graudenz and P.~M.~Zerwas,
  Nucl.\ Phys.\ B {\bf 453}, 17 (1995)
  [hep-ph/9504378].
  
\bibitem{Alloul:2013bka} 
  A.~Alloul, N.~D.~Christensen, C.~Degrande, C.~Duhr and B.~Fuks,
  Comput.\ Phys.\ Commun.\  {\bf 185}, 2250 (2014)
  [arXiv:1310.1921 [hep-ph]].
  
\bibitem{Degrande:2011ua} 
  C.~Degrande, C.~Duhr, B.~Fuks, D.~Grellscheid, O.~Mattelaer and T.~Reiter,
  Comput.\ Phys.\ Commun.\  {\bf 183}, 1201 (2012)
  [arXiv:1108.2040 [hep-ph]].
    
\bibitem{Alwall:2011uj} 
  J.~Alwall, M.~Herquet, F.~Maltoni, O.~Mattelaer and T.~Stelzer,
  JHEP {\bf 1106}, 128 (2011)
  [arXiv:1106.0522 [hep-ph]].
  
\bibitem{Sjostrand:2006za} 
  T.~Sjostrand, S.~Mrenna and P.~Z.~Skands,
  JHEP {\bf 0605}, 026 (2006)
  [hep-ph/0603175].
  
\bibitem{Cacciari:2008gp} 
  M.~Cacciari, G.~P.~Salam and G.~Soyez,
  JHEP {\bf 0804}, 063 (2008)
  [arXiv:0802.1189 [hep-ph]].
  
\bibitem{Cacciari:2011ma} 
  M.~Cacciari, G.~P.~Salam and G.~Soyez,
  Eur.\ Phys.\ J.\ C {\bf 72}, 1896 (2012)
  [arXiv:1111.6097 [hep-ph]].
  
\bibitem{Ravindran:2002dc} 
  V.~Ravindran, J.~Smith and W.~L.~Van Neerven,
  Nucl.\ Phys.\ B {\bf 634}, 247 (2002)
  [hep-ph/0201114].
  
\bibitem{Field:2002pb} 
  B.~Field, J.~Smith, M.~E.~Tejeda-Yeomans and W.~L.~van Neerven,
  Phys.\ Lett.\ B {\bf 551}, 137 (2003)
  [hep-ph/0210369].
  
\bibitem{Beenakker:2002nc} 
  W.~Beenakker, S.~Dittmaier, M.~Kr\"amer, B.~Plumper, M.~Spira and P.~M.~Zerwas,
  Nucl.\ Phys.\ B {\bf 653}, 151 (2003)
  [hep-ph/0211352].
  
\bibitem{Dawson:2002tg} 
  S.~Dawson, L.~H.~Orr, L.~Reina and D.~Wackeroth,
  Phys.\ Rev.\ D {\bf 67}, 071503 (2003)
  [hep-ph/0211438].
  
\bibitem{Frederix:2011zi} 
  R.~Frederix, S.~Frixione, V.~Hirschi, F.~Maltoni, R.~Pittau and P.~Torrielli,
  Phys.\ Lett.\ B {\bf 701}, 427 (2011)
  [arXiv:1104.5613 [hep-ph]].
  
\bibitem{Martin:2009iq} 
  A.~D.~Martin, W.~J.~Stirling, R.~S.~Thorne and G.~Watt,
  Eur.\ Phys.\ J.\ C {\bf 63}, 189 (2009)
  [arXiv:0901.0002 [hep-ph]].
  
\bibitem{Khachatryan:2014rra} 
  V.~Khachatryan {\it et al.}  [CMS Collaboration],
  arXiv:1408.3583 [hep-ex].
  
\bibitem{Aad:2014kra} 
  G.~Aad {\it et al.}  [ATLAS Collaboration],
  JHEP {\bf 1411}, 118 (2014)
  [arXiv:1407.0583 [hep-ex]].
  
\bibitem{ATLAS:2012maq} 
  ATLAS Collaboration, 
  \href{http://cds.cern.ch/record/1497732/files/ATLAS-CONF-2012-166.pdf}{http://cds.cern.ch/record/1497732/files/ATLAS-CONF-2012-166.pdf}  
  
\bibitem{Chatrchyan:2013xna} 
  S.~Chatrchyan {\it et al.}  [CMS Collaboration],
  Eur.\ Phys.\ J.\ C {\bf 73}, 2677 (2013)
  [arXiv:1308.1586 [hep-ex]].
  
\bibitem{Barr:2009jv}
  A.~J.~Barr, B.~Gripaios and C.~G.~Lester,
  JHEP {\bf 0911} (2009) 096
  [arXiv:0908.3779 [hep-ph]].
  
\bibitem{Konar:2009qr} 
  P.~Konar, K.~Kong, K.~T.~Matchev and M.~Park,
  JHEP {\bf 1004}, 086 (2010)
  [arXiv:0911.4126 [hep-ph]].
  
\bibitem{Bai:2012gs} 
  Y.~Bai, H.~C.~Cheng, J.~Gallicchio and J.~Gu,
  JHEP {\bf 1207}, 110 (2012)
  [arXiv:1203.4813 [hep-ph]].
  
\bibitem{ATL-COM-PHYS-2014-549}   
ATLAS Collaboration, \href{http://cds.cern.ch/record/1708859/files/ATL-COM-PHYS-2014-549.pdf}{http://cds.cern.ch/record/1708859/files/ATL-COM-PHYS-2014- 549.pdf}  

\bibitem{ATL-PHYS-PUB-2013-011} 
ATLAS Collaboration, \href{http://cds.cern.ch/record/1604505/files/ATL-PHYS-PUB-2013-011.pdf}{http://cds.cern.ch/record/1604505/files/ATL-PHYS-PUB-2013- 011.pdf}

\bibitem{Akerib:2013tjd}
  D.~S.~Akerib {\it et al.}  [LUX Collaboration],
  Phys.\ Rev.\ Lett.\  {\bf 112} (2014) 9,  091303
  [arXiv:1310.8214 [astro-ph.CO]].
  
\bibitem{Ackermann:2013yva} 
  M.~Ackermann {\it et al.}  [Fermi-LAT Collaboration],
  Phys.\ Rev.\ D {\bf 89}, 042001 (2014)
  [arXiv:1310.0828 [astro-ph.HE]].
  
\bibitem{Buchmueller:2014yoa} 
  O.~Buchmueller, M.~J.~Dolan, S.~A.~Malik and C.~McCabe,
  JHEP {\bf 1501}, 037 (2015)
  [arXiv:1407.8257 [hep-ph]].
  
\bibitem{Goodman:2010ku} 
  J.~Goodman, M.~Ibe, A.~Rajaraman, W.~Shepherd, T.~M.~P.~Tait and H.~B.~Yu,
  Phys.\ Rev.\ D {\bf 82}, 116010 (2010)
  [arXiv:1008.1783 [hep-ph]].
   
\bibitem{Cotta:2012nj} 
  R.~C.~Cotta, J.~L.~Hewett, M.~P.~Le and T.~G.~Rizzo,
  Phys.\ Rev.\ D {\bf 88}, 116009 (2013)
  [arXiv:1210.0525 [hep-ph]].
  
\bibitem{Crivellin:2015wva} 
  A.~Crivellin, U.~Haisch and A.~Hibbs,
  Phys.\ Rev.\ D {\bf 91}, 074028 (2015)
  [arXiv:1501.00907 [hep-ph]].
  
\bibitem{Ellis:2013yxa} 
  J.~Ellis, D.~S.~Hwang, K.~Sakurai and M.~Takeuchi,
  JHEP {\bf 1404}, 004 (2014)
  [arXiv:1312.5736 [hep-ph]].
  
\bibitem{Demartin:2014fia} 
  F.~Demartin, F.~Maltoni, K.~Mawatari, B.~Page and M.~Zaro,
  Eur.\ Phys.\ J.\ C {\bf 74}, 3065 (2014)
  [arXiv:1407.5089 [hep-ph]].

\bibitem{Harris:2014hgav2} 
  P.~Harris, V.~V.~Khoze, M.~Spannowsky and C.~Williams,
  Phys.\ Rev.\ D {\bf 91}, no. 5, 055009 (2015)
  [arXiv:1411.0535v2 [hep-ph]].
  
\bibitem{Alioli:2010qp} 
  S.~Alioli, P.~Nason, C.~Oleari and E.~Re,
  JHEP {\bf 1101}, 095 (2011)
  [arXiv:1009.5594 [hep-ph]].
  
\end{thebibliography}
\end{document}